\documentclass[11pt]{article}
\input{setup.sty}

\title{Axion Wormholes and the AdS/CFT Factorization Problem}
\author[a]{Jesse Held,} \emailAdd{jheld@ucsb.edu}
\author[b]{Molly Kaplan,} \emailAdd{molly.kaplan@inria.fr}
\author[a]{Donald Marolf,} \emailAdd{marolf@ucsb.edu}
\author[c]{Zhencheng Wang} \emailAdd{zcwang1@illinois.edu}

\affiliation[a]{Department of Physics, University of California, Santa Barbara, CA 93106, USA}
\affiliation[b]{Laboratoire de Physique de l’École Normale Supérieure, Mines Paris, Inria, CNRS, ENS-PSL, Centre Automatique et Systèmes (CAS),
Sorbonne Université, PSL Research University, Paris, France}
\affiliation[c]{Department of Physics, University of Illinois Urbana-Champaign, Urbana, IL 61801, USA}

\abstract{This work investigates the relevance of Euclidean and complex axion wormholes to the AdS/CFT factorization problem.  We use a framework that defines bulk gravitational path integrals by integrating over a real Lorentz-signature contour and then, as needed, perhaps further analytically continuing the resulting functions of boundary conditions. For technical reasons we focus on the case of 2+1 bulk dimensions.  The AdS boundary conditions (in any dimension) require us to impose Dirichlet boundary conditions on the standard Euclidean axion $\chi_E$. Fixing its asymptotic values on two boundary spheres to $\pm \chi_{E,\infty}$, we find
such wormholes to be subdominant to a UV-sensitive endpoint contribution for $\chi_{E, \infty}$ near the real axis, and that (with our conventions) they become dominant only for $\chi_{E, \infty}$ near the negative imaginary axis. Furthermore, such wormholes are irrelevant to our computation for ${\rm Im} \chi_{E, \infty} >0$ (in the sense that the associated ascent contour fails to intersect the contour of integration).  The relevance of the wormhole saddle for real positive $\chi_{E, \infty}$ is in fact a matter of choice, as the saddle then lies on a Stokes' line at which the relevant intersection number changes from zero to one.
}

\begin{document}
%\input{editionlegend.tex}

% Title Page
\maketitle

\section{Introduction}
\label{sec:intro}

\begin{figure}[!t]
    \centering
    \includegraphics[width=0.85\linewidth]{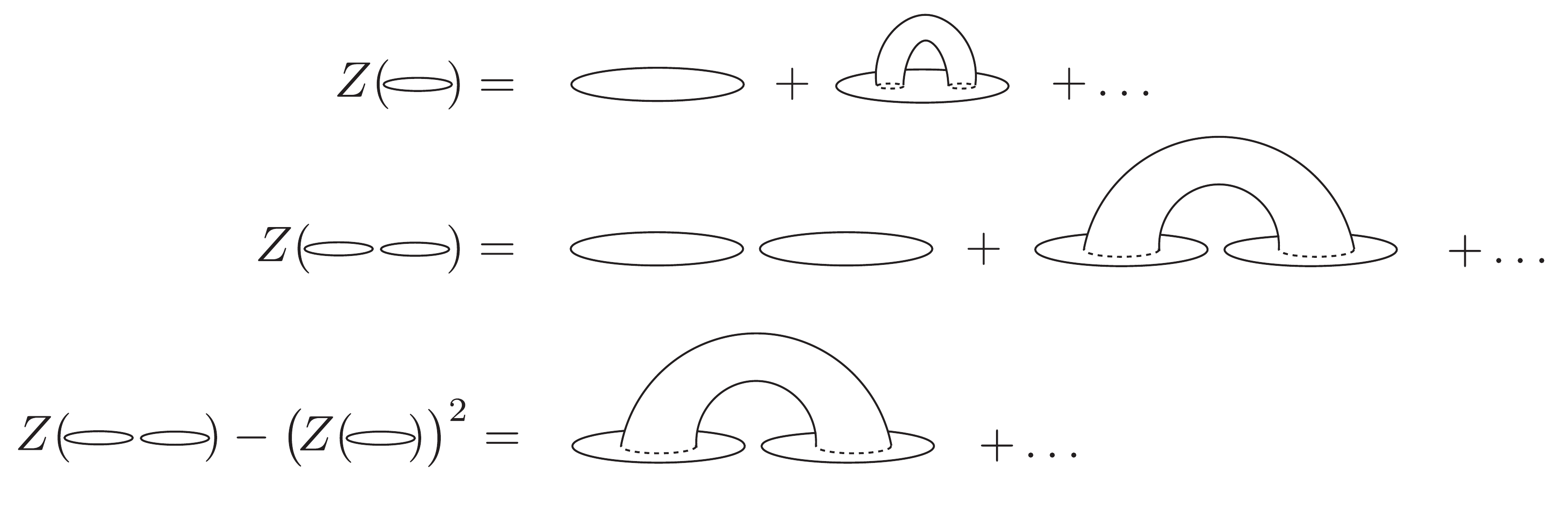}
    \caption{Gravitational path integrals are naturally taken to sum over possible topologies.  As illustrated in the bottom line, wormhole topologies can lead to failures of factorization. }
    \label{fig:PartitionFunctionVariance}
\end{figure}

There has been much recent study of the fact \cite{Lavrelashvili:1987jg,Hawking:1987mz,Hawking:1988ae,Coleman:1988cy,Giddings:Original_Axion_Wormhole,Giddings:1988cx} that contributions to gravitational path integrals from spacetime wormholes can lead to S-matrices, boundary correlators, and boundary partition functions that fail to factorize over disconnected boundaries.
Here we use the term spacetime wormhole to mean any bulk geometry that connects two boundary components that are not connected through the boundary.
A cartoon example of this phenomenon is shown in figure \ref{fig:PartitionFunctionVariance}.  For examples of such studies, see e.g. \cite{Maldacena:2004rf} and references thereto.

However,  there remain important questions as to when, and to what extent, spacetime wormholes do in fact contribute.  We will focus here on understanding the proper treatment of on-shell wormholes, which are thought to provide non-trivial saddles for the gravitational path integral.  However, we will see in the end that this provides insight into contributions from off-shell wormholes as well.  

Spacetime wormholes are generally discussed in the context of Euclidean path integrals.  Indeed, smooth on-shell spacetime wormholes are not expected to exist in Lorentz signature\footnote{The double cone of \cite{Saad:2018bqo} is an interesting Lorentz-signature spacetime wormhole.  But the real section is not smooth.} with real matter fields due to the tendency of non-trivial topology to collapse to singularities \cite{Friedman:1993ty,Galloway:1999bp}.  Investigations of contributions from spacetime wormholes are thus greatly hampered by fundamental questions regarding how such path integrals should be formulated.  

In particular, we remind the reader that, since the Euclidean gravitational action is unbounded below on the space of real Euclidean metrics \cite{Gibbons:1978ac}, the path integral cannot be formulated as an integral over this space; see also recent comments in \cite{Horowitz:2025zpx}.  This is often called `the conformal factor problem' since the problem can be seen by multiplying any fixed metric by a spacetime-dependent scalar.  While it is generally assumed that one should resolve this problem by instead integrating over some other (more convergent) contour in the space of complex metrics, there is as yet no broadly-accepted choice of contour in the general context\footnote{Gibbons, Hawking, and Perry proposed that, for asymptotically flat spacetimes, one write the metric in the form $\Omega^2 \tilde g$ where $g$ has vanishing Ricci scalar, and that one should then integrate over real $\tilde g$ and imaginary $\ln \Omega$. However, this proposal was deemed unsatisfactory because not all asymptotically flat metrics can be written in this form.}.  We emphasize that this is far more than a technicality:  Given any saddle, it is always possible to choose a contour (possibly with finite endpoints) that defines an integral for which that particular saddle gives the dominant contribution in the semiclassical limit.  But it is generally also possible to choose a contour for which the given saddle does not contribute to the associated integral in any meaningful way.  As a result, without making a choice for the defining contour of integration, the contribution of any given saddle is impossible to determine.

Since the real Euclidean contour cannot be used to define the gravitational path integral, it has been proposed by many authors \cite{Hartle:2020glw,Schleich:1987fm,Mazur:1989by,Giddings:1989ny,Giddings:1990yj,Marolf:1996gb,Dasgupta:2001ue,Ambjorn:2002gr,Feldbrugge:2017kzv,Feldbrugge:2017fcc,Feldbrugge:2017mbc,Brown:2017wpl,Marolf:2020rpm,Colin-Ellerin:2020mva,Colin-Ellerin:2021jev, Marolf:Thermodynamics}
that the defining contour should instead run over the space of real Lorentz-signature metrics.  While the associated path integral is then highly oscillatory, such integrals can be defined and controlled with sufficient care.  Indeed, in standard quantum field theory contexts, the derivation of the Lorentzian path integral from the canonical formalism endows it with enough structure for the path integral to both converge and to yield physically meaningful results.  

Subtleties of the gravitational case will be addressed in section \ref{sec:Background} below.
In particular, we will follow a version of this proposal that was described in \cite{Marolf:2020rpm,Colin-Ellerin:2020mva,Colin-Ellerin:2021jev,Marolf:Thermodynamics}, and which includes Lorentzian metrics with certain codimension-2 conical singularities.  Such singularities allow for the inclusion of topology-changing processes and, as we will see, also allow the inclusion of spacetime wormholes. 
%Following approach of \cite{Marolf:Thermodynamics}, this then allows one to use Lorentz-signature path integrals to compute partition functions describing gravitational thermodynamics (and which have also been traditionally studied in Euclidean signature);
%As noted above, despite our interest in Euclidean wormholes we wish to formulate our problem in terms of a sum over {\it Lorentz}-signature geometries.  We will do so by following the paradigm described in \cite{Marolf:Thermodynamics}, which used Lorentz-signature path integrals to compute partition functions describing gravitational thermodynamics (and which have also been traditionally studied in Euclidean signature);
This approach was shown in \cite{Marolf:Thermodynamics} to reproduce standard physically-satisfying results for black-hole thermodynamics, and in \cite{Colin-Ellerin:2021jev} and \cite{Held:2024qcl} to reproduce standard results for gravitational R\'enyi and HRT entropies, though many other processes remain to be investigated.  See \cite{Chen:2025leq} for extensions of this approach relevant to charged or rotating spacetimes; see also \cite{Mahajan:2025bzo,Singhi:2025rfy,Ailiga:2025osa} as well as \cite{Colin-Ellerin:2020mva,Colin-Ellerin:2021jev,Held:2024qcl} for applications to gravitational R\'enyi entropies, \cite{Dittrich:2024awu,Padua-Arguelles:2025koj} for related analyses using Regge calculus, and \cite{Chen:2025jqm} for recent application of this paradigm to understanding the phase of de Sitter partition functions.  This paradigm will be reviewed in section \ref{sec:Background} below.

This paper will focus on using the above proposal to study the particular case of asymptotically-AdS axion wormholes in 2+1 dimensions.  Axion wormholes are the subject of both historical interest \cite{Giddings:Original_Axion_Wormhole} and continued investigation in the recent literature     \cite{Hertog:AxionWormholeUnstable, Loges:2022nuw, andriolo:Axion_wormholes_massive_dilaton,Loges:2023ypl,aguilar:Axion_dS_Wormholes,Hertog:Axion-Saxion_Wormholes_Stable,loveridge:Axion_Wormhole_Alternate_Topologies,Marolf:2025evo}.  This is in part due to their appearance in string theory, and in particular in top-down models of AdS/CFT \cite{Arkani-Hamed:2007cpn}.
Axions are also well-known for the fact that, with standard conventions, they have the `wrong sign' for their kinetic term in Euclidean signature.  This is typically described as being related to the pseudo-scalar nature of axions, with the fact that under time-reversal ($t\rightarrow -t$) an axion $\chi$ transforms as $\chi \rightarrow -\chi$ motivating the transformation $\chi \rightarrow i\chi$ under $t\rightarrow it$.    However, this then leads to the question of whether the Euclidean axion should be integrated over a real contour or over an imaginary contour.  By starting with an integral over real fields in Lorentz signature, we will find clarity on this point as well.

Our analysis will focus on the question of whether (and to what extent) axion wormholes lead to non-factorization of AdS/CFT partition functions.    In particular, since axion wormholes are typically taken to be spherically symmetric, we will analyze axion wormhole contributions to AdS$_{d+1}$/CFT$_d$ partition functions associated with taking the AdS$ _{d+1}$ boundary to be two copies of the Euclidean sphere $S^d$.  As observed in \cite{Maldacena:2004rf}, in contrast with the wormholes studied there (and also in contrast with those of  \cite{Marolf:2021kjc}), such axion wormholes have exact zero modes that lead to potential divergences at the one-loop level.  However, since that will not directly affect our leading-order semiclassical analysis, we defer discussion of this issue to section \ref{sec:Discussion}.

In setting up our analysis,  it will be useful to define our problem in a manner that is as directly physical as possible. Let us therefore recall that the Euclidean sphere $S^d$ is the analytic continuation of the de Sitter space $dS_{d}$.  In particular, on the CFT side of the correspondence, the partition function on $S^d$ is naturally interpreted as representing the partition function $\Tr e^{-\beta_{\mathrm{dS}} H_{\mathrm{sp}}}$ where $H_{\mathrm{sp}}$ is the CFT Hamiltonian on the static patch of de Sitter space and where $\beta_{\mathrm{dS}} = 1/T_{\mathrm{dS}}$ is the inverse de Sitter temperature. In fact, we will need to place our CFT on a pair of such static patches so that the associated partition function at $\beta_{\mathrm{dS}}$ defines the desired two-boundary Euclidean partition function $Z[S^d \sqcup S^d]$ associated with two such spheres.  We can then write our 
$Z[S^d \sqcup S^d]$ as a Lorentz-signature gravitational path integral by following the procedure used to study one-boundary partition functions in \cite{Marolf:Thermodynamics}.  

We will study the resulting integral for a simple bulk theory containing only Einstein-Hilbert gravity (with a negative cosmological constant) and a minimally-coupled (and explicitly massless) pseudoscalar axion $\chi$.  However, we emphasize that
a Lorentz-signature minimally-coupled axion in such a model is indistinguishable from a Lorentz-signature minimally-coupled scalar. 
The fact that our analysis is defined by the Lorentz-signature formulation thus means that we must find identical results for both axion and scalar fields.  Indeed, from our perspective, calling $\chi$ an axion is merely a convention that connects to interesting past literature.   

We require each boundary of the gravitating bulk spacetime to be asymptotically (locally) anti-de Sitter.
Since the axion is massless, it is essential to fix its value at each such boundary.  Any other boundary condition would violate the CFT unitarity bounds (see e.g. \cite{Minwalla:1997ka} for discussion of such unitarity bounds in the context of AdS/CFT and, in particular, \cite{andrade:BeyondUnitarityBound} for an analysis from the bulk perspective).  It will be convenient to refer to the first $S^d$ as the `left boundary' and the 2nd $S^d$ as the `right boundary.' We thus denote the axion values at the left and right boundaries by $\chi_L, \chi_R$, both of which we take to be constant over their respective spheres.  The partition functions are then functions of $\chi_L, \chi_R$.  In particular, $\chi_L, \chi_R$ would be interpreted as coupling constants in any CFT dual. One can alternatively and equivalently describe the axion by a $d$-form gauge-potential (see e.g. \cite{Hertog:AxionWormholeUnstable,Hertog:Axion-Saxion_Wormholes_Stable}), but this will not be critical for our discussion here. 

We emphasize that there can be no on-shell wormholes with real fields without a surprising spontaneous breaking of symmetry.  In particular, if the time-translation symmetry of the boundary de Sitter static patch is preserved in the bulk, any wormhole would be static, and would thus also be traversable.  But the asymptotically-locally AdS version \cite{Galloway:1999bp} of topological censorship \cite{Friedman:1993ty} forbids such solutions in theories that respect the null energy condition.  Nevertheless, wormhole saddles can exist when the fields are complex (since the null energy condition is generally violated in that context).  Our task is then to evaluate the relevance of such complex saddles for various values of $\chi_L, \chi_R$.  
In particular, it can be interesting to attempt to analytically continue the path integral to complex values of $\chi_L, \chi_R$ and to study the resulting behavior.  Indeed, as explained in section \ref{ssec:EOSWormholes}, the Euclidean axion $\chi_E$ discussed in the abstract is in fact $\chi_E=i\chi$ in terms of our Lorentzian axion $\chi$. In particular, the abstract used $\chi_L= i \chi_{E,\infty}$ and $\chi_R= -i \chi_{E,\infty}$.

In the end, we will find the connected part of our partition function to be controlled by an on-shell axion wormhole only when the boundary values $\chi_L, \chi_R$ are analytically continued rather far around an associated Riemann surface.  In other cases, the connected part is instead controlled by a UV-sensitive endpoint contribution associated with wormholes that pinch off.  Nevertheless, since any analytic function that vanishes on an interval on the real line must vanish everywhere, our discussion of semiclassical saddles after analytically continuing $\chi_L, \chi_R$ means that the above endpoint contributions cannot vanish for any UV-completion that provides only perturbative corrections to such IR computations.

Before we begin, it is important to note that contributions from axion wormholes were previously analyzed from a Lorentzian path integral perspective in \cite{Loges:2022nuw}. That work was motivated by concerns similar to those expressed above.
Interestingly, and in contrast to our results reported above, their analysis argued that on-shell axion wormholes {\it do} in fact control certain computations associated with pairs of Euclidean spheres without performing excessive analytic continuations.  

However, there are several important differences between our approach and that of \cite{Loges:2022nuw}. The first is that the model studied in \cite{Loges:2022nuw} had no cosmological constant and, as a result, the associated wormholes were asymptotically flat.  Furthermore, this asymptotic flatness then allowed \cite{Loges:2022nuw} to impose boundary conditions that fixed the axion charge (``Neumann boundary conditions'') instead of fixing the value of the axion.  This is all rather different than our asymptotically AdS setting described above.  Indeed, we will find evidence that the choice of boundary conditions plays a critical role.  

\begin{figure}[t]
    \centering
    \includegraphics[width=0.5\linewidth]{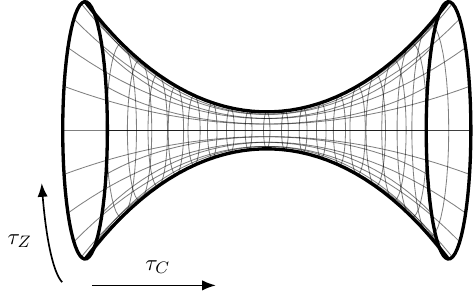}
    \caption{A simple cartoon of a Euclidean wormhole that contributes to $Z[S^d\sqcup S^d]$.  Our approach writes $Z[S^d\sqcup S^d]$ as a Lorentzian path integral where the timelike direction runs along the boundary (as indicated by the arrow marked $\tau_Z$), while that of \cite{Loges:2022nuw} uses a timelike direction transverse to the boundary (indicated by the arrow labeled $\tau_C$).  It is not yet clear if these choices define the same $Z[S^d\sqcup S^d]$.  We hope to return to this question in future work.}
    \label{fig:WickRotations}
\end{figure}

Yet another potentially-important difference is that, motivated by the above formulation in terms of Hamiltonians for static patches on a de Sitter boundary, on the defining contour for our path integral we take the timelike direction to run along our boundaries.  In contrast, the analysis of \cite{Loges:2022nuw} takes the timelike direction to run along the wormhole so that, in their Lorentz signature description, the wormhole becomes a spacetime describing the evolution of a closed spherical cosmology; see figure \ref{fig:WickRotations}.  Their choice of contour for the path integral thus differs from ours and could, in principle, lead to different results.  We will return to this point in section \ref{sec:Discussion} below.

We begin our discussion in section \ref{sec:Background} below with a brief review of the Lorentzian path integral computation of gravitational partition functions from \cite{Marolf:Thermodynamics}.  Section \ref{sec:overview} then provides an overview of the remaining elements of the strategy we will use to analyze axion wormhole contributions. In particular, our analysis will make important use of constrained wormholes, which are saddles for a constrained variational principle in which certain degrees of freedom are held fixed.  In this way our analysis is conceptually similar to that of \cite{Cotler_2021:ConstrainedInstantons}, though we work in Lorentz signature and use a rather different choice of constraint.  In particular, our constraint will be manifestly covariant.
Section \ref{sec:overview} will also discuss the UV regulation required by the fact that the static patch of de Sitter space has a boundary at the horizon, as well as a trick that allows an essentially-analytic treatment in the special case of
2+1 bulk dimensions.  Because of the utility of this trick, we will focus on the 2+1 case below, though we see no reason why the final results would be qualitatively different in higher dimensions.
%With the right choice of constraint, such spacetime wormholes can provide constrained saddle-points even in Lorentz-signature.
%One may then more explicitly study the remaining finite-dimensional integral over the constraint parameters to learn about the full path integral. 

The specific model to be studied will be introduced in section \ref{sec:OSWormholes}.    Section \ref{sec:OSWormholes} will also 
review the on-shell Euclidean wormholes in this model as well as related singular solution in Lorentz-signature.
Such solutions (even the singular ones) are important ingredients for using a cut-and-paste procedure to construct the  constrained wormholes mentioned above.   The stage is then set for  
the desired investigation of the path integral over off-shell wormholes in section \ref{sec:ConstrainedWormholes}.    This analysis makes use of Picard-Lefschetz theory, for which we provide a brief review in appendix \ref{app:PL}.  The analysis also draws from  technical results in appendices \ref{sec:details}, \ref{app:spurious}, and \ref{sec:AlternateParamatrization}. We close in section \ref{sec:Discussion} with a summary and final discussion.  This discussion will include comparison with Euclidean techniques and will draw in part from an analysis of Euclidean constrained wormholes performed in appendix \ref{sec:EuclideanCWormholes}.   

\section{Review: Partition functions from Lorentzian Path Integrals}
\label{sec:Background}

As stated in the introduction, our analysis will focus on bulk path integrals that compute a gravitational path integral  that one may think of as being defined by a partition function $Z(\beta_{sp}) = \Tr  e^{-\beta_{sp} H_{sp}}$ associated with a pair of dual CFTs, which in our case will live on the static patch of de Sitter space.  It is common to think of partition functions on any static background in terms of Euclidean path integrals, and thus as describing physics in Euclidean signature.  However, since the Hamiltonian is also an interesting operator in the Lorentz-signature theory, it must alternatively be possible to write $Z(\beta)$ (for any $\beta$) as a CFT path integral in Lorentz signature.  One way to do so was described in \cite{Marolf:Thermodynamics}, which also discussed the gravitational dual.  We review this discussion below, making a few minor additions along the way.

The essential idea is to write the above CFT partition function in terms of the operators $e^{-iHT}$ for real $T$. These unitary operators are naturally associated with Lorentz-signature path integrals.  It is straightforward to write partition functions in this way when $H$ is bounded below by some  energy $E_0$ (which, for clarity and simplicity, we take to be slightly below the actual ground state energy).  In that context, we may freely replace the operator $e^{-\beta H}$ with the product $e^{-\beta H}\, \theta(H-E_0)$ where $\theta$ denotes the step function that satisfies $\theta(x) =1$ for $x> 0$ while $\theta(x) = 0$ for $x < 0$.  Since the function $e^{-\beta E}\, \theta(E-E_0)$ is square-integrable, it admits a well-defined Fourier transform $f_\beta(T)$.  We may thus write an  operator relation of the form
\begin{equation}
\label{eq:expFT}
e^{-\beta H} = e^{-\beta H}\theta(H-E_0)  = \int_{-\infty}^{\infty} dT \, f_\beta(T) e^{-iHT},
\end{equation}
where a calculation with the above conventions gives
        \begin{equation}
            f_\beta(T)= \frac{e^{iTE_0-\beta E_0}}{2\pi i(T+i\beta)}.
        \end{equation}
Note that this weighting function has a pole at $T=-i\beta$.  Since $H \ge E_0$, one can easily check \eqref{eq:expFT} by using the factor of $iTE_0$ in $f_\beta$ and the bound $H\ge E_0$ to close the contour of integration in the lower-half of the complex $T$-plane and then using this pole to evaluate the integral.
Taking the trace of \eqref{eq:expFT} then immediately yields
\begin{equation}
\label{eq:ZFT}
Z(\beta) = \Tr \int_{-\infty}^{\infty} dT \, f_\beta(T) e^{-iHT}.
\end{equation}

We now wish to understand the gravitational dual of \eqref{eq:ZFT}.  The proposal of \cite{Marolf:Thermodynamics} is motivated by supposing that one can interchange the trace and the $T$-integral in \eqref{eq:ZFT} to write $Z(\beta)$ in terms of $\Tr e^{-iHT}$.  This latter object is naturally given by an integral over all real Lorentz-signature bulk spacetimes where the appropriate boundary time\footnote{Since we consider asymptotically (locally) AdS spacetimes, an appropriate notion of boundary time can be measured by a conformally-rescaled boundary metric.  The key point is then that the boundary conditions vary with $T$ only by changing the period of some time coordinate.} is periodic with period $T$ (and with caveats and details to be discussed below). While it is not at all clear that such an interchange is allowed for real $T$ (and, indeed, even whether $\Tr e^{-iHT}$ is a sensible quantity for real $T$), it is certainly justified if we give $T$ a small negative imaginary part\footnote{Since the above discussion started by stressing the conformal-factor problem of Euclidean quantum gravity, the reader may at first feel that this issue will again be raised by giving $T$ an imaginary part.  But the situation is not at all the same.  For real Euclidean metrics, the Euclidean integrand $e^{-S_E}$ is both real and positive.  There is thus no possibility that divergences will cancel.  In contrast, we now consider a definite real value of the Lorentzian time $T$ deformed by adding a small negative imaginary part.   As one would expect, and as we will see below, the corresponding gravitational path integral fails to converge absolutely, which might be considered to be a remnant of the conformal factor problem.  However, there are also oscillatory phases that provide opportunities for cancellations.  Indeed, as we will discuss, there is evidence that this path integral gives meaningful answers when the integrals are performed in the proper order. }.  Ref \cite{Marolf:Thermodynamics} thus writes
\begin{equation}
\label{eq:Zgrav}
Z(\beta) = \int {\cal D}g {\cal D}\phi \  f_\beta(T) \ e^{iS_{bulk}},
\end{equation}
where $\phi$ is a collective label for any matter fields and where one integrates over Lorentz-signature spacetimes with periodic boundary time (where $T$ is now just the name for this period and is one of the variables associated with the measure ${\cal D}g$).  

We will discuss what is meant by \eqref{eq:Zgrav} in more detail below. However, despite the fact that we have emphasized that the $T$ integral must be performed before evaluating the trace in \eqref{eq:ZFT}, it will be useful to introduce notation for the formal object associated with the fixed-$T$ Lorentzian bulk path integral 
\begin{equation}
\label{eq:ZgravT}
Z(B_T) = \int_{\partial M = B_T} {\cal D}g {\cal D}\phi  \ e^{iS_{bulk}},
\end{equation}
where \eqref{eq:ZgravT} integrates over all bulk fields with boundary conditions specified by a fixed Lorentzian boundary metric ${\cal B}_T$ in which time has period $T$ (and where we now explicitly indicate not just the period $T$ of boundary time but the entire boundary geometry ${\cal B}_T$).  In practice, we will in fact wish to also fix a collection of bulk parameters $\nu$ (e.g., the lengths or areas of certain bulk surfaces) and to study the constrained path integrals
\begin{equation}
\label{eq:ZgravTnu}
Z(B_T;\nu) = \int_{\partial M = B_T; \nu} {\cal D}g {\cal D}\phi  \ e^{iS_{bulk}},
\end{equation}
where the notation indicates that we now sum only over bulk spacetimes with the given parameters.  With these additional constraints, the objects $Z({\cal B}_T;\nu)$ will be well-defined (at least at leading semiclassical order).  We will then compute \eqref{eq:Zgrav} by writing it in the form
\begin{equation}
\label{eq:Zgrav2}
Z(\beta) = \int d\nu \int_{-\infty}^\infty dT   \  f_\beta(T)\  Z(B_T;\nu) ,
\end{equation}
with the $T$-integral being performed before the $\nu$-integral as described above.

\begin{figure}[!h]
    \centering
    \includegraphics[width=0.2\linewidth]{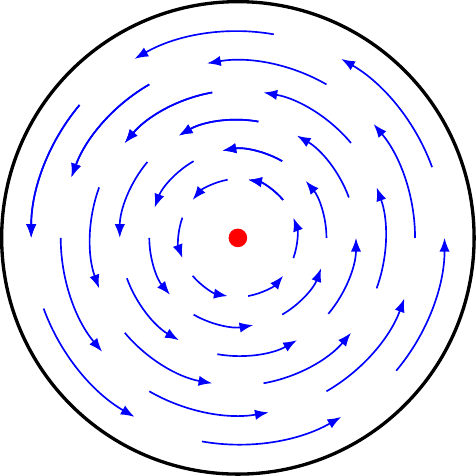}
    \caption{  In two dimensions, the cylinder is the only orientable connected manifold that admit an everywhere non-vanishing vector field that is tangent to its boundaries. Here we show a disk with a vector field that is tangent to the boundary and which exhibits the expected zero. }
    \label{fig:VectorDisk}
\end{figure}

As described in \cite{Marolf:Thermodynamics}, even at the level of a bulk effective field theory, there are two related issues to be addressed in defining the path integrals \eqref{eq:Zgrav} and \eqref{eq:Zgrav2}.  The first is that we will often be interested in settings in which there are no {\it smooth} Lorentz-signature metrics that satisfy the stated boundary conditions.  Here it is illustrative to consider a simple example in 1+1 dimensions, where we also consider a theory that allows only orientable manifolds and we then take the boundary to be a timelike circle (with period $T$).  Now, any manifold that admits a smooth Lorentz-signature metric must also admit a smooth non-vanishing vector field (associated with the choice of timelike direction).  But the only connected orientable two-dimensional manifold that admits an everywhere non-vanishing vector field that is tangent to a non-trivial boundary is the cylinder\footnote{Note that one could glue two copies of such a manifold together to make a closed manifold with an everywhere non-vanishing vector field.  The Poincar\'e-Hopf theorem then states that, since it is orientable, the resulting manifold-without-boundary must be a torus.  Allowed manifolds with boundary must thus be obtainable by cutting such a torus in half.}, which is excluded by our requirement that there be only a single $S^1$ boundary; see figure \ref{fig:VectorDisk}. 

\begin{figure}[!h]
    \centering
    \includegraphics[width=0.3\linewidth]{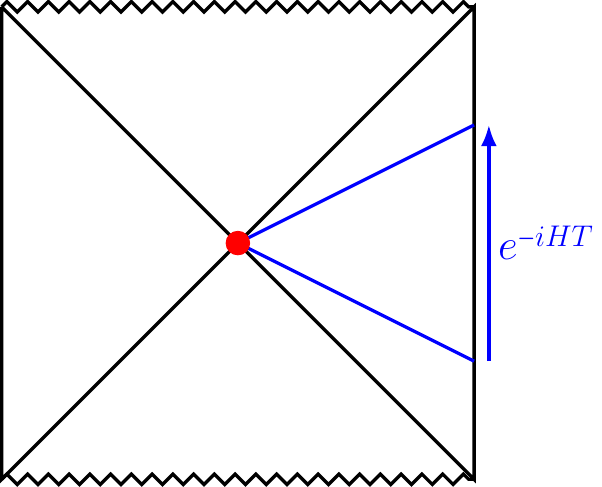}
    \caption{An AdS-Schwarzschild black hole with two surfaces of constant killing time shown in blue. The surfaces are related by $e^{-iHT}$ for some $T$. Cutting out the corresponding wedge of the spacetime and identifying the two surfaces yields a spacetime with a codimension-2 singularity at the would-be horizon bifurcation surface and which we may call a Lorentzian conical defect.  The defect is located at the would-be bifurcation surface of the original black hole horizon (red dot).}
    \label{fig:BlackHoleQuotient}
\end{figure}

However, one can easily construct metrics that one might describe as being of Lorentz signature with codimension-2 conical singularities.  Indeed, the proposal of \cite{Marolf:Thermodynamics} is to allow a class of such metrics constructed by a cut-and-paste procedure from smooth Lorentz-signature metrics.  A simple example of such a metric is obtained by considering any static black hole solution, slicing open the exterior region along two surfaces of constant Killing time, and identifying points on the two cuts that are related by a time-translation; see figure \ref{fig:BlackHoleQuotient}. We refer the interested reader to \cite{Marolf:Thermodynamics} for details of the more general construction.

It now remains to define the bulk action on spacetimes with such singularities.  The two-dimensional versions of such singularities were studied in
\cite{Louko:1995jw} (see also \cite{Neiman:2013ap}) by considering them to be limits of smooth {\it complex} metrics.
As argued in \cite{Colin-Ellerin:2020mva}, the higher dimensional case is essentially the same.
For the Einstein-Hilbert action, the result obtained in
\cite{Colin-Ellerin:2020mva} may then be transcribed as in \cite{Marolf:Thermodynamics} into the convenient form
\begin{equation}
\label{eq:defectSEH}
S_{EH} = \frac{1}{16\pi G_N} \int_M \sqrt{-g}R : = \lim_{\epsilon \rightarrow 0} \left( \frac{1}{16\pi G_N} \int_{M\setminus U_\epsilon} \sqrt{-g}R  - \frac{1}{8\pi G_N} {\cal P} \int_{\partial U_\epsilon} \sqrt{|h|}K  \right) + i \left(\frac{\cal N}{4}-1\right)\frac{A_\gamma}{4G_N},
\end{equation}
where $U_\epsilon$ is a small neighborhood around the singular codimension-2 surface $\gamma$, $A_\gamma$ is the area of $\gamma$, $G_N$ is Newton's constant, $\cal P$ denotes the principal part of the associated integral (whose integrand has simple poles where $\partial U_\epsilon$ becomes null), and ${\cal N}$ denotes the number of orthogonal null geodesics to $\gamma$ at each point of $\gamma$.  For example, Minkowski space (with no singularities) has ${\cal N}=4$ while the example in figure \ref{fig:BlackHoleQuotient} has ${\cal N}=0$.   See \cite{Colin-Ellerin:2020mva} and \cite{Marolf:Thermodynamics} for further discussion.  In contrast, the action for a minimally-coupled matter field receives no special contributions from the defect $\gamma$ and can be computed by integrating the usual Lagrange density over the non-singular region.

Let us focus on the final term in \eqref{eq:defectSEH}, which is proportional to the area $A[\gamma]$.  Note that this term is imaginary.  This is to be expected from the use of a complex regulator and the observation that no real regulator will suffice.  The sign of this imaginary term was chosen in \cite{Louko:1995jw} by choosing the imaginary parts of the complex metrics used to regulate the singularity so that standard matter path integrals converged; i.e., by using the basic idea behind what is often called the Kontsevich-Segal-Witten criterion \cite{Kontsevich:2021dmb,Witten:2021nzp} (though this basic idea was also used even earlier, e.g. in \cite{Halliwell:1989dy}).

As we will see, the spacetimes of interest in this work are analogues of those shown in figure \ref{fig:BlackHoleQuotient} and, correspondingly, again have ${\cal N}=0$.  We thus also now specialize to this case.  We also note that, later, we will focus on 2+1 dimensional spacetimes where the `area' $A_\gamma$ is  dimensionally a length.  As a result, the quantity called $A_\gamma$ here will later be denoted by $2L_\gamma$ (where the factor of $2$ is a convention related to the details of the case we study).

Having set ${\cal N}=0$,  we see from \eqref{eq:defectSEH} that the integrand $e^{iS}$ of the Lorentzian-path-integral  takes the form $e^{A_\gamma/4G_N}$ times a pure phase.  The magnitude of the integrand thus grows at large $A_\gamma$ and the integral fails to converge absolutely.   This is the second of the issues to be addressed that were foreshadowed above.

The important point is that, despite the lack of absolute convergence,  it has been argued in various examples \cite{Marolf:Thermodynamics,Held:2024qcl} that, by first performing the purely oscillatory integral defined by holding $A_\gamma$ (or $2L_\gamma$) fixed and integrating over all metric parameters, one then obtains a convergent integral over $A_\gamma$ (2$L_\gamma$); i.e that one may apply \eqref{eq:Zgrav2} with $\nu= A_\gamma$ (or, equivalently, with $\nu= 2L_\gamma$).    We will thus proceed assuming this to be true in the general case, and we will see that our computations are consistent with this assumption.  Indeed, while some dependence on the order of integrations may be expected when the integrand is oscillatory, our results below will be consistent with the hypothesis that the only important such orderings are that the integral over $A_\gamma$ (or $2L_\gamma$) be performed after the integral over the parameter $T$, and that the integral over $T$ be performed sufficiently late so that its contour can be closed in the lower half-plane (i.e., so that the associated notion of `energy' is effectively bounded from below by the chosen value of $E_0$).  The former condition appears to be a natural bulk analogue of the CFT statement noted above that the integral over $T$ should be performed before taking the trace over CFT states, and the latter condition is natural given our use of \eqref{eq:expFT}.  It may be that both conditions can be further loosened, though
we leave further investigation of such issues for future work.

The particular choice made in \cite{Marolf:Thermodynamics} was to first fix  both $T$ and $A_\gamma$ (or $2L_\gamma$) while performing the integrations over all other variables using the stationary phase approximation.  In this approximation, the constrained integral defines a constrained variational problem for which saddle points satisfy all of the
Einstein's equations but one (since stationarity with respect to $A_\gamma$ is not required\footnote{Since $T$ parameterizes boundary conditions, the constraint on $T$ does not remove bulk equations of motion.}).  The single exception allows  Lorentzian conical defects at $\gamma$ that are precisely the form discussed above (and, in particular, where the `conical deficit' is constant along the defect).  The black hole spacetimes with periodic Lorentzian time described in figure \ref{fig:BlackHoleQuotient} are thus saddle-points of this constrained variational principle and, furthermore, they are saddles that lie on the defining (Lorentzian) contour of integration.  

Now, since the associated integral is performed at fixed $A_\gamma$ (or at fixed $2L_\gamma$), the associated integrand is proportional to a pure phase.  As argued in \cite{Marolf:Thermodynamics},  under such conditions a saddle on the defining contour {\it must} contribute to the integral; see appendix \ref{app:PL} for a detailed review. It is then interesting to also assume the above saddles to dominate the constrained integrals at each fixed $T, A_\gamma$ (or $2L_\gamma$).  After doing so, it was straightforward for \cite{Marolf:Thermodynamics} to study the remaining $T$ and $A_\gamma$ integrals in detail and to show that they reproduce standard results for black hole thermodynamics.

\section{Overview and Strategy}
\label{sec:overview}

Our actual implementation of the above paradigm will involve a number of technical steps.  We thus dedicate this section to providing a bird's-eye view of our procedure before diving into the details in sections \ref{sec:OSWormholes} and \ref{sec:ConstrainedWormholes}.  In particular, after some general introduction, section \ref{sec:obndy} will discuss a UV divergence and the way we choose to regulate it.  We will then describe our choice of bulk constraints ($\nu$) in section \ref{sec:obulk}.

To begin, let us
recall that we wish to formulate our problem as a study of the Hamiltonian associated with taking each bulk spacetime to have a pair of  asymptotically locally AdS boundaries where each boundary metric is that of the static patch of de Sitter (dS) space. We will work in the theory of Einstein-Hilbert gravity with a minimally-coupled massless axion $\chi$ and with no other matter fields.  For the reasons given in the introduction, we also impose Dirichlet boundary conditions that fix the asymptotic values of the axion to some constant at each boundary.

Our problem is thus 
of the form described in section \ref{sec:Background}, but for the special case where the boundary $B$ of the bulk spacetime, and thus the spacetime $B$ on which a dual CFT might live,  is taken to be disconnected.  In particular, in a sense that we will describe, we take $B$ to be the disjoint union $B= {\cal B}_1 \sqcup {\cal B}_2$ of a pair ${\cal B}_1,{\cal B}_2$ of two $dS_2$ static patches.  This is just a special case of the general formalism above.  

In particular, as described in section \ref{sec:Background}, while we wish to study partition functions associated with choosing some inverse temperature $\beta$, we will express those partition functions as integrals over quantities $Z(B_T;\nu)$ defined by Lorentz-signature boundary spacetimes $B_T$ in which time has period $T$ (and by also imposing further constraints $\nu$ on the class of bulk spacetimes over which we integrate). To study wormholes at fixed time-period $T$, we thus require the boundary to be two disconnected copies of a connected boundary spacetime ${\cal B}_T$.  We denote this by writing $B_T = {\cal B}_T \sqcup {\cal B}_T$.  We then wish to study the `connected' partition function 
\begin{equation}
Z^\text{\tiny c}[{\cal B}_T \sqcup {\cal B}_T] :=
Z[{\cal B}_T \sqcup {\cal B}_T;\nu]  - 
Z[{\cal B}_T]^2.
\end{equation}

As usual, such connected partition functions may be computed by summing only over bulk spacetimes in which the two boundaries lie in the same connected component.    After doing so,  we will integrate over $T$ (as in \eqref{eq:Zgrav2}) to construct the desired connected UV-regulated partition function
\begin{eqnarray}
Z^\text{\tiny c}(\beta) &:=& Z({\cal B} \sqcup {\cal B}, \beta)- [Z({\cal B}, \beta)]^2 \\ &=&\int_{-\infty}^\infty dT   \  f_\beta(T)\  Z^\text{\tiny c}[{\cal B}_T \sqcup {\cal B}_T] :=\int d\nu \int_{-\infty}^\infty dT   \  f_\beta(T)\  Z^\text{\tiny c}[{\cal B}_T \sqcup {\cal B}_T;\nu],
\label{eq:Zgrav2c}\end{eqnarray}
where the notation on the first line indicates the relevant static boundary geometry ${\cal B}$ or ${\cal B} \sqcup {\cal B}$ as well as the inverse temperature $\beta$.  The final equality in \eqref{eq:Zgrav2c} simply indicates that the expression on the left-hand-side of the final line is in fact defined by  first imposing constraints $\nu$ that yield a well-defined connected constrained fixed-$T$ partition function $Z^\text{\tiny c}[{\cal B}_T^\lambda \sqcup {\cal B}_T^\lambda;\nu]$, performing the $T$ integral over such quantities, and by then performing the $\nu$ integral at the very end.  In passing to the 2nd line we have also used the relation
\begin{eqnarray}
\label{eq:Zgrav2cextra}
 \int_{-\infty}^\infty dT   \  f_\beta(T)\ F(T) = F(-i\beta),
\end{eqnarray}
where $F$ is any function that is analytic in the lower half plane and which also allows the $T$ contour in the integral on the right-hand-side to be closed in the lower half plane.  In particular, we assumed this to be true for both $F(T) = [Z({\cal B}_T)]^2$ and $F(T) = Z({\cal B}_T \sqcup {\cal B}_T)$.

\subsection{A UV divergence and a regulator}
\label{sec:obndy}

\begin{figure}
    \centering
    \includegraphics[width=0.3\linewidth]{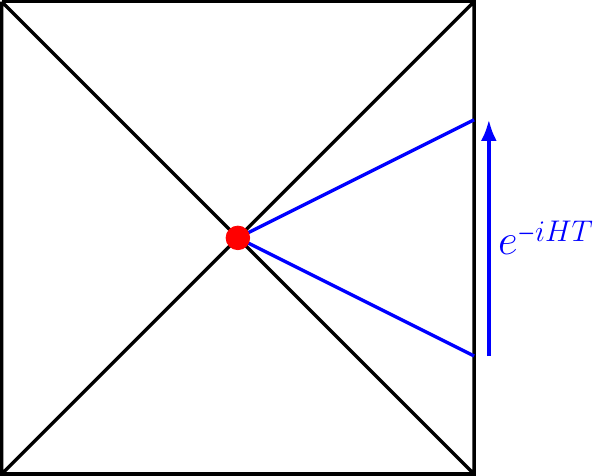}
    \caption{The Penrose diagram for a de Sitter spacetime with two surfaces of constant static time shown in blue. The surfaces are related by $e^{-iHT}$ for some $T$. Cutting out the corresponding wedge and identifying the two surfaces yields a spacetime with a codimension-2 singularity located at the would-be cosmological horizon bifurcation surface of the original spacetime (red dot).}
    \label{fig:placeholder1}
\end{figure}

Recall now that our Lorentz-signature boundary spacetimes ${\cal B}_T$ are to be defined by the static patch of $dS_d$. In particular, each ${\cal B}_T$ is constructed by identifying the $t=0$ slices of such a $dS_d$ static patch with its image under a static-patch time translation by a (dimensionless) proper time $T$ as measured along the central geodesic; see figure \ref{fig:placeholder1}.  This construction is analogous to that shown in figure \ref{fig:BlackHoleQuotient}, though it is now used on the boundary rather than in the bulk.

For finite $T$, this identification necessarily leads to a Lorentzian conical defect on the boundary.  As a result, our boundary metrics are not smooth, and the standard counter-terms need not suffice to define a good variational principle, or even to render the action finite.  Indeed, since the timelike direction on the boundary must circle this defect, continuity requires any associated bulk spacetime to have a corresponding defect that stretches to touch ${\cal B}_T$ and which thus has infinite size. In particular, in the 2+1 context on which we focus, the associated defect length $L_\gamma$ is necessarily infinite and thus leads to novel divergences in the action.  Similarly, from a dual CFT perspective one expects divergences in the CFT partition function since the CFT now propagates on a spacetime with a singular metric.

\begin{figure}
    \centering
    \includegraphics[width=0.85\linewidth]{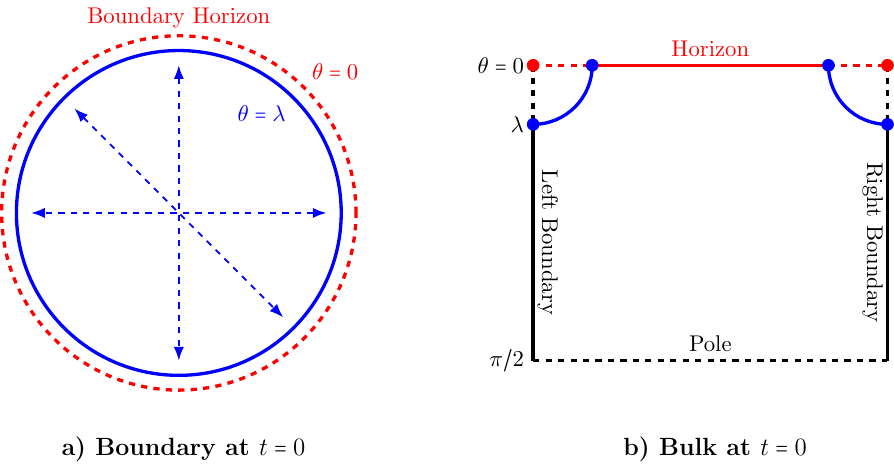}
    \caption{Constant $t$ slices of our UV regulated boundary and bulk (say, at $t=0$). (a)  On the boundary, the $t=0$ slice of the static patch of $dS_d$ is a hemisphere, which we represent by a disk.  Our coordinate $\theta$ vanishes on an $S^{d-2}$ that we represent as a dashed red circle.  The UV regulator excises the region outside the $S^{d-2}$ at $\theta =\lambda$ (blue circle).  It also identifies antipodal points on this $S^{d-2}$ (as indicated by the dashed arrows).  (b)  The bulk at $t=0$ stretches between the left and right asymptotic boundaries.  Each vertical slice is the $t=0$ slice of the static patch of $dS_d$.   Each vertical slice runs from the pole  (dashed line at bottom) at $\theta=\pi/2$ to the $dS$ horizon (solid red line at top).  In particular,  the dashed red line (solid blue line) from panel (a) on (say) the left boundary has become just the red dot (blue dot) on the left edge of panel (b). The region excised in (a) is represented by the dashed black line, while solid black lines indicate the remaining regulated boundaries.  Each generic point in the figure represents a suppressed $S^{d-2}$, though each point on the blue curve represents a suppressed ${\mathbb RP}^{d-2}$ that reflects the antipodal identifications performed on the asymptotic boundaries at $\theta=\lambda$. The dashed red line in (b) lies outside our regulated spacetime and indicates the would-be extension of the horizon in the limit $\lambda \rightarrow 0$ in which the regulator is removed. Thus the blue and (dashed) red loci in panel (a) are the boundaries of the correspondingly colored loci in panel (b).}
    \label{fig:placeholder}
\end{figure}

We will deal with this issue by modifying the boundary spacetime ${\cal B}_T$ near the defect, viewing this as a UV-regularization of the divergent problem.  In particular, for a $dS_d$ boundary, each  slice of constant Killing time in the static patch can be described as a $(d-1)$-dimensional hemisphere
\begin{equation}
\label{eq:slicemetric}
ds^2_{S^{d-1}} = d\theta^2 + \cos^2 \theta d\Omega_{d-2}^2,    
\end{equation}
where we take $\theta = 0$ at the bifurcation surface of the de Sitter horizon and $\theta = \pi/2$ at the pole of the hemisphere  so that we have  $\theta \in [0,\pi/2]$.  \ As in \cite{marolf:EntanglementPhaseTransition}, we then simply excise the annular region ($S^{d-2} \times I$) given by $\theta \le \lambda$ for some small $\lambda$ from each Killing slice of the boundary and then identify points on the $S^{d-2}$ boundary of the remaining disk ($D^{d-1}$) that are related by the antipodal map\footnote{As usual, this choice of regulator is not unique. While excising the region $\theta\le \lambda$ creates a boundary $\partial {\mathcal B}_T$ for ${\mathcal B}_T$, and thus also for the bulk spacetime, there are many possible choices of boundary condition at $\partial {\mathcal B}_T$.  For example, one could treat it as the anchor point for an end-of-the-world brane with some tension $\mu$. Our antipodal identification can be viewed as a higher-dimensional analogue of periodic boundary conditions, which in practice is identical to using an end-of-the-world brane with $\mu=0$ due to the ${\mathbb Z}_2$ symmetry of our solutions under the antipodal map on each $S^{d-2}$.}; see panel (a) in figure \ref{fig:placeholder}.  Each Killing slice of the resulting regulated boundary ${\cal B}_T^\lambda$ is then an ${\mathbb RP}^{d-1}$. See also panel (b) of figure \ref{fig:placeholder} for a sketch of the corresponding bulk spacetimes. 

We will be most interested in the case where $d=2$ and ${\mathbb RP}^{d-1}= S^1$  is orientable (so that ${\cal B}_T^\lambda$ is orientable as well), though the theory we will study is in fact also compatible with non-orientable boundaries.  We then define the partition function of interest to be given by the $\lambda \rightarrow 0$ limit of those defined by the boundaries ${\cal B}_T^\lambda$.

\subsection{Constraints for bulk wormholes}
\label{sec:obulk}

We plan to proceed as outlined in section \ref{sec:Background} by focusing on the Lorentzian path integral 
\begin{equation}
Z^\text{\tiny c}[{\cal B}_T^\lambda \sqcup {\cal B}_T^\lambda;A_\gamma] :=
Z[{\cal B}_T^\lambda \sqcup {\cal B}_T^\lambda;A_\gamma]  - 
Z[{\cal B}_T^\lambda;A_\gamma]^2,
\end{equation}

In particular, we focus on contributions having a single defect\footnote{After introducing our regulator for a finite value of $\lambda$, even with periodic Lorentzian time $t$ there is also a sector with no defects at all.  In this sector the bulk has topology ${\mathbb RP}^{d-1} \times S^1 \times {\mathbb R}$, where the $S^1$ is the time circle and ${\mathbb R}$ is the direction that runs through the wormhole from one boundary to another.  But since there is no defect, the Lorentzian action is real and $e^{is}$ is a pure phase. Contributions from these spacetimes are thus infinitely suppressed in the limit $\lambda \rightarrow 0$ relative to the spacetimes shown in figure \ref{fig:placeholder} as the latter are weighted by $e^{\frac{A_\gamma}{4G_N}}$ with diverging $A_\gamma$.} at which the local causal structure takes the form shown in figure \ref{fig:placeholder1}; i.e., using the notation of \eqref{eq:defectSEH}, with ${\cal N}=0$ orthogonal null surfaces at the defect.    This is the sector studied in detail in \cite{Marolf:Thermodynamics} in the context of black hole thermodynamics and which, as reviewed in section \ref{sec:Background}, reproduces expected results.  Furthermore, other sectors must be subdominant unless the $U(1) \times {\mathbb Z}_2\times {\mathbb Z}_2$ symmetry of the boundary conditions is spontaneously broken.\footnote{Since the boundary spacetime has only a single defect with ${\cal N}=0$, any defect which extends to the boundary will also have ${\cal N}=0$.  The question is then whether there are other defects in addition. 
If so, they will necessarily break the $U(1) \times {\mathbb Z}_2\times {\mathbb Z}_2$ symmetry of our setup to a smaller group.} Here we simply assume this does not occur, though it would clearly be useful to study  contributions from symmetry-breaking sectors more carefully in the future.

Using \eqref{eq:defectSEH}, we see that  $Z^\text{\tiny c}_{A_\gamma}[{\cal B}_T^\lambda\sqcup {\cal B}_T^\lambda]$ sums over spacetimes 
with Lorentzian action $S = -i \frac{A_\gamma}{4G_N} +S_R$ where $S_R$ is real.  Since $A_\gamma$ is fixed, the integrand is then $e^{\frac{A_\gamma}{4G_N}}$ times an oscillating phase.  It is thus natural to suppose that, while evaluating the oscillatory integral may require some care, doing so will give a meaningful result.

One might then ask if we can evaluate $Z^\text{\tiny c}_{A_\gamma}[{\cal B}_T^\lambda\sqcup {\cal B}_T^\lambda]$  using stationary phase methods in parallel with the treatment in \cite{Marolf:Thermodynamics}.  However, when the stationary phase approximation was invoked at this stage in \cite{Marolf:Thermodynamics}, a key point was that there were {\it real} saddles which lay on the defining contour of integration for the path integral.  As reviewed in section \ref{sec:Background} above, general arguments then showed that such saddles necessarily contributed to the desired partition function. 

However, in this work we wish to study axion wormholes with de Sitter boundaries.  As mentioned in the introduction, if the de Sitter symmetry is maintained in the bulk, such wormholes are then static in the de Sitter static patch and are thus traversable.  But with real fields in Lorentz signature such  
wormhole saddles are forbidden by topological censorship.  There are thus no real saddles for $Z^\text{\tiny c}_{A_\gamma}[{\cal B}_T^\lambda\sqcup {\cal B}_T^\lambda]$.

On the other hand, it turns out that there are Lorentzian wormhole saddles for the constrained variational principle in which we \emph{also} fix the codimension-1 volume of the surface on which the axion vanishes. 
As a result, for any fixed value of this codimension-1 volume, one can in principle perform the remaining integrals by using the stationary phase approximation, and one knows in particular that such real constrained saddles contribute\footnote{\label{foot:Stokes}The argument holds rigorously for real values of $T$ and for real axion boundary conditions.  However, we will in fact be interested in imaginary $T=-i\beta$ and, in some cases, in complex axion boundary conditions.  As a matter of principle, one should thus be aware of the potential for interesting Stokes' phenomena which might cause the desired constrained saddle to lose relevance.  However, since the axion wormhole saddles that we wish to study are special cases of these constrained saddles, and since the main message of our paper is that the standard Euclidean axion wormhole saddles are not in fact relevant, such additional Stokes' phenomena would only add further weight to this conclusion (though they could weaken the conclusion of section \ref{ssec:AnalyticContinue} that similar wormholes do contribute after further analytic continuation).  We thus leave the question of such Stokes' phenomena open.  However, it would be interesting to study this more carefully.}.  After doing so, we are left with three integrals to be studied in more detail (including the integrals over both the defect area $A_\gamma$  and the period $T$ of Lorentzian time).  In practice, we expect the integral over the volume of the $\chi=0$ surface to be of greatest interest, since it is this integral whose complex saddles will lead to complex on-shell wormholes.

Of course, to make progress we must be able to actually find the desired constrained wormholes.  Unfortunately, our introduction of the conical defect breaks de Sitter symmetry and thus makes this task more difficult.  It is a cohomogeneity-2 problem that would generally require non-trivial numerics.

Nevertheless, in the special case where the bulk has 2+1 dimensions, one can make progress by using a special trick.  Starting with a solution with de Sitter symmetry, we can compactify the static-patch time coordinate as described earlier to create a defect.  But in 2+1 dimensions we can also change the strength of this defect by making a boundary conformal transformation (and by simultaneously changing the period of identification of a bulk angular coordinate).  This can then generate a one-parameter family of solutions labeled by a renormalized notion of the (infinite) defect area $A_\gamma$.  While regulating the boundary as in figure \ref{fig:placeholder} renders $A_\gamma$ finite, and while this then means that boundary conformal transformations cannot in fact change $A_\gamma$ at finite $\lambda$, we will argue in section \ref{sssec:singreg} that we can first apply it to the unregulated spacetimes (with $\lambda=0$) and that we can then build an approximate solution to the finite-$\lambda$ problem through a cut-and-paste procedure.  We will then perform various checks in appendix \ref{sec:details} to argue that as $\lambda \rightarrow 0$ this approximation becomes sufficiently accurate to extract the desired physics.

For the above reason, we will henceforth focus on the 2+1 case, and we will correspondingly rename $A_\gamma$ as $2L_\gamma$ (since it is dimensionally a length, and since in 2+1 dimensions the unregulated defect has two disconnected components [though the regulated defect is a single connected $S^1$]).  Similarly, we will now speak of the codimension-1 {\it area} of the surface $\chi=0$.  Note that this is the case $d=2$ in the above conventions, so that the $S^{d-2}$ suppressed in figure \ref{fig:placeholder} is just a pair of points (an $S^0$).  As a result, for the rest of the paper it will be useful to use slightly different conventions in which $\theta$ in \eqref{eq:slicemetric} runs over $[0,\pi]$ and there is simply no $d\Omega_{d-2}^2 = d\Omega^2_0$ term in the line element.  The UV-regulator introduced above now identifies the circle $\theta= \lambda$ on each boundary geometry with the corresponding  circle $\theta = \pi-\lambda$, so that each regulated boundary has the topology of a torus.

We close this section by mentioning that the desired constrained wormholes with de Sitter symmetry are straightforward to construct.  Since we constrain only properties of the surface $\chi=0$, our constrained wormholes must satisfy the full equations of motion away from this surface.  We thus need only find all solutions of the {\it original} equations of motion with de Sitter symmetry (perhaps allowing them to become singular when one moves sufficiently deep into the bulk), cut these solutions along the surface $\chi=0$, and glue them together in appropriate pairs to make the required constrained wormholes.  The next section is dedicated to describing the solutions that arise both with and without imposing the various constraints.

\section{Wormholes,  wormhole-like solutions, and constrained wormholes in Axion Gravity}
    \label{sec:OSWormholes}
    We will now study wormholes and wormhole-like solutions of three-dimensional axion gravity.  We wish to use Lorentz-signature solutions with de Sitter symmetry to construct constrained wormholes via the cut-and-paste construction procedure mentioned above.  But since there are no smooth wormholes in Lorentz signature, and since Euclidean axion wormholes will already be familiar to many readers, we begin by reviewing Euclidean axion wormholes with spherical symmetry  in section \ref{ssec:EOSWormholes}.  The spherical symmetry is relevant since spherical cross-sections Wick-rotate to the desired de Sitter cross-sections in Lorentz signature.  We will then discuss the desired Lorentzian wormhole-like solutions in section \ref{ssec:LOSWormholes} and construct constrained wormholes in section \ref{ssec:ConstrainedWormholesConstruction}.

        \subsection{Euclidean Axion Wormholes }
            \label{ssec:EOSWormholes}

    We begin with a general discussion of the role of the axion in stabilizing Euclidean wormholes.  This will then set the stage for us to  present the details of the Euclidean axion wormhole as a point of reference for the Lorentzian discussion in section \ref{ssec:LOSWormholes}. 
            
            We remind the reader that we wish to study axion gravity in 3 dimensions defined by the Einstein-Hilbert action with a negative cosmological constant and with a minimally coupled massless axion.  We also remind the reader that the standard convention is to take the kinetic term of the Euclidean axion to be {\it negative}.  As remarked earlier, for us this is merely a convention.  To emphasize this, we denote the Euclidean axion below by $\chi_E$.  The unusual sign in the Euclidean kinetic term then implies $\chi_E = i \chi$, where $\chi$ is the standard Lorentzian axion (which of course has a positive kinetic term).  The Euclidean action thus takes the form

            \begin{equation}
                S_E = -\frac{1}{16\pi\Gn}\int_\M d^3x \sqrt{g}\Big(R+\frac{2}{\ell^2} +\frac{1}{2}\partial_\mu\chi_E\partial^\mu\chi_E\Big) - S_{E,\tiny \partial\M}.
                \label{eq:EAction1}
            \end{equation}

The term $S_{E,\tiny \partial\M}$ is to contain boundary terms appropriate to our choice of boundary conditions.             
We wish to impose asymptotically locally AdS (AlAdS) boundary conditions on the metric.  In particular, we will fix what is commonly called the `boundary metric' to be a unit two-sphere. As noted earlier, the axion requires a Dirichlet boundary condition that fixes the values of $\chi_E$ to be $\chi_E^+$ on the `right' boundary and $\chi_E^-$ on the `left' boundary.  The point is that, since $\chi_E$ is massless, taking Neumann boundary conditions would define a dual-CFT operator of dimension zero which would violate unitarity bounds \cite{Klebanov:1999tb}.  Correspondingly, one can show directly that it would lead to ghosts in the bulk theory \cite{andrade:BeyondUnitarityBound} that we wish to avoid.  

If for simplicity we now take $\chi_E^\pm = \pm \chi_{E,\infty}$, the boundary conditions on our bulk fields take the form:

            \begin{equation}
                g_{\mu\nu}dx^{\mu}dx^{\nu}|_{\tiny \partial\M_{L,R}} = \frac{\ell^2}{z_{L,R}^2}\Big(d\theta^2 +\sin^{2}(\theta) \,d\tau^2\Big) +\mathcal{O}(z_{L,R}^0), \quad \chi_E(x)|_{\tiny \partial\M_{L,R}} = \pm\chi_{E,\infty},
                \label{eq:EuclideanBCs}
             \end{equation}
            Here $\partial\M_{L,R}$ represent the left and right regulated boundaries and $z_{L,R}$ are an appropriately defined Fefferman-Graham coordinates that respectively vanish at $\partial\M_{L,R}$.  Without loss of generality, we will take $\chi_{E,\infty}$ to be positive for the rest of this section.
        
            With these boundary conditions, the appropriate boundary terms consist simply of the usual Gibbons-Hawking-York boundary term as well as the familiar counter terms for asymptotic locally AdS spacetimes \cite{Chapter19},

            \begin{equation}
                S_{\tiny E, \partial\M} = \frac{1}{8\pi\Gn}\int_{\tiny \partial\M}d^2x\,\sqrt{|h|}\Big(\,K -\frac{1}{\ell} + \frac{\delta^2}{2\ell}\log(\delta)\mathcal{R}^{(2)}\Big),
                \label{eq:ESbndy}
            \end{equation}
            where $\partial\M = \partial\M_L \sqcup \partial\M_R$ and $h,\mathcal{R}^{(2)}$ are the determinant and Ricci scalar of the (divergent) induced metric on the regulated boundary at $z_L=z_R=\delta$.

            We seek  stationary points of the action \eqref{eq:EAction1} which preserve the spherical symmetry of the boundary conditions.  We thus impose the bulk metric ansatz

            \begin{equation}
                ds^2 = dr^2 +\ell L(r)\big(d\theta^2 +\sin^2(\theta)d\tau^2\big),\quad \chi_E(x)=\chi_E(r),
                \label{eq:EAnsatz}
            \end{equation}
            where the coordinate $\tau$ takes values in $[0,2\pi)$. We can determine the bulk axion profile by noting that the action enjoys a shift symmetry $\chi\rightarrow\chi+\Lambda$. The associated conserved current is $j_\mu=\partial_\mu\chi_E$, which we can integrate over each $r=\text{const.}$ surface to obtain a conserved charge $q_E$. We normalize this charge to satisfy

            \begin{equation}
                4\pi q_E =\int_{\tiny \Sigma_r} d^2x \,\sqrt{|h|}\,n_r^\mu\, j_\mu,
                \label{eq:EqDef}
            \end{equation}
            where $n_r^\mu$ is the positive-directed unit normal to the $r=\text{const.}$ surface $\Sigma_r$. Since our ansatz requires $\chi_E$ to be homogeneous on each constant-$r$ surface,  we conclude from the above expression that we have

            \begin{equation}
                j_\mu=\partial_\mu \chi_E = \frac{q_E\,\delta_\mu^r}{\ell L(r)}.
            \end{equation}
            This is indeed a solution to the equation of motion $\nabla^2\chi=0$ obtained by varying the action with respect to $\chi$.

            It is now straightforward to determine the geometry.  As has been discussed in many places,  the $rr-$component of Einstein's equations acts as a radial version of the Hamiltonian constraint.  As a result, it gives a first-order ODE that takes the form

            \begin{equation}
                \big(\partial_r L\big)^2 = \
                \frac{4L}{\ell^2}\Big(\ell-\frac{q_E^2}{4L}+L\Big). 
                \label{eq:Eeom}
            \end{equation}
            The (positive) $\Z_2$-symmetric solutions to this equation of motion are

            \begin{equation}
                L(r) = \frac{\ell}{2}\Bigg(\sqrt{1+\frac{q_E^2}{\ell^2}}\cosh\Big(\frac{2r}{\ell}\Big)-1\Bigg).
                \label{LEuclidean}
            \end{equation}

\begin{figure}[h!]
                \centering
                \includegraphics[width=0.5\linewidth]{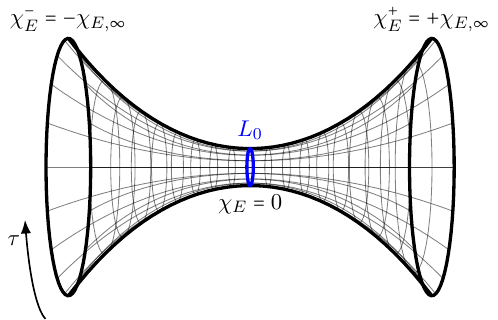}
                \caption{A constant $\theta$ section of the Euclidean axion wormhole. The axion profile  $\chi_E$ is a $\mathbb{Z}_2$-odd function taking values $\pm\chi_{E,\infty}$ at the left and right asymptotic boundaries and vanishing at the throat of the wormhole (blue circle). The two-dimensional throat has proper area $4\pi\ell L_0$.}
                \label{fig:EOSAxionWormhole}
            \end{figure}

A sketch of a constant-$\theta$ cross-section of such a solution is shown in figure \ref{fig:EOSAxionWormhole}.
            The function $L(r)$ is everywhere positive for real $q_E$ (which we will assume for now). It thus gives us a connected wormhole geometry on which the minimum of $L(r)$ is $L_0 = \frac{\ell}{2}\Big(\sqrt{1+\frac{q_E^2}{\ell^2}}-1\Big)$. By symmetry, this minimum occurs at the surface where $\chi_E$ vanishes.

            We would like to express all parameters in terms of the boundary conditions we have imposed, and to this end we can relate the conserved axion flux $q_E$ to $\chi_{E,\infty}$ by integrating the $r$-component of the associated conserved current out to the boundary. To do so, we again make use of the fact that the $\Z_2$ symmetry of the solution forces the axion to vanish at the throat of the wormhole. We find

            \begin{equation}
                q_E=-\ell \tan(\chi_{E,\infty}).
            \end{equation}
            The above expression should be strictly positive for positive choices of $\chi_{E,\infty}$. This informs us that (again, for positive $\chi_{E,\infty}$) we need to restrict our choice of boundary conditions to the interval $\chi_{E,\infty} \in (\frac{\pi}{2}, \pi)$ so that the total change in the axion between the two asymptotic boundaries is $\Delta\chi_E\in(\pi,2\pi)$. This observation is consistent with previous results \cite{Arkani-Hamed:2007cpn}.  Indeed, as noted in e.g. \cite{marolf:AdS_Euclidean_Wormholes}, it is a common property of Euclidean wormholes that they exist only at sufficiently large sources and, in this sense, the phase diagram of AdS Euclidean wormholes is generally similar to that of the well-known Hawking-Page transition \cite{Hawking:1982dh}.

            \subsection{Solutions of Lorentzian Axion Gravity}
                \label{ssec:LOSWormholes}
                Having examined  Euclidean wormhole solutions, we can now turn to studying related solutions of Lorentzian axion gravity. In the Lorentzian theory, our axion $\chi$ is indistinguishable from an ordinary scalar field and, in particular, it has positive kinetic energy. The action takes the form

                \begin{equation}
                    S= \frac{1}{16\pi\Gn}\int_{\tiny\M} d^{3}x\,\sqrt{|g|}\,\Big(R+\frac{2}{\ell^2} -\frac{1}{2}\partial_\mu\chi\partial^\mu\chi\Big) \,+ S_{\tiny \partial\M}.
                    \label{eq:LAction1}
                \end{equation}

                As in the Euclidean theory, we must impose Dirichlet boundary conditions on both the metric and the axion. 
                As described in the above sections, we take the boundary metric to be the static patch of 1+1 de Sitter space.  In particular, for now, we work with the full static patch without yet compactifying the time direction.  Our boundary conditions may thus be written
                
                \begin{equation}
g_{\mu\nu}dx^{\mu}dx^{\nu}|_{\tiny \partial\M_{L,R}} = \frac{\ell^2}{z^2_{L,R}}\Big(d\theta^2 -\sin^{2}(\theta) \,dt^2\Big) +\mathcal{O}(z_{L,R}^0), \ \ \
                    \chi(x)|_{\tiny \partial\M_{L,R}} = \chi_{L,R},
                \label{eq:LMetricBC1}
                \end{equation}
                where, as in the Euclidean case,  $z_{L,R}$ is an appropriate Fefferman-Graham-like coordinate that vanishes respectively at  $\partial\M_{L,R}$.
The boundary terms are now just the Lorentzian analogue of \eqref{eq:ESbndy}; i.e., 
                \begin{equation}
                S_{\tiny \partial\M} = \frac{1}{8\pi\Gn}\int_{\tiny \partial\M}d^2x\,\sqrt{|h|}\Big(\,K -\frac{1}{\ell} + \frac{\delta^2}{2\ell}\log(\delta)\mathcal{R}^{(2)}\Big),
                \label{eq:LSbndy}
            \end{equation}
            where each term is again evaluated on a regulated boundary at $z_{L,R}=\delta$.
            
            The equations of motion are also direct analogues of the Euclidean case. We will solve them using the analogous ansatz
              \begin{equation}
                ds^2 = dr^2 +\ell L(r)\big(d\theta^2 -\sin^2(\theta)dt^2\big),\quad \chi(x)=\chi(r).
                \label{eq:L2Ansatz}
            \end{equation}
We again make use of the conserved current
            \begin{equation}
            \label{eq:Lconsj}
                j_\mu =\partial_\mu\chi= \frac{ q}{\ell L(r)}\delta_\mu^r.
            \end{equation}
and the $rr-$component of Einstein's equations:
            \begin{equation}
                \big(\partial_r L)^2=\frac{4L}{\ell^2}\Big(\ell+\frac{q^2}{4L}+L\Big).
                \label{eq:LEOM}
            \end{equation}
            This differs from the Euclidean equation of motion \eqref{eq:Eeom} only by a sign in front of the term containing $q^2$. The Lorentzian sign prohibits the right-hand-side from vanishing for real $q$.  Thus $L(r)$ must be monotonic and  (as expected) there can be no real Lorentzian wormholes.

            Nevertheless, one can formally find a solution satisfying wormhole boundary conditions by analytically continuing the Euclidean solution found above. A useful analytic continuation takes the form

            \begin{equation}
                L(r)= \frac{\ell}{2}\Big(\sqrt{1-\frac{q^2}{\ell^2}}\cosh\big(\frac{2r}{\ell}\big)-1\Big).
                \label{eq:LA(r)}
            \end{equation}
            This solution is real-valued for $|q|<\ell$.  We will focus on this case below. There is also a real solution for $|q|> \ell$ which is obtained from \eqref{eq:LEOM} by shifting $\frac{r}{\ell} \rightarrow \frac{r}{\ell}  + \frac{i \pi}{4}$, though it has $L\rightarrow -\infty$ at one boundary and so fails to satisfy \eqref{eq:LMetricBC1} by a sign.
            
\begin{figure}[h!]
                \centering
                \includegraphics[width=0.55\linewidth]{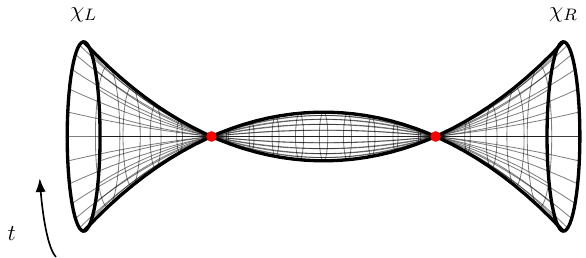}
                \caption{A constant $\theta$ section of our formal solution to the equations of motion of Lorentzian axion gravity. Here we take $t$ periodic in order to represent $L(r)$ by the size of the circle at each $r$.  Starting at left, constant $r$ slices degenerate to zero size (red dot) before re-expanding (with negative $L(r)$) in a region with spacelike $t$ and timelike $\theta$. Such surfaces degenerate again at another value of $r$ before re-expanding into a region geometrically-identical to the first but which connects to the right asymptotic boundary. The associated axion lives on a Riemann surface with logarithmic branch points at values of $r$ where $L(r)=0$.  If we take the axion to be real at the left boundary on some sheet of the Riemann surface, it will generally be complex on other sheets and in the middle and right regions.}
                \label{fig:LOSWormholes}
            \end{figure}

            Even for $|q|<\ell$, \eqref{eq:LA(r)} does not define a smooth Lorentzian metric. Indeed, the metric has curvature singularities at the two values of $r$ at which $L=0$.    Our formal solution instead contains three smooth pieces, with one piece attached to each asymptotic boundaries and third piece in the middle that is separated from the others by curvature singularities; see figure \ref{fig:LOSWormholes} (where we have taken $t$ to be periodic in order to visually depict the function $L(r)$ as the size of the relevant circle).  Furthermore, the middle piece has negative $L(r)$.

            We see from \eqref{eq:Lconsj} that the axion will diverge logarithmically at the loci $r^*_{\pm}$ where the metric degenerates.  As a result, if we require $\chi_L$ to be real and then extend our solution for the axion to the right beyond the first singularity ($r_-$), the axion becomes complex.  Indeed, depending on how one chooses to go around the branch points at $r=r^*_{\pm}$ it generally continues to be complex after crossing the second singularity at $r_+$.

  \subsection{Constrained Axion Wormholes}
        \label{ssec:ConstrainedWormholesConstruction}

        We are now ready to construct the desired dS-symmetric constrained Lorentzian wormholes.  As noted above, we wish to fix the area of a codimension-1 surface deep in the bulk. In order to specify this surface in a diffeomorphism-invariant manner, we take it to be the surface $\chi=0$.
        
        We can impose this constraint by using a Lagrange multiplier $\mu$ and adding the following term to our Lorentzian action \eqref{eq:LAction1}:

        \begin{eqnarray}
            S_\mu &=& \frac{\mu}{16\pi\Gn}\Bigg[\frac{1}{2}\int_{\cal M} \sqrt{|h|}\, d^2x\wedge d \left[\text{sgn}(\chi)\right]  -2T \ell L_0\Bigg] \\ &=& \frac{\mu}{16\pi\Gn}\Bigg[\int_{-\infty}^\infty dr\int_{\tiny\Sigma_r}d^2x\sqrt{|h|}\,\frac{1}{2}\partial_r\text{sgn}(\chi)  -2T \ell L_0\Bigg],
        \label{eq:Smu}
        \end{eqnarray}
        where $\sgn(\chi)$ is the sign of $\chi$, $\sqrt{h}\, d^2x$ the appropriately-oriented area 2-form on each $\chi=constant$ surface and we have now periodically identified the coordinate $t$ with period $T$ in order to give the $\chi=0$ surface a finite area.  Here we first write $S_\mu$ in a form that is manifestly invariant under diffeomorphisms and then, on the final line, we use our metric ansatz to rewrite it in a more convenient form using the surfaces $\Sigma_r$ at each value of our radial coordinate $r$.  As usual, this term will modify the usual equations of motion.  To cleanly determine the correct modifications, let us note that the quantity $\partial_r\sgn(\chi)$ gives a delta-function localized at the surface $\chi=0$, which we take to be at some\footnote{Strictly speaking, we should allow a general surface at this stage that might not lie at a constant value of $r$.  However, given our metric ansatz, the current treatment turns out to be sufficient.} $r_0$.  It is therefore useful to parameterize the variations of $\chi$ by using $\delta \chi(r)$ for $r\neq r_0$ and by also using $\delta r_0$.

With this understanding, the trace of the Einstein equations and the equations resulting from varying the axion yield

        \begin{equation}
            \begin{split}
                R=-\frac{6}{\ell^2}+\frac{1}{2}\partial_\mu\chi\partial^\mu\chi-2\mu\delta(r-r_0)\ \ \ \ \ \ \ \ \ \ \ \ \ \ \ \ \ \   \\
                    \begin{cases}
                        \nabla^2\chi=0 & r\neq r_0\\
                        \displaystyle \lim_{\epsilon\rightarrow0^+}(\partial_\mu\chi\partial^\mu\chi+\mu\sgn(\chi)\nabla_\mu n^\mu_r)|_{r_0-\epsilon}^{r_0+\epsilon}  = 0& r=r_0,
                    \end{cases}
            \end{split}
            \label{eq:CWEOMS}
        \end{equation}
         where, again, $n^\mu_r$ is the positively-directed unit normal to a constant $r$ surface. The second term in the final equation can be rewritten as the sum of the limits at $r_0$ (taken from  the left and right sides) of the traced extrinsic curvature  of constant surfaces of constant $r$. In addition, using \eqref{eq:Lconsj}, the first term can be written as $L(r_0)^{-2}$ times the difference in the squared axion flux ($q^2$) between the left and right sides of the surface $r=r_0$. (Note that there is no equation requiring continuity of $q$ at $r_0$.) Defining $L_0 = L(r_0)$, we may thus rewrite the final equation in \eqref{eq:CWEOMS} in the form:

        \begin{equation}
\frac{q_{\tiny{R}}^2-q_{\tiny{L}}^2}{L_0^2}+\mu\big(K_{\tiny{R}}-K_{\tiny{L}}\big)=0,
            \label{eq:AxionSource}
        \end{equation}
        where $q_{\tiny{L,R}}$ are axion fluxes to the left and right of $r_0$ and $K_L, K_R$ the limits at $r_0$ of the traced extrinsic curvature of constant $r_0$ surfaces, but where we now compute $K_L,K_R$ using the normal that is directed \emph{from} $r=r_0$ \emph{toward} the relevant $L,R$ asymptotic boundary  (rather than necessarily using the direction of increasing $r$).
        
        Suppose for a moment that the solutions to these modified equations of motion are $\mathbb{Z}_{2}$ symmetric. Then we immediately find $K_{\tiny{L}}=K_{\tiny{R}}$, whence \eqref{eq:AxionSource} requires  $q_{\tiny{R}}=q_{\tiny{L}}\equiv q$. It then remains only to work out the value of the Lagrange multiplier $\mu$. To do so, recall that as in the standard analysis   of the Israel junction conditions (see e.g. \cite{Poisson:2009pwt}) the Ricci scalar has a delta-function contribution $R_{\chi=0}$ localized at the $\chi=0$ surface which takes the form
        \begin{equation}    R_{{\tiny{\chi=0}}}=2(K_{\tiny{R}}+K_{\tiny{L}})\delta(r-r_0).
        \end{equation}
        Using the trace of Einstein's equations then yields

        \begin{equation}
            \mu=-\big(K_{\tiny{R}}+K_{\tiny{L}}\big).
            \label{eq:musolve}
        \end{equation}

\begin{figure}[h!]
        \centering
\includegraphics[width=0.8\linewidth]{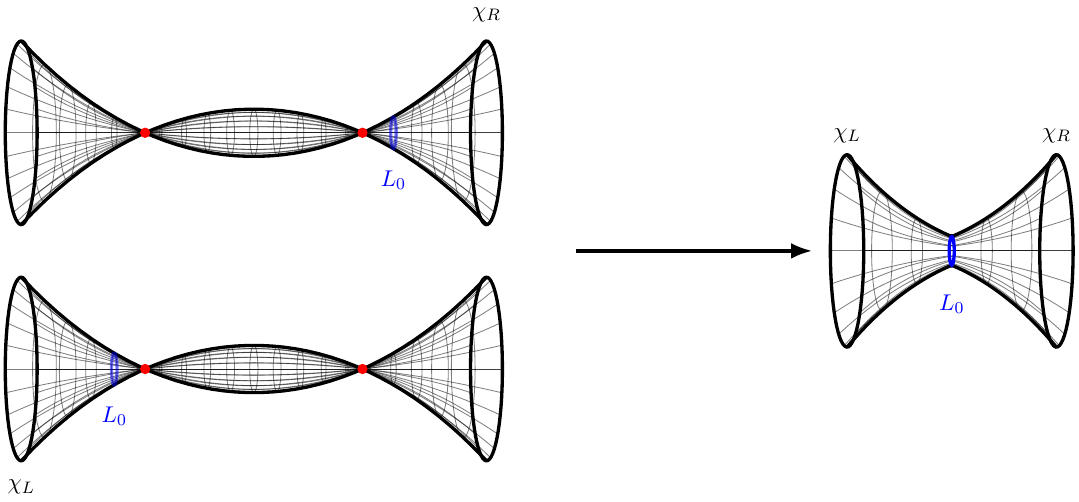}
        \caption{The construction of our Lorentzian constrained wormhole with parameter $L_0$.  We start with a pair of degenerate formal solutions, where we again draw the time direction as periodic. In each case, $\chi$ vanishes at the surface where $L(r)=L_0$.  In the first (second) solution, this occurs at the left-most (right-most) surface where $L(r)=L_0$ and, furthermore, the axion takes the desired value $\chi_L$ ($\chi_R$) at the left (right) boundary.  Cutting each solution where $\chi=0$ yields a smooth piece and a singular piece.  Discarding the latter, we glue the two smooth pieces together across the $\chi=0$ surface to form a single connected geometry in which $\chi$ runs continuously from $\chi_L$ to $\chi_R.$}
        \label{fig:CutAndPaste}
    \end{figure}

 Since the bulk equations of motion away from the $\chi=0$ are manifestly unchanged by our addition of the fixed area constraint, the solution in such regions is again characterized by the metric and axion profile \eqref{eq:L2Ansatz} with the previously-stated solution for $L(r)$:

        \begin{equation}
            L(r)= \frac{\ell}{2}\Big(\sqrt{1-\frac{q^2}{\ell^2}}\cosh\big(\frac{2r}{\ell}\big)-1\Big).
                \label{eq:LA(r)2}
        \end{equation}
        Thus, as  advertised, our ${\mathbb Z}_2$-symmetric constrained wormholes are obtained by taking two copies of a solution described in section \ref{ssec:LOSWormholes},  cutting the solutions along the constant-$r$ surface with $L=L_0$ that occurs before reaching any singularities, and then gluing the two non-singular pieces together along the cut; see figure \ref{fig:CutAndPaste}. 
Note that, due to the shift symmetry of $\chi$, it is not necessary to cut the solution from section \ref{ssec:LOSWormholes} along a $\chi=0$ surface.  One can instead cut along any surface and then simply shift the value of $\chi$ to set $\chi=0$ on the cut. Doing so generates a large set of constrained-wormhole solutions from which one can then select the ones that satisfy the desired boundary conditions and which have the desired value of $L_0$.   A similar cut-and-paste construction can also be applied to the $L>0$ regions of the $|q|>\ell$ solutions mentioned below \eqref{eq:LA(r)}.

\subsection{Singularities and regulators}
\label{sssec:singreg}

\begin{figure}[h!]
                    \centering
                    \includegraphics[width=0.9\linewidth]{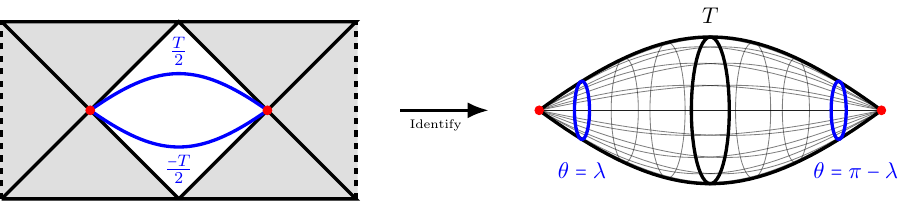}
                    \caption{Our boundary spacetime, and in fact each constant $r$ slice of the bulk, is constructed by identifying two time-slices of the static patch of $\text{dS}_2$ (left). The resulting geometry (right) takes the shape of a (American) football. On either end are codimension-2 conical defects at what  were the horizon bifurcation points in $\text{dS}_2$. We regulate these defects by cutting off the geometry at constant $\theta=\lambda,\pi-\lambda$ surfaces shown in blue on the right side of the figure, and by identifying these two surfaces. }
                    \label{fig:LBCs}
                \end{figure}

We have now constructed a two-parameter family of ${\mathbb Z}_2$-symmetric (and dS$_{1+1}$-symmetric) constrained wormholes labeled by the asymptotic value of the axion $\chi_\infty$ and by the value $L_0$ of the metric variable $L$ at the surface $\chi=0$.  As described in section \ref{sec:overview}, we in fact wish to identify the Lorentz-signature boundary-static-patch time coordinate $t$ with finite period $T$. 
One class of constrained wormholes with such boundary conditions can clearly be constructed from the constrained wormholes of section \ref{ssec:ConstrainedWormholesConstruction} by identifying the bulk static patch under the time translation $t\rightarrow t+T$.  This creates a Lorentzian conical defect on the bifurcation surface of the dS$_{1+1}$ that (in the uncompactified geometry) lies at each value of $r$; see figure \ref{fig:LBCs}.  Since each such bifurcation surface is a pair of points (an $S^0$), the result is a pair of defect lines that thread the wormhole and reach the asymptotic boundaries.  The resulting constrained wormholes are then labeled by $\chi_\infty, L_0$ and $T$.

However, as described in section \ref{sec:Background}, this is not yet a sufficient family of constrained solutions for the desired study of our Lorentzian path integral.  Instead, we need to find such a solution for each possible area (here, length) $L_\gamma$ of each Lorentzian conical defect.   Since the defects stretch to infinity, $L_\gamma$ is in fact infinite in our family of solutions.  But with an appropriate regulator it will become finite, and in the above solutions it will be a fixed function of $\chi_\infty$ and $L_0$ (it will not depend on $T$).

We thus need to generalize our solutions by introducing an additional free parameter that allows $L_\gamma$ to vary independently of $\chi_\infty$ and $L_0$.  We can do so by employing the boundary-conformal-transformation trick mentioned at the end of section \ref{sec:overview}. 
Recall in particular that we have fixed the boundary metric to be
\begin{equation}
\label{eq:Bg}
ds^2_{\partial {\cal M}_R} = d\theta^2 -\sin^2 (\theta) dt^2,
\end{equation}
and that we have compactified the $t$ coordinate with period $T$. If we now introduce an arbitrary $\alpha \in {\mathbb R}^+$, and if we define a new time coordinate $t_\alpha = t/\alpha$, the boundary metric becomes  
\begin{equation}
ds^2_{\partial {\cal M}_R} = d\theta^2 -\alpha^2 \sin^2 (\theta) dt^2_\alpha,
\end{equation}
where $t_\alpha$ has period $\alpha T$.  The point is then that, in two dimensions, any two metrics with the same topology are related by a Weyl transformation and a change of coordinates.  It is thus possible to introduce a function $\theta_\alpha(\theta)$ such that we have
\begin{equation}
ds^2_{\partial {\cal M}_R} = \Omega^2(\theta) \left( d\theta_\alpha^2 - \sin^2 (\theta) dt^2_\alpha\right),
\end{equation}
which is just an overall conformal factor
$\Omega^2(\theta)$ times the desired boundary metric \eqref{eq:Bg}.  As in \cite{marolf:EntanglementPhaseTransition}, the desired transformation turns out to be
 \begin{equation}
            \begin{split}
                &\qquad\quad \Omega =\alpha\sin(\theta)\frac{1+\tan^{2/\alpha}(\theta/2)}{2\tan^{1/\alpha}(\theta/2)}\\
                &\qquad\qquad\theta=2\arctan(\tan^\alpha({\theta}_\alpha/2)),
            \end{split}
            \label{eq:S2toDefectS2}
        \end{equation}
where $\theta_\alpha$ again takes values in $[0,\pi]$.

As a result, constrained wormholes satisfying the desired boundary conditions can be constructed from those in section \ref{ssec:ConstrainedWormholesConstruction} by first renaming the bulk static patch coordinate to $t_\alpha$, then identifying the bulk static patch on each constant-$r$ slice under $t_\alpha \rightarrow \alpha T$, and finally inverting the above boundary conformal transformation (i.e., by changing between corresponding sets of Fefferman-Graham coordinates in the bulk) to use the boundary conformal frame \eqref{eq:Bg}.  This procedure will change 
$L_\gamma$ since our regulator will be determined by the choice of boundary conformal frame or, equivalently, by the choice of Fefferman-Graham coordinate $z$ in the bulk, and since this has now been altered.  Thus $L_\gamma$ becomes a function of $\alpha, \chi_\infty$ and $L_0.$  
We will be able to determine the precise dependence of $L_\gamma$ on $\alpha, L_0$ and $T$ after introducing a definite regulator below to make $L_\gamma$ finite.
We will then invert this relation to write $\alpha$ in terms of $L_\gamma,\chi_\infty, L_0$.  Note that the asymptotic values $\pm\chi_\infty$ of our axion are manifestly unaffected by this transformation\footnote{In the language of AdS$_{d+1}$/CFT$_d$, the point is that boundary sources for our axion have conformal dimension zero.  This is always the case for massless bulk axions (or scalars), for which the dual operator always has dimension $d$.}.

Doing so  yields a family of constrained wormholes labeled by independent parameters $L_\gamma,\chi_\infty, L_0, T$ as desired. We emphasize that, with the above conventions, the bulk static patch time of \eqref{eq:L2Ansatz} is now called $t_\alpha$ and has period $\alpha T$.  As a result, the area of the $\chi=0$ surface is 
\begin{equation}
\label{eq:AL001}
{\cal A}_0 =  2\alpha\ell T L_0.   
\end{equation} 

However, as noted in the strategic overview in section \ref{sec:overview}, a remaining issue is that our boundary metrics are not smooth.  As a result, the standard boundary counter-terms will not suffice to make the action finite, nor are we guaranteed that they will make our action into a valid variational principle.  

Now, we suspect that both of these issues can be dealt with by simply adding an additional counter-term to the action to cancel divergences associated with the infinite length of the Lorentzian defect. Furthermore, we suspect that there is a natural way to formulate this counter-term so that one can meaningfully compare the resulting actions for boundary metrics with different periods for the boundary-static-patch-time (as is needed if we wish to integrate over this period).  If we had such a formalism in hand,  then we would have found the desired class of solutions and we could proceed with our analysis of the semiclassical approximation.
However, rather than show this to be true in general, we will proceed by introducing a UV regulator near the boundary de Sitter horizon as described in section \ref{sssec:singreg}.  This regulator will transform our problem into one with a smooth boundary metric and will thus yield a finite action and a good variational principle.

The regulator we use takes the form of a simple surgery on each boundary metric.  Choosing some small $\lambda \in {\mathbb R}_+$, we merely restrict the range of the boundary coordinate $\theta_\alpha$ to $[\lambda, \pi-\lambda]$ and then identify the circle $\theta_\alpha = \lambda$ with  $\theta_\alpha = \pi - \lambda$ on the same boundary.  This turns our boundary manifold into a non-singular Lorentzian torus.

\begin{figure}
    \centering
    \includegraphics[width=0.4\linewidth]{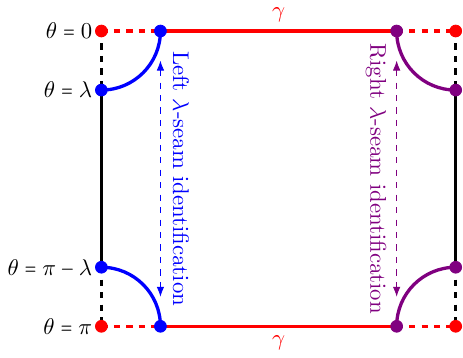}
    \caption{The left and right $\lambda$-seam identifications  are defined by choosing a pair of surfaces anchored to $\theta = \lambda, \pi-\lambda$ at each boundary and which cut off the associated piece of the defect $\gamma$ (red lines) at either $\theta=0$ or $\theta=\pi$. The left such surfaces have been drawn as blue lines while the right such surfaces are purple lines.  The surfaces are chosen to preserve the ${\mathbb Z}_2$ symmetry $\theta \rightarrow \pi-\theta$.  This symmetry is then used to identify each pair of surfaces as shown by the blue and purple dashed arrows.}
\label{fig:RegulatingIdentification3d}
\end{figure}

We now perform a related surgery on our constrained wormholes as follows.  In each such spacetime, and for each boundary, we choose a convenient codimension-1 surface that is anchored on the boundary to the surface $\theta_\alpha = \lambda$ and which cuts off the defect at $\theta=0$, and we choose a corresponding surface anchored to  $\theta_\alpha =\pi-\lambda$ so that we preserve the ${\mathbb Z}_2$ symmetry $\theta \rightarrow \pi-\theta$; see figure \ref{fig:RegulatingIdentification3d}.  At this stage these surfaces are largely arbitrary, though we will soon describe a class of choices that turns out to be useful.  In any case, after making such a choice we remove the regions `outside' these surfaces (i.e., we remove the regions in which the defects extend to infinity).  Having done so, we then then identify the two chosen surfaces that are anchored to the left boundary, and similarly on the right; see figure \ref{fig:RegulatingIdentification3d}.  We refer to the resulting identifications as the right and left `$\lambda$-seams'.  

The result is then a constrained wormhole spacetime that satisfies the new regulated boundary conditions.  Unfortunately, however, it will generally fail to be a constrained {\it saddle} because the extrinsic curvature can experience a jump across the junction where the regulating surfaces were identified. Our task is therefore to choose regulating surfaces on which the extrinsic curvature is as small as possible and, in particular, where the extrinsic curvature vanishes in the limit $\lambda \rightarrow 0.$

It is thus useful to note that, in the limit $\lambda \rightarrow 0$, each surface will be very close to the AdS boundary.  As a result, the spacetime along the surface will be well-approximated by empty AdS space.  Furthermore, in empty AdS it is straightforward to find a surface with {\it exactly} vanishing extrinsic curvature that is anchored to the boundary on any circle.  Indeed, in empty AdS this surface is related by an AdS isometry to the surface $\theta=\pi/2$ which is manifestly invariant under the ${\mathbb Z}_2$-symmetry $\theta \rightarrow \pi- \theta$.  As a result, we may expect our construction at small $\lambda$ to give a good approximation to the actual $\lambda$-regulated saddle
so long as we choose the regulating surfaces to be close to the surface that would have vanishing extrinsic curvature in empty AdS space.

In particular, appendix \ref{sec:details} studies a specific choice of this surface defined by taking the surface to have the same coordinate form as the surface of vanishing extrinsic curvature in empty global AdS (in the above coordinates).  It then  verifies that the jump in extrinsic curvature vanishes as $\lambda \rightarrow 0$ for a given bulk wormhole (with fixed $T, \alpha, L_0$) in the limit $\lambda\rightarrow 0$.  It then combines this result with a first-order Taylor-series approximation to estimate the difference in action $\Delta S$ between our wormhole and the actual solution to the $\lambda$-regulated problem.  Doing so indicates that $\Delta S$ also vanishes in this limit, as desired.  We may therefore proceed to use the action of our cut-and-paste solution as an approximation to the action of the desired saddles at small $\lambda$.

Importantly, the parameter $L_\gamma$ is now finite.  In particular, our surgery has transformed what before was a pair of infinite line defects into a single defect with topology $S^1$ as shown in figure \ref{fig:RegulatingIdentification3d}.  If one starts at the $\chi=0$ surface $r=r_0$ and at $\theta=0$, one can trace the defect up to the   extremal surface junction associated with our UV cutoff near the left boundary where $\theta=0$ is identified with $\theta=\pi$, across the junction and back to $r_0$ at $\theta=\pi$, and then similarly across the junction near the right boundary to return again to $r_0$ at $\theta=0$.
It will be convenient to henceforth define this $S^1$ defect to have proper length $2L_\gamma$.

\section{Constrained Wormholes and the Lorentzian Path Integral}
    \label{sec:ConstrainedWormholes}

Having constructed the desired constrained Lorentzian wormholes above, it now remains only to compute their action as a function of $L_0, L_\gamma$, $\chi_\infty$, and $T$ and to integrate over the parameters $L_0, L_\gamma$, and $T$.  Results for the action and expressions for $q,\alpha,r_0$ in terms of $L_0, L_\gamma, \chi_\infty, T$ are described in section \ref{ssec:AP}.  The integrals over 
$L_0, L_\gamma$, and $T$ are then analyzed in sections \ref{ssec:SPCD} and \ref{ssec:AnalyticContinue}.

Before proceeding, we should emphasize that our study below is performed at each finite value of the regulator $\lambda$.  Now, due to the surgery associated with $\lambda$, each boundary manifold has become a torus.  As a result, even with periodic Lorentzian time $t$, there can be completely smooth wormhole geometries in which no defect appears.  However, the action of any such geometry is real, making our integrand a pure phase.  This is in sharp contrast to the exponential enhancement $e^{A_\gamma/4G}=e^{L_\gamma/2G}$ noted above when there is a non-trivial ${\cal N}=0$ defect $\gamma$.  In particular, as $\lambda 
\rightarrow 0$ this leads to infinite suppression of the defect-free geometries relative to the natural class of defect-geometries in which $L_\gamma$ grows with $\ln \lambda$.  We will therefore simply ignore contributions from defect-free geometries, though we will check below that there is indeed a range of $L_\gamma$ where our integrand is exponentially enhanced.

\subsection{Action and parameters}
\label{ssec:AP}

To begin, recall that our constraint at the $\chi=0$ surface allows the 
extrinsic curvature of constant-$r$ surfaces to be discontinuous at $r_0$.  As is familiar from studies of the Israel junction conditions (see e.g. \cite{Poisson:2009pwt}), this discontinuity leads to a delta-function in the bulk Ricci scalar.  The result is that the contribution of this delta-function to the Einstein-Hilbert action is precisely the difference between Gibbons-Hawking-York boundary terms evaluated on either side of the junction. 

It is also useful to separate out the contribution to the Einstein-Hilbert action from the codimension-2 delta-function on the Lorentzian defect.  Recall that, after performing the surgery described in section \ref{sssec:singreg}, we have an $S^1$ defect of length $2L_\gamma$.  We will then define $2S_\gamma$ to be the sum of the contributions from the terms in \eqref{eq:defectSEH} associated with the area $A_\gamma$ (here $2L_\gamma$) and with the extrinsic curvature at $\partial U_\epsilon$.  The former term is imaginary and (since ${\cal N}=0$) gives just $-2i \frac{L_\gamma}{4G_N}$.  The latter term is real and is manifestly proportional to $L_\gamma \alpha T$ (since the bulk time coordinate $t$ in \eqref{eq:L2Ansatz} has period $\alpha T$).  The remaining coefficient is computed in appendix \ref{sec:details} and yields
\begin{equation}
S_\gamma = -\frac{L_\gamma}{8\pi\Gn}(\alpha T+2\pi i).
\end{equation}

However, our metrics are smooth at all other points in our spacetime. 
For $\Z_2$-symmetric constrained wormholes, we may thus write the action in the  form
        \begin{equation}
            \begin{split}
                S_{\tiny{CWH}}=&\frac{2}{16\pi\Gn}\Bigg(\int_{\M_{\tiny{R}}} d^3x\,\sqrt{|g|}\,\Big(R_\text{\tiny{reg.}}+\frac{2}{\ell^2}-\frac{1}{2}\partial_\mu\chi\partial^\mu\chi\Big) 
                \\&+2\int_{\partial\M_{\tiny{R}}}d^2x\,\sqrt{|h|}\,K 
                +2\int_{\chi=0}d^2x\,\sqrt{|h|}\,K +16\pi\Gn S_{CT,{\tiny{R}}}\Bigg) + 2 S_\gamma,
            \end{split}
            \label{eq:CWHAction}
        \end{equation}
where   $R_\text{\tiny{reg.}}$ is the regular part of the Ricci curvature obtained by removing the singular contributions from the codimension-1 and codimension-2 defects discussed above, and where
$S_{CT,{\tiny{R}}}$ is given by the usual counter-terms at the right asymptotic boundary $\partial \M_R$.  Here we integrate only over the half wormhole $(\M_{\tiny{R}})$ to the right of the $\chi=0$ surface, but we have multiplied the result by $2$ to obtain the action for the full wormhole.  Note that there is no explicit contribution from the Lagrange multiplier term $S_\mu$ since our spacetimes satisfy the associated constraint and thus have $S_\mu=0.$ 

The remaining terms in the above action are evaluated explicitly in appendix \ref{sec:details}.  The result takes the form 
        \begin{equation}
        \begin{split}
            S_\text{\tiny{CWH}} &= \frac{-\alpha T}{8\pi\Gn}\Bigg[2 L_\gamma \bigg(1+\frac{2\pi i}{\alpha T}\bigg) +2\bigg(2r_0+\frac{\ell+\ell\log(\lambda/2)}{\alpha}\bigg)\\
            &\qquad +\ell\bigg(\log\big(1-\frac{q^2}{\ell^2}\big)+2\alpha\log(\lambda/2)\bigg) +2\sqrt{\ell^2-q^2}\sinh\Big(\frac{2r_0}{\ell}\Big)\Bigg].
            \end{split}
            \label{eq:CWHAction_Explicit}
        \end{equation} 
For simplicity, we have expressed  the above action in terms of the axion flux $q$, the $r$-coordinate  $r_0$ of the $\chi=0$ surface (as defined by \eqref{eq:LA(r)}, and the conical defect angle $2\pi\alpha$.  But both $\alpha$ and $q$ are in fact determined by the parameters $L_0, L_\gamma, T$ and the boundary condition $\chi_R = - \chi_L = \chi_\infty$.  

As in section \ref{ssec:EOSWormholes}, it is straightforward to integrate \eqref{eq:Lconsj} from $r_0$ to infinity to relate $q$ to $\chi_\infty$ and $r_0$.  Solving for $r_0$ yields
        \begin{equation}
        \label{eq:r0solve}
            r_0= \ell \arctanh\Bigg(\frac{q \tanh\Big(\frac{\chi_\infty}{2}+\arctanh\big(\frac{\ell+\sqrt{\ell^2-q^2}}{q}\big)\Big)}{\ell+\sqrt{\ell^2-q^2}}\Bigg).
        \end{equation}
        With this understanding
        the expression \eqref{eq:CWHAction_Explicit} is valid for constrained wormholes with any value of $q$, and in particular for those built from both the $|q|<\ell$ and the $|q|>\ell$ solutions described below \eqref{eq:LA(r)}, though for the latter this $r_0$ has a non-zero imaginary part (equal to $-\ell\pi/4$). Indeed, while the imaginary part of $r_0$ is discontinuous at $q=\ell$, the action \eqref{eq:CWHAction_Explicit} is both finite and continuous there (as one may check by using the additional equations below).

Note that we also have the condition
        \begin{equation}
            L(r_0)= -\frac{\ell}{2}\Big(1-\sqrt{1-\frac{q^2}{\ell^2}}\cosh\big(\frac{2r_0}{\ell}\big)\Big)=L_0,
        \end{equation}
which can be solved for the axion flux $q$ to find
        \begin{equation}
            \label{eq:qresult} q=\sqrt{2L_0\big(L_0+2\ell+L_0\cosh(\chi_\infty)\big)}\sinh\bigg(\frac{\chi_\infty}{2}\bigg)+L_0\sinh(\chi_\infty).
        \end{equation}
It now remains only to express $\alpha$ in terms of $L_0, L_\gamma$ and $T$. We can do this simply by explicitly computing the geometric length $L_\gamma$ of the defect $\gamma$ and solving for $\alpha$.   To do so, let us recall that the defects in our bulk solution are at $\theta_\alpha=\lambda$ and $\theta_\alpha=\pi-\lambda$ and, before identifying the surfaces along the $\lambda$-seam,  run from one boundary to the other. Using the fact that our constrained wormholes are $\mathbb{Z}_2$ symmetric about the the surface at $r=r_0$, we simply integrate the line element from $r=r_0$ to $z\approx \lambda_\alpha := \theta_\alpha\Big|_{\theta=\lambda}$   to find 

\begin{eqnarray}
    \label{eq:alphasolve} 
    L_\gamma &&= 2 \int_{\theta=0} dr  = -2\bigg(r_0 +\frac{\ell}{4}\log\big(1-\frac{q^2}{\ell^2}\big)+\alpha\ell\log(\lambda/2)\bigg)+\mathcal{O}(\lambda_\alpha)\\
            \rm{and\ thus} \ \ \alpha &&= \frac{\frac{1}{2}L_\gamma+r_0+\frac{1}{4}\ell\log\big(1-\frac{q^2}{\ell^2}\big)+O(\lambda_\alpha)}{-\ell \log\big(\frac{\lambda}{2}\big)}.
            \label{eq:alphasolve2}
\end{eqnarray}
One may again use \eqref{eq:r0solve} to check that expression is again continuous at $q=\ell,$

         We have thus expressed all parameters in \eqref{eq:CWHAction_Explicit} in terms of $T$, $L_0$, $L_\gamma$, the boundary condition $\chi_\infty$, and the regulator $\lambda$.  A particularly useful application of \eqref{eq:alphasolve2} is to write $S_{CWH}$ in the form

        \begin{equation}
            S_\text{\tiny{CWH}} = - A L_\gamma^2 + B L_\gamma+ C,
            \label{eq:CWHAction_Explicit2}
        \end{equation}
        where  we have $A = \frac{T}{-16\pi G_N \ell \log (\frac{\lambda}{2})}$ (which is positive for real $T>0$ since $\lambda$ is small) and the coefficients $B,C$ contain all of the $L_0$-dependence.          In particular, 
    \begin{equation}
        B=\frac{T}{16\pi G_N\ell \log(\lambda/2)}\left(4r_0+\ell \log(1-\frac{q^2}{\ell^2}) +2\sqrt{\ell^2-q^2}  \sinh(\frac{2r_0}{\ell}) \right) -\frac{i}{2 G_N}
    \end{equation}  
        \begin{equation}
        \begin{split}
        C=\frac{T\left(4r_0+\ell \log(1-\frac{q^2}{\ell^2})\right)}{64\pi G_N\ell \log(\lambda/2)}  \left(4r_0+\ell \log(1-\frac{q^2}{\ell^2}) +4\sqrt{\ell^2-q^2}  \sinh(\frac{2r_0}{\ell}) \right) -\frac{\ell T\big(1+\log(\lambda/2)\big)}{4\pi G_N}\label{eq:C}
        \end{split}
    \end{equation} 
    For later purposes, it will be useful to note that in the limit $\lambda\rightarrow 0$
 with fixed $r_0$ and $q$ (or, equivalently with fixed $\chi,L_0)$ both $B$ and $C$ are dominated by the final term in each line.  In this limit, for all complex values of $\chi_\infty$ we thus find the $L_\gamma$ saddle  to be 
 \begin{equation}
 L_\gamma^*=\frac{B}{2A}=\frac{-4\pi \ell \log(\frac{\lambda}{2})}{iT} +\dots 
 \label{eq:Lgamma*}
 \end{equation}
 at fixed $T,L_0$.  In particular, since $\lambda$ is small, $L_\gamma^*$ is real and positive for $T=-i\beta$ with real positive $\beta$.

    \subsection{Stationary Points and Contour Deformation}
\label{ssec:SPCD}
           
With the explicit form of the on-shell action for our constrained wormholes in hand, we are now prepared to analyze the remaining integrals over $T,L_0,L_\gamma$.  Before doing so, let us pause to note that our Lagrange multiplier term \eqref{eq:Smu} in fact fixes the area of the $\chi=0$ surface which, for the spacetimes described above, is in fact $2\alpha L_0 \ell T$.   However, since $\alpha$ is expressed through \eqref{eq:alphasolve} as a function of $L_\gamma, L_0$, the boundary condition $\chi_\infty$ and the regulator parameter $\lambda$, integrating over 
$T,L_0,L_\gamma$ does indeed integrate over all values of this area.
Furthermore, since we work only at the leading semiclassical order, the detailed choice of integration measure will not affect our results.    We will thus henceforth use the simple measure $dL_\gamma dT dL_0$, where we integrated $T$ over the entire real line while both $L_0$ and $L_\gamma$ are integrated over ${\mathbb R}^+$.  

Taking the independent parameters to be $T,L_0,L_\gamma$ has the convenient property that \eqref{eq:alphasolve} allows the action \eqref{eq:CWHAction_Explicit} to be written as an explicit 
(if complicated) analytic functions of $L_\gamma, L_0$, and $T$.  However, the interested reader may consult appendix \ref{sec:AlternateParamatrization} for a study using the perhaps more natural alternative parameterization 
in terms of $L_\gamma$, $T$, and $\tilde L_0 = \alpha L_0$ (and using the corresponding measure $d L_\gamma dT d\tilde L_0$) and which finds equivalent results by using  somewhat more numerics.

We should also discuss the order in which we will perform the remaining integrals.  We argued in section \ref{sec:Background} that the $T$-integral should be performed before the integral over $L_\gamma$.  We also suggested that the order of the other integrations should not be important so long as the $T$-integral is performed late enough that the associated energy has become bounded below.   We will thus choose here to integrate first over $T$, followed by $L_0$ and finally $L_\gamma$.  That other orders of integration do indeed give the same result (so long as the $T$-integral is performed before the $L_\gamma$ integral) is then argued in section \ref{sec:Discussion}.   However, as we will see shortly, if the $T$-integral is performed first  the double integral over $L_0,L_\gamma$ turns out to converge absolutely, so the final two integrals can be performed in any order.

\begin{figure}[h!]
\begin{center} 
\includegraphics[width=0.4\linewidth]{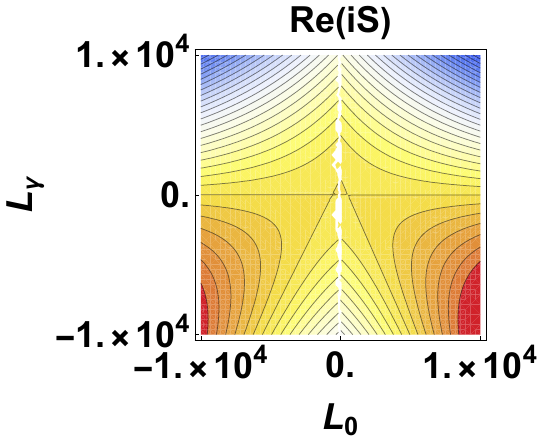}
    \caption{A contour plot of $\text{Re}(iS)$ in the $L_0,L_\gamma$ plane for $\chi_\infty= 1/2$ real $\beta=2\pi$. We have taken $\ell=1,\,8\pi\Gn=1,\, \lambda=10^{-2^8},\,L_\gamma=2^8.$ Warmer colors indicate more positive values of $\text{Re}(iS)$ while cooler colors represent more negative values. The fact that $\text{Re}(iS)$ decays in the upper right quadrant indicates that the double integral over real positive $L_0,L_\gamma$ converges absolutely in agreement with the analytic analysis. }
    \label{fig:A0AgammaAsymptotics}
\end{center}
\end{figure}

In particular, let us recall from sections  \ref{sec:Background} and \ref{sec:overview} that our remaining integrals take the form

        \begin{equation}
            \begin{split}
Z_2^\text{\tiny c}(\beta) : =  Z({\cal B}^\lambda \sqcup {\cal B}^\lambda,\beta)- [Z({\cal B}^\lambda,\beta)]^2 &=\int_{L_0,L_\gamma \in {\mathbb R}^+} dL_0\,dL_\gamma\int_{T\in{\mathbb R}} dT \,f_\beta(T)\,Z^\text{\tiny c}[{\cal B}_T^\lambda \sqcup {\cal B}_T^\lambda ; L_0, L_\gamma]\\
                &\approx\int_{L_0,L_\gamma \in {\mathbb R}^+} dL_0\,dL_\gamma\int_{T\in{\mathbb R}} dT \,f_\beta(T)\,\exp\Big[iS_\text{\tiny CWH}[T,L_\gamma,L_0]\Big],
            \end{split}  
            \label{eq:ZvarInt}
        \end{equation}
where $f_\beta (T)= \frac{1}{2\pi i}\frac{e^{-E_0(\beta-iT)}}{T+i\beta}$ for any (perhaps $\lambda$-dependent) lower bound 
$E_0$ 
on the energies of the gravitational system\footnote{The approximation in the final step of \eqref{eq:ZvarInt} ignores both loop corrections and non-perturbative corrections.  However, as noted in section \ref{sssec:singreg}, our Dirichlet axion boundary conditions are invariant under arbitrary changes of boundary conformal frame.  They are thus in particular invariant under arbitrary conformal isometries of the boundary sphere.   This gives rise to a set of exact zero modes associated with the action of arbitrary conformal isometries which should multiply the final expression by the infinite volume of this group.  We have suppressed this factor as the quantity of physical interest will be the connected partition function per unit volume of this group.}.  We should therefore check that the appropriate notion of energy is in fact bounded below in our context.   Since $S_{CWH}$ is linear in $T$, this can be done by writing
       \begin{equation}
       \label{eq:SCWHEform}
           S_\text{CWH}= -ET-i\frac{2L_\gamma}{4\Gn}.
       \end{equation}
       Hamilton-Jacobi theory then guarantees that this $E$ is indeed the energy of our constrained wormhole (as defined by the time-translation invariance of the action that includes the Lagrange multiplier term $S_{\mu}$)\footnote{Though, since this time translation is defined to act on {\it both} the right and left boundaries, our $E$ is in fact twice the value of the boundary Hamiltonian associated with the de Sitter static patch on either boundary separately.}.
Furthermore, using expressions \eqref{eq:CWHAction_Explicit2}-\eqref{eq:C}, one may check that $E$ is bounded below as desired.    In particular, since $A>0$ this property is manifest at each fixed $L_0$.  Furthermore,   since we have already noted that \eqref{eq:CWHAction_Explicit} is continuous in $q$, it is also continuous in $L_0$.  Thus $E$ must be continuous in $L_0$ as well.   To complete the argument that $E$ is bounded below, one thus need only expand \eqref{eq:r0solve}, \eqref{eq:alphasolve}, \eqref{eq:qresult}, and \eqref{eq:CWHAction} at large $L_0$ to write
\begin{equation}
\label{eq:Easympt}
\begin{split}
E  &=\frac{2L_\gamma + 4r_0 +\ell \log\Big(1-\frac{q^2}{\ell^2}\Big)}{-16\pi\Gn\ell\log(\lambda/2)}\Bigg[\frac{1}{2}L_\gamma+r_0 +\frac{1}{4}\ell\log\Big(1-\frac{q^2}{\ell^2}\Big)+\sqrt{\ell^2-q^2}\sinh\Big(\frac{2r_0}{\ell}\Big)\Bigg] +\frac{\ell\big(1+\log(\lambda/2)\big)}{4\pi\Gn}\\
&= \frac{-\cosh(\chi_\infty)}{4\pi\Gn\log(\lambda/2)} L_0\log\big(\frac{L_0}{\ell}\big) - \frac{\cosh(\chi_\infty)\Big(L_\gamma + 2r_0+i\pi\frac{\ell}{2} + \ell\log\big(\sinh(\chi_\infty)\big)\Big)}{4\pi\Gn\ell\log(\lambda/2)}L_0 +\mathcal{O}({L_0}^0),
\end{split}
\end{equation}
where we note that $r_0 + i \pi \frac{\ell}{4}$ approaches a positive real $\chi_\infty$-dependent constant as $L_0 \rightarrow +\infty$ with positive real $\chi_\infty$. Noting that the $L_0\log (L_0/\ell)$ term dominates at large $L_0$, and noting also that its coefficient is positive for small $\lambda$, it follows that $E$ is bounded below as desired; see also figure \ref{fig:A0AgammaAsymptotics} for numerical confirmation.

 There is a sense in which localizing the $T$ integral at the pole $T=-i\beta$ implements the usual Wick rotation $t\rightarrow -i\beta$ with the quantity $\beta E - \frac{L_\gamma}{4G_N}$ playing the role of the usual Euclidean action (in terms of which our integrand is $e^{-S_E}$). However, in our present context there is no room for ambiguity or guess-work about what the `right' Wick rotation might be.  In particular, since the metric and axion profiles on a $t=constant$ slice of any constrained wormhole were already independent of $T$, neither of these profiles can change under this notion of Wick rotation.  As a result, both profiles remain manifestly real.  In particular, the parameters $L_0$ and $\chi_\infty$ remain real.  Picking apart the computations already performed above, we thus see that the axion kinetic term has contributed {\it positively} to the Euclidean action $S_E := - i S_{CWH}\Big|_{T=-i\beta} = \beta E - \frac{L_\gamma}{4G_N}$.  This should not be a surprise since, given our Lorentzian starting point, any treatment of our axion will be indistinguishable from the corresponding treatment of a minimally-coupled scalar.  Here we have focused on partition functions defined by real $\chi_\infty$, though we will return to address the possibility of analytically continuing $\chi_\infty$ to complex values after discussing the $L_0,L_\gamma$ integrals below.

After performing the $T$ integral we are left with
        \begin{equation}
        Z_2^\text{\tiny c}(\beta) 
        =\int_{L_0,L_\gamma \in {\mathbb R}^+} dL_0\,dL_\gamma \exp\Big[-\beta E[L_\gamma,L_0]+\frac{L_\gamma}{4\Gn}\Big]
               =\int_{L_0,L_\gamma \in {\mathbb R}^+} dL_0\,dL_\gamma e^{-S_E}.
            \label{eq:ZvarIntEuclidean}
        \end{equation}
In particular, since we found above that $E \rightarrow +\infty$ whenever $L_0$ and/or\footnote{From figure \ref{fig:A0AgammaAsymptotics} it may not be clear that this is the case when we take $L_0\rightarrow \infty$ with $L_\gamma=0$.  However, this follows from the analytic expansion \eqref{eq:Easympt} for $E$ in which the leading $L_0\log L_0$ term is independent of $L_\gamma.$} $L_\gamma$ tend to $+\infty$, the double integral \eqref{eq:ZvarIntEuclidean} converges absolutely and we are free to perform the $L_0, L_\gamma$ integrals in either order.

We choose to integrate over $L_0$ before integrating over $L_\gamma$.  Since our Lagrange multiplier term \eqref{eq:Smu} fixes the area of the $\chi=0$ surface, and since induced geometry is canonically conjugate to extrinsic curvature, it is natural to expect such $L_0$-saddles to arise when the traced extrinsic curvature $K_R$ vanishes at the $\chi=0$ surface -- a condition we naturally denote by writing $K\Big|_{\chi=0}=0$.\footnote{{Since we computed $S_{CWH}$ using only an approximation to the actual solution to our variational problem,  we should in fact find $K\Big|_{\chi=0}=0$ at values of $L_0$ for which $\frac{\partial S_E}{\partial L_0}=0$ only in the limit $L_\gamma \rightarrow\infty,\ \lambda\rightarrow0$. However we have checked numerically that at fixed finite $\alpha$ and $\lambda$ the variation of the action (with respect to $L_0$) at the point in the $L_0$ plane at which $K\Big|_{\chi=0}=0$ is nevertheless $\mathcal{O}(\frac{\alpha}{16\pi\Gn})\times10^{-15}$ for a wide range of $\lambda\ll1$ {and moderate values of $L_\gamma$}.}}  We will focus on such saddles in this section.  However, as discussed in appendix \ref{app:spurious}, the fact that the $\chi=0$ area is proportional to $\alpha L_0$ (and that we choose to parameterize our constrained wormholes using $L_\gamma,L_0$) can potentially lead to other spurious saddles.  Nevertheless, appendix
\ref{app:spurious} also verifies that such spurious saddles are irrelevant to the analysis below.  In addition, it is possible to check our results by using an alternate (but more complicated) parametrization in terms of $L_\gamma$ and $\tilde L_0=\alpha L_0$ for which no such spurious saddles can arise.  For the interested reader, an analysis using this more complicated parameterization is presented in appendix \ref{sec:AlternateParamatrization}.

Now, as seen already in section \ref{ssec:LOSWormholes}, for real $\chi_\infty$ there can be no stationary point of the Euclidean action $S_E$
on the positive real $L_0$ axis\footnote{While section \ref{ssec:LOSWormholes} discussed only wormholes without a conical singularity on $\gamma$, the conical singularity allowed by our current discussion can be removed by taking $t$ to range over the entire real line.}.   One may correspondingly check that there is no constrained wormhole configuration with positive $L_0$ with vanishing extrinsic curvature at $\chi=0$. However the geometry in the left panel of figure \ref{fig:CutAndPaste} suggests that there should be a saddle at some negative real value of $L_0$.

All of these conclusions are readily verified by explicit differentiation of $S_{CWH}$.  Furthermore,  since the integrand is real and the integral converges absolutely, it is clear that the integrand decreases monotonically with $L_0$.  As a result, at each non-negative real $L_\gamma$ the $L_0$-integral is dominated by the endpoint contribution at $L_0=0$ and no saddle point can play a significant role. In particular, no ascent contour from any stationary point can cross the positive $L_0$  axis.\footnote{If it did, the symmetry with respect to complex conjugation would require that it crosses at a saddle; i.e., that it lies on a Stokes ray.  But we have seen that there are no saddles at positive real $L_0$. Readers interested in a brief review of the relevant facets of Picard-Lefschetz theory should consult appendix \ref{app:PL}.}  

Since the $L_0$ integral is dominated by $L_0=0$ for all $L_\gamma$, the semiclassical limit of this integral is given by $e^{-S_E}$ evaluated at $L_0=0$. Using \eqref{eq:CWHAction_Explicit2}-\eqref{eq:C} we find
\begin{equation}
\label{eq:zeroaction}
S_E \Big|_{L_0=0} = - \frac{\beta}{16\pi G_N\ell \log\big(\frac{\lambda}{2}\big)}L_\gamma^2 - \frac{1}{2 G_N}L_\gamma + \frac{\beta\ell\Big(1+\log\big(\frac{\lambda}{2}\big)\Big)}{4\pi G_N}.
\end{equation}
As noted above, the action is quadratic in $L_\gamma$ and the coefficient of $L_\gamma^2$ is positive.  We thus see again that the integral of $e^{-S_E}$ converges absolutely.  Furthermore, in the limit $\lambda\rightarrow0$ the coefficient of the linear term becomes $-1/2G_N<0$.  Thus, in this limit, the saddle-point value $L_\gamma^*$ of $L_\gamma$ is positive and lies on our integration contour for all $\chi\in {\mathbb C}$.  Indeed, from the Gaussian nature of \eqref{eq:zeroaction} it is clear that this saddle dominates the integral over $L_\gamma$ at $L_0=0$; i.e., it is the dominant parts of the endpoint contribution at $L_0=0$. Specifically, We find
\begin{equation}
\label{eq:zeroaction2}
S_E \Big|_{L_0=0,L_\gamma=L_\gamma^*} = \frac{2\ell\beta}{8\pi\Gn}+\frac{\ell\Big(\frac{4\pi^2}{\beta}+\beta\Big)\log\big(\frac{\lambda}{2}\big)}{4\pi\Gn},
\end{equation}
which turns out to coincide precisely with twice the action of the smooth disconnected solution given by global Euclidean AdS with a constant axion when its action is evaluated in the corresponding conformal frame. We will comment further on this point in section \ref{sec:Discussion}.
However, since it was evaluated by setting $L_0=0$, the detailed form of \eqref{eq:zeroaction2} is sensitive to potential UV corrections.

The dominance of configurations with $L_0 =0$ should not be a surprise.  Since our problem was entirely formulated in Lorentz signature, and since our Lorentzian axion action \eqref{eq:LAction1} is indistinguishable from that of a minimally-coupled scalar, our final results must be indistinguishable as well.  But for scalar fields there is no reason to expect Euclidean wormholes to be important for real values of $\chi_\infty.$  Indeed, we now see that the standard Euclidean AdS-axion wormhole described in section \ref{ssec:EOSWormholes} can be relevant only for imaginary values of $\chi_\infty;$ i.e., for a problem in which we analytically continue $Z_2^{\tiny c}(\beta)$ to imaginary values of  $\chi_\infty.$  We will soon study this mathematically-interesting problem in section \ref{ssec:AnalyticContinue} after providing some further physical motivation below.

In particular, let us return to the UV-sensitivity of the final result \eqref{eq:zeroaction2}.  We emphasize that higher curvature corrections and other UV effects could thus significantly change the result, and there is no reason to believe that our final result above is physically correct.  One might therefore ask if the physically correct UV completion might in fact make $Z_2^c(\beta)$ vanish.

One way to explore this idea is to analytically continue our path integral to complex values of $\chi_\infty$.  The logic here is that if $Z_2^c(\beta)$ were to vanish for all real $\chi_\infty,$ then its analytic continuation must also vanish for all complex $\chi_\infty.$  This would then mean that we should find no UV-safe saddles that dominate for any complex value of $\chi_\infty$ (or, at least, that for every complex $\chi_\infty$ the sum of all formally-dominant UV-safe saddles vanishes exactly).   But we will see below that this is not the case.

%        In principle, we should perform the integral over $L_0$ and then analyze the resulting integral over $L_\gamma$, again using the positive-real axis as the integration contour. However, after performing the $L_0$ integral the $L_\gamma$ integral will be precisely Gaussian and so can be done exactly. We will not do such an analysis explicitly, however, since it has little bearing on whether or not semiclassical wormholes contribute to the integral, which is out primary interest.
        
%         Based on the analysis above we find that, unlike in the setup of \cite{Loges:2022nuw}, our Lorentzian path integral is not controlled by a complex stationary point which corresponds to the Euclidean axion wormhole solution. Rather, the variance of the gravitational thermal partition function seems to be controlled by the degenerate endpoint contribution of the $L_0$ integral $L_0=0$.
         
    \subsection{Analytic continuation of $Z_2^{\tiny c}(\beta)$ to complex values of $\chi_\infty$} 
    \label{ssec:AnalyticContinue}

As we have seen, the usual so-called Euclidean axion wormholes can be relevant to computations of $Z_2^{\tiny c}(\beta)$ only after analytic continuation of $Z_2^{\tiny c}(\beta)$ to imaginary values of $\chi_\infty$.  We now explore this analytic continuation, both to address historical interest in the relevance of such axion wormholes and also to probe the question mentioned at the end of the previous section as to whether modifications of the theory that affect {\it only} the UV might perhaps lead to $Z_2^{\tiny c}(\beta)$ vanishing identically.

\begin{figure}[t!]
            \centering
            \begin{minipage}{0.375\textwidth}             \includegraphics[width=0.8\linewidth]{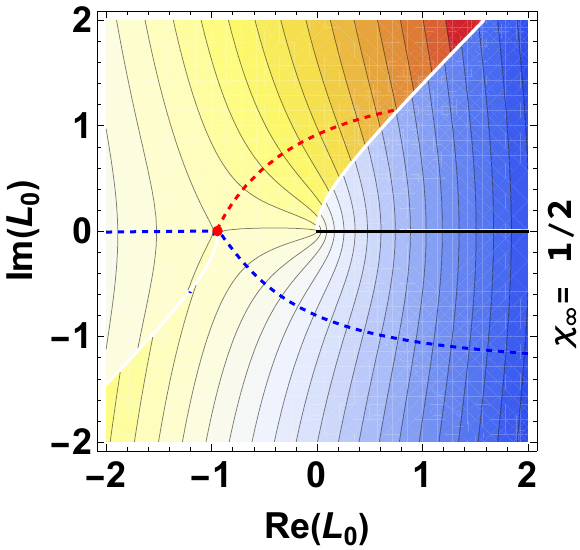}
            \end{minipage}
            \begin{minipage}{0.375\textwidth}
\includegraphics[width=0.8\linewidth]{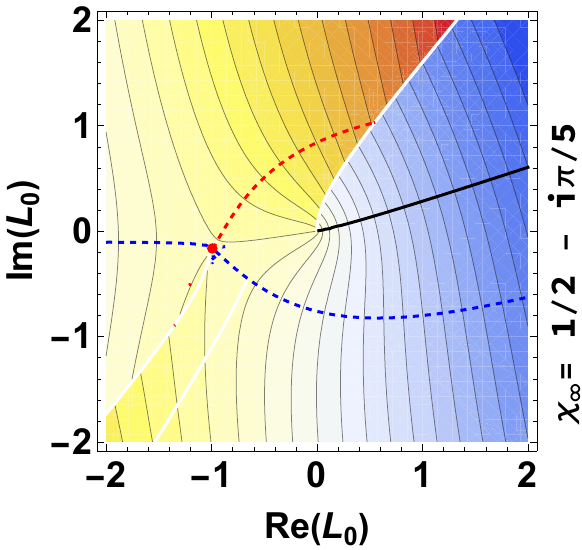}
            \end{minipage}
            \begin{minipage}{0.375\textwidth}
\includegraphics[width=0.8\linewidth]{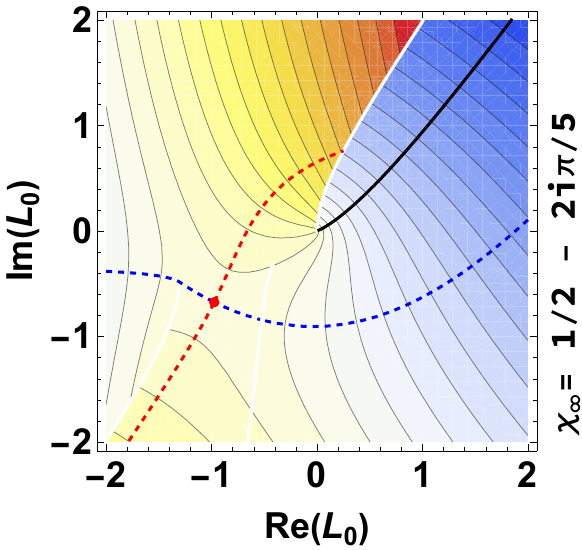}
            \end{minipage}
            \begin{minipage}{0.375\textwidth}  \includegraphics[width=0.8\linewidth]{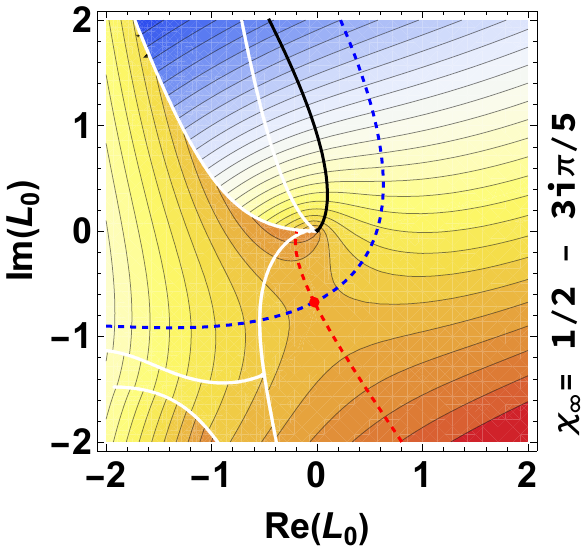}
            \end{minipage}
            \begin{minipage}{0.375\textwidth}\includegraphics[width=0.8\linewidth]{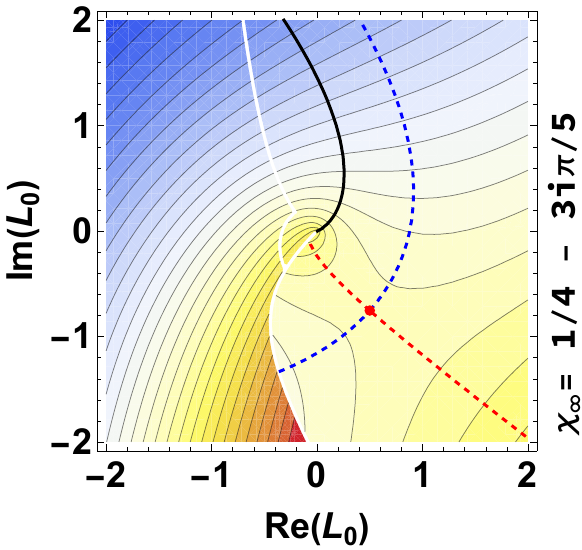}
            \end{minipage}  
            \begin{minipage}{0.375\textwidth}\includegraphics[width=0.8\linewidth]{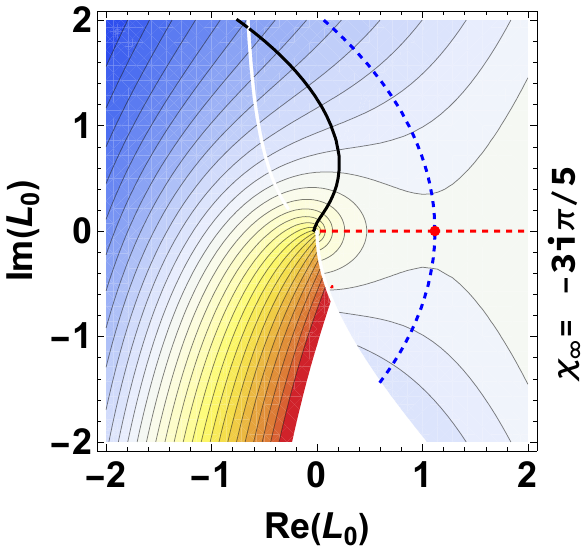}
            \end{minipage}  
            \caption{A sequence of contour plots showing $\text{Re}(iS_{CWH})$ in the complex $L_0$ plane for various $\chi_\infty \in {\mathbb C}$ with $\text{Im} \ \chi_\infty \leq0$ and for fixed parameters $\ell=1,\,8\pi\Gn=1,\, \lambda=10^{-2^8},\,L_\gamma=2^8,\,\beta=2\pi=iT $.
 Reading the plots from left to right across each row in turn (i.e., in the same pattern that one would use to read lines 
 of text), $|\text{Im} \ \chi_\infty|$ increases
while $|\text{Re} \ \chi_\infty|$ decreases. Warmer colors indicate larger  $\text{Re}(iS)$ while cooler colors represent smaller values. Red dots mark the saddle with vanishing $K|_{\chi=0}$. Dashed red (blue) curves are the associated steepest ascent (descent) contours  from this saddle.  Solid black lines show examples of valid integration contours for $Z_2^{\tiny c}$.  The blank region in the center of the bottom of the last panel indicates that $\text{Re}(iS_{CWH})$ is very large.}
            \label{fig:ActionProgression}
        \end{figure}

The analytic continuation of our action $S_{CWH}$ to various complex values of $\chi_\infty$ is shown in figure \ref{fig:ActionProgression} below as a function of $L_0$.  Here $S_{CWH}$ is evaluated at $T=-i\beta$ for the stated fixed values of $\beta$ and $L_\gamma$, but other values of $\beta, L_\gamma$ give similar results.   We find that 
taking $\text{Im}(\chi_\infty)>0$ would tend to move the  stationary point with vanishing $K\Big|_{\chi=0}$ onto a sheet of the Riemann surface for $S_{CWH}$ that differs from the one on which the original contour is defined, and that it would thus make it exceedingly difficult for this saddle to contribute to $Z_2^{\tiny c}(\beta)$.  We therefore display results only for  $\text{Im}(\chi_\infty)<0$.

Starting from real positive $\chi_\infty$, we may study the effect of increasing the magnitude of the imaginary part of $\chi_\infty$ and decreasing the real part of $\chi_\infty$.
In figure \ref{fig:ActionProgression}, this corresponds to moving left-to-right across each successive row in turn -- i.e., moving in the same pattern as one reads lines of text.  We find that this progression moves the saddle point counter-clockwise around the $L_0=0$ endpoint of the contour so that it approaches the real positive $L_0$ axis.  Indeed, as one might expect, the saddle (red dot) with vanishing $K\Big|_{\chi=0}$  reaches the real positive $L_0$ axis when $\chi_\infty$ becomes purely imaginary (shown in the lower right panel). 

\begin{figure}[h!]
    \centering
    \begin{minipage}{0.35\textwidth}
        \includegraphics[width=\linewidth]{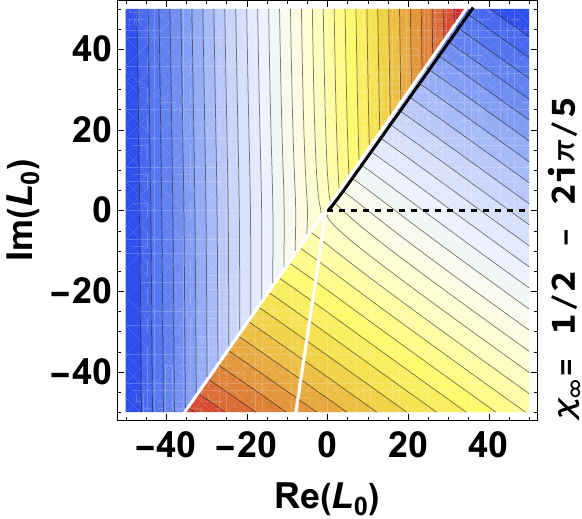}
    \end{minipage}
    \begin{minipage}{0.35\textwidth}
        \includegraphics[width=\linewidth]{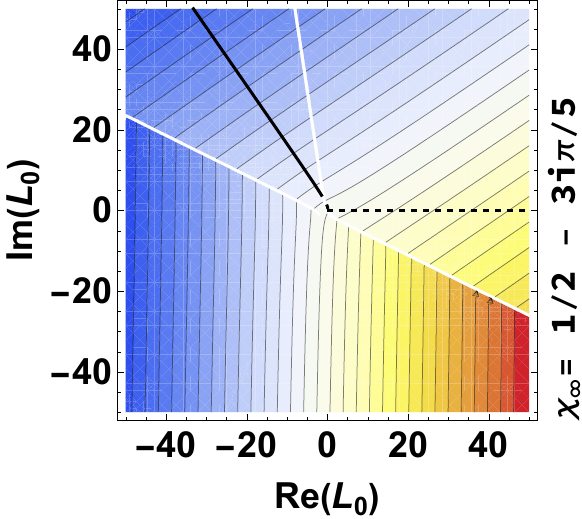}
    \end{minipage}
    \caption{Zoomed-out versions of panels 3 and 4 from figure \ref{fig:ActionProgression} showing that the original contour of integration (dashed black lines) remains convergent for $\chi_\infty=\frac{1}{2}-2i\pi/5$ but would give a divergent integral for $\chi_\infty=\frac{1}{2}-3i\pi/5$.}
    \label{fig:AsymptoticExamples}
\end{figure}

However, as suggested by the last 3 plots in figure \ref{fig:ActionProgression}, the integral over the positive real axis fails to converge at some of our values of $\chi$.    This is particularly clear in figure \ref{fig:AsymptoticExamples}, which provides further zoomed-out versions of panels 3 and 4 from figure \ref{fig:ActionProgression} showing that the original contour of integration (dashed black line in each panel of figure \ref{fig:AsymptoticExamples}) remains convergent at $\chi_\infty= \frac{1}{2}-2i\pi/5$ but gives a divergent integral for  $\chi_\infty= \frac{1}{2}-3i\pi/5$.         As a result, analytic continuation of the integral representation of $Z_2^{\tiny c}(\beta)$ requires us to deform the contour of integration\footnote{As usual, by Cauchy's theorem, no continuous deformation of the contour at fixed $\chi_\infty$ can change $Z_2^{\tiny c}(\beta)$.  Furthermore, any continuous deformation of this contour with changes of $\chi_\infty$ that preserves exponential convergence of the integral will provide an analytic extension since, in the case, derivatives with respect to $\chi_\infty$ also converge exponentially so that the Cauchy-Riemann equations for the integrand also imply the Cauchy-Riemann equations for $Z_2^{\tiny c}(\beta)$. We are thus free to choose {\it any} suitably convergent contour along which to evaluate our integral, so long as it is obtained by smoothly deforming the original contour while maintaining convergence at every step. The final result will be independent of this choice.  Examples of such contours are shown in figure \ref{fig:ActionProgression}.}.  In each panel, the solid black line shows a valid choice of contour  given by such continuous deformations of the original defining contour at real $\chi_\infty.$  Thus, while the red dot saddle has moved to the positive real axis in the final frame, the integration contour has moved away toward the upper left. 

\begin{figure}[h!]
            \centering
            \includegraphics[width=0.4\linewidth]{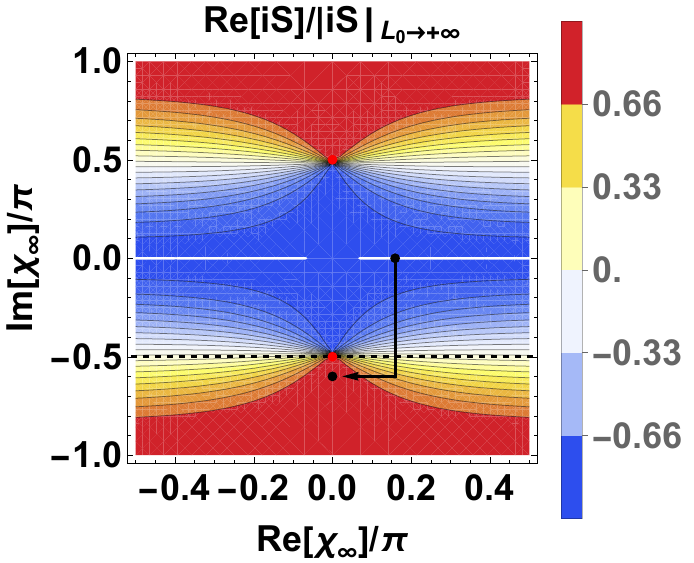}
            \caption{A contour plot showing the ratio $\frac{\text{Re}(iS_{CWH})}{|iS_{CWH}|}$ as a function of general complex $\chi_\infty$ plane for positive real $\beta = iT$ evaluated at 
            $L_0 = 10^6 \ell$.  The results agree with the asymptotics implied by the analytic result \eqref{eq:Easympt}
            in the limit where we take $L_0$ to $+\infty$ along the positive real axis. The integral of $e^{iS_{CWH}}$ over the positive real $L_0$ axis should converge for values of $\chi_\infty$ such that the above limit is negative (blue regions in plot), and will diverge for those $\chi_\infty$ that make the above limit positive (yellow, orange, and red regions in plot).  The limit vanishes along the dashed black line.  The solid black line represents the path through the complex $\chi_\infty$ plane corresponding to the sequence of images in figure \ref{fig:ActionProgression}. We again take $\ell=1,\,8\pi\Gn=1,\, \lambda=10^{-2^8},\,L_\gamma=2^8,\,\rm{and}\,\beta=2\pi$.}
            \label{fig:ActionAsymptotics}
        \end{figure}

The above claims about the need to deform the contour of integration can in fact be established analytically.  Since $S_{CWH}$ does not diverge at any finite values of $L_\gamma, L_0, \beta$, and $\chi_\infty$, the original integration contour needs to be deformed if (and only if), for some $\beta$ and some $\chi_\infty$, the limit of $S_{CWH}$ diverges at large positive $L_0.$  But this asymptotic behavior is easily read off from our expansion \eqref{eq:Easympt} and agrees with the numerics shown in figure   
\ref{fig:ActionAsymptotics} (which also agrees with those of figure \ref{fig:AsymptoticExamples}).

Returning now to figure \ref{fig:ActionProgression}, we can again ask if the saddle with vanishing $K\Big|_{\chi=0}$ makes an important  contribution to $Z_2^{\tiny c}$ at each value of $\chi_\infty$.  The answer is that it generally does not.  In particular, in every case the steepest ascent contour associated with the stationary point fails to intersect the interior of our integration contour. In general, it misses the contour entirely.  However, when $\chi_\infty$ becomes purely imaginary in the final frame (so that the stationary point corresponds to the usual Euclidean axion wormhole),  we find a marginal case in which the steepest ascent contour from the saddle intersects the contour of integration precisely at the endpoint $L_0=0$.   As discussed in appendix \ref{app:PL}, we may then say that this stationary point lies on a Stokes' ray, though it is clearly subdominant to the endpoint contribution at $L_0=0$ (since $L_0=0$ lies on the ascent contour from the saddle). Here we again refer the interested reader to appendix \ref{app:PL} for a review of the relevant aspects of Picard-Lefschetz theory\footnote{For the first few panels in figure \ref{fig:ActionProgression}, one may also note that the integrand is larger in magnitude at the red dot than at any point along the integration contour.  But in the semiclassical approximation, the integral is bounded by the maximum value of the magnitude of the integrand (up to a subleading one-loop correction factor).  As a result, the saddle cannot be relevant.  Similarly, in the final 3 panels the saddle is clearly smaller in magnitude than the endpoint contribution at $L_0=0$ (and could thus be swamped by either perturbative or non-perturbative corrections to that contribution).}. 

\begin{figure}
            \centering
            \begin{minipage}{0.4\textwidth}
                \includegraphics[width=\linewidth]{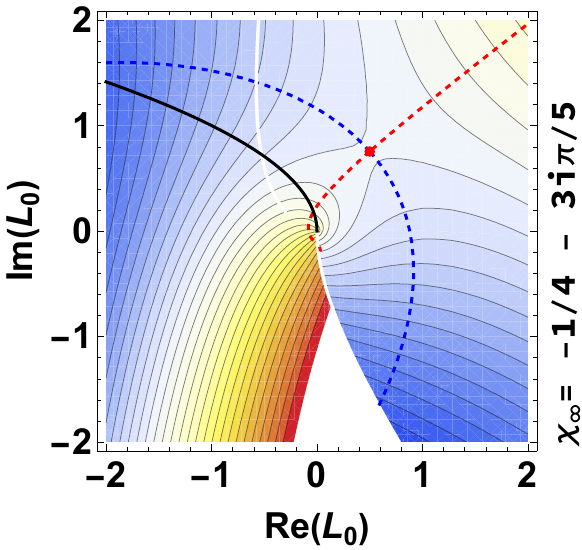}
            \end{minipage}
            \begin{minipage}{0.4\textwidth}
                \includegraphics[width=\linewidth]{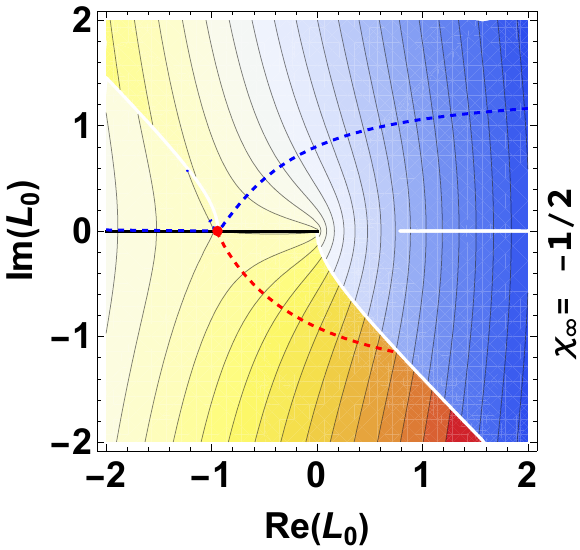}
            \end{minipage}
            
            \caption{Contour plots of $\text{Re}(iS)$ in the complex $L_0$ plane for $\chi_\infty=-1/4 -3/5\,\pi i$ (left) and $\chi_\infty=-1/2$ (right). We have taken $\ell=1,\,8\pi\Gn=1,\, \lambda=10^{-2^8},\,L_\gamma=2^8,\,\beta=2\pi $. The contour of integration is shown in black, while the stationary point associated with $K|_{\chi=0}=0$ is denoted by the red dot. The associated steepest ascent and descent contours are denoted by the dashed red and blue lines respectively. Warmer colors indicate larger values of $\text{Re}(iS)$ while cooler colors represent smaller values. Since the ascent contour intersects the interior of the black line in each panel, the saddle contributes to the integral with non-zero weight.}
            \label{fig:WormholeContributes}
        \end{figure}

        Since the case of imaginary $\chi_\infty$ was marginally relevant, when we further analytically continue $\chi_\infty$ to values with  \emph{negative} real-part we find that the steepest ascent contour from the saddle with vanishing $K\Big|_{\chi=0}$ now unambiguously {\it does} intersect the contour of integration. However, at first the contribution from the endpoint of the integration contour still dominates over that of the stationary point. Such a case is illustrated in the left panel of figure \ref{fig:WormholeContributes}. 
        On the other hand, if one pushes the analytic continuation of $\chi_\infty$ even further (e.g., to where $\chi_\infty$ is again real but now negative as shown at right in figure \ref{fig:WormholeContributes}), the contribution from our $K\Big|_{\chi=0}=0$ saddle does in fact dominate over the contribution from the endpoint. As noted above, this argues that -- at least in a certain sense -- despite the UV sensitivity of \eqref{eq:zeroaction2}, corrections confined to the UV cannot set $Z_2^{\tiny c}=0$ at more than isolated values of $\chi_\infty$.
        
\section{Discussion}
    \label{sec:Discussion}

The goal of our work was to explore contributions from axion wormholes to gravitational path integrals defined by integrating over real Lorentz-signature metrics in the sense of \cite{Colin-Ellerin:2020mva,Marolf:2020rpm,Marolf:Thermodynamics}.  In particular, we wished to explore their potential contributions to the factorization problem of AdS/CFT. We chose to study partition functions associated with what, in an AdS/CFT context, would be called CFTs on the static patch of de Sitter spacetime.  At the appropriate temperature $T=1/\beta$, such partition functions may be called CFT partition functions $Z[S^d]$ on the Euclidean sphere $S^d$.

Since our main interest was in exploring the use of such techniques for evaluating the relevance of Euclidean and complex saddles, we chose a simple model where the bulk theory contains  only Einstein-Hilbert gravity with a minimally-coupled (massless) axion.  In this model, and especially in Lorentz signature where our path integral is defined, this axion is indistinguishable from a minimally-coupled massless scalar and must lead to equivalent results.  
We focused on the case $d=2$ which admits a technical trick that allowed us to proceed analytically. We expect higher dimensional cases to behave similarly, though the details remain to be explored.

While the path integral integrates over an infinite family of parameters, we focused on the integrals over parameters called $T, L_0, L_\gamma$, which respectively describe the period of Lorentzian time, the size of the wormhole at the surface where the axion vanishes, and the length of a Lorentzian conical defect that arises from taking time to be periodic.  The fact that there are real Lorentz-signature {\it constrained} wormhole saddles when one imposes constraints that fix  $T, L_\gamma$  and $L_0$ allows us to argue on general grounds that the remaining integrals can legitimately be done by using the stationary phase approximation for real values of the remaining parameters (at least for real values of $T,\chi_\infty$; see below for further comments on the case of complex $T,\chi_\infty$). 

Of the $T,L_0, L_\gamma$ integrals that we analyze explicitly, the integral over $L_0$ is of most interest.   However, some comments on the $T$-integral are also in order.  In particular, we expect to perform the $T$ integral by closing the contour of integration in the lower half-plane and using the pole in $f_\beta(T)$.  This necessarily involves analytic continuation of the path integral to general complex $T$. As a result, we should ask if the above argument for the relevance of our constrained-wormhole saddles remains valid at general complex $T$ in the lower half-plane.  A partial answer is that, so long as enough of the integrals in our path integral have been performed to render the total energy positive for real fields, the remaining integrals should be convergent for all complex time-periods $T$ in the lower half-plane.  Thus a real saddle will continue to lie on the defining contour and will contribute\footnote{Except, perhaps, when its ascent contour is tangent to the defining contour.  One can in principle check for such exceptions via a perturbative analysis.  We refrain from doing so here since such an exception would only amplify our basic message that the usual Euclidean axion wormholes saddles do not in fact contribute.}.

Perhaps critically, our partition functions were defined by taking the metric to be asymptotically (locally) anti-de Sitter and by fixing the axion on each of two boundaries (called left and right) to take definite values $\chi_L,\chi_R$.  In an asymptotically (locally) AdS context, Dirichlet boundary conditions for the axion are in fact required to avoid ghosts \cite{andrade:BeyondUnitarityBound}.  

For real $\chi_L,\chi_R$, we found that on-shell wormholes (saddles) play no significant role.  Instead, within the class of spacetimes we investigate, the connected partition function on two spheres,
\begin{equation}
Z^{\tiny c} [S^d \sqcup S^d] =:  Z[S^d \sqcup S^d] - \left(Z[S^d]\right)^2,
\end{equation}
is fully described in the semiclassical limit by an endpoint contribution associated with degenerate wormholes.

However, with our conventions the asymptotically-AdS version of the traditional Euclidean axion wormholes of \cite{Giddings:1987cg} turn out to have imaginary $\chi_R,\chi_L$.  One may also be motivated to study complex $\chi_R,\chi_L$ by the fact that the endpoint contributions that dominate for real $\chi_R,\chi_L$ are highly sensitive to UV modifications of the theory and so cannot be trusted to make physical predictions.  One might thus ask if they can be canceled by UV modifications.  One way to explore that question is to note that if $Z^{\tiny c} [S^d \sqcup S^d]$ vanishes at all real $\chi_L,\chi_R$, then its analytic continuation to complex 
$\chi_L,\chi_R$ must vanish identically as well.  As a result, if one finds that the analytic continuation of the gravitational path integral to some complex $\chi_L,\chi_R$ is both well-defined and dominated by a smooth macroscopic semiclassical on-shell wormhole, it shows that such complete cancellations are forbidden so long as the supposed UV modifications remain negligible in the IR at complex $\chi_L,\chi_R$.

The result of section \ref{ssec:AnalyticContinue} was that analytic continuation of $Z^{\tiny c} [S^d \sqcup S^d]$  to complex $\chi_L, \chi_R$ is well-defined at our level of analysis, but that computing it via the gravitational path integral required deforming the contour of integration as one analytically continues $\chi_L,\chi_R$; see figures \ref{fig:ActionProgression} and  \ref{fig:WormholeContributes} in section \ref{ssec:AnalyticContinue}.   In particular, AdS axion wormholes with real Euclidean geometries are smooth saddles when $i(\chi_R-\chi_L)\in(\pi,2\pi)$. For such cases, we found that  $Z^{\tiny c} [S^d \sqcup S^d]$ is still dominated by the endpoint contribution of degenerate wormholes.  However, there is a sense in which the smooth Euclidean wormhole is marginally relevant as a subleading contribution.  In particular, for such cases the smooth Euclidean wormholes lie on a Stokes ray, so that whether or not they should be included depends on which non-perturbative corrections one chooses to associate with the dominant (endpoint contribution).  One then finds that further analytic continuation of $\chi_L, \chi_R$ to real `negative' values then gives connected partition functions  $Z^{\tiny c} [S^d \sqcup S^d]$ that are in fact dominated by smooth semiclassical wormholes.  As a result, we find that 
even for real $\chi_L, \chi_R$ our $Z^{\tiny c} [S^d \sqcup S^d]$ cannot be set to zero by UV modifications that are negligible in the IR for all complex boundary conditions.\footnote{Unless our analytic continuation in $\chi_{L,R}$ leads to additional Stokes' phenomena related to the integrals over the infinite set of degrees of freedom that we have not tracked in detail: see again footnote \ref{foot:Stokes}. Such Stokes' phenomena could potentially cause the constrained saddles discussed in section \ref{sec:ConstrainedWormholes} to become irrelevant at some complex $\chi_L,\chi_R$. It would be interesting to address this further in future work, though it would not affect the main result of this paper that the usual Euclidean axion wormhole saddles do {\it not} contribute for the associated (purely imaginary) values of $\chi_L,\chi_R$.}

Our results thus support the general expectations \cite{Maldacena:2004rf} that 
connected gravitational partition functions do indeed fail to vanish due to contributions from spacetime
wormholes.  However, whether or not a given Euclidean saddle contributes appears to depend on details (as seen in our study of imaginary $\chi_L,\chi_R$ where the wormhole lies on a Stokes' line).    We will return to this issue in forthcoming works that investigate Euclidean wormholes in other dimensions and supported by other types of matter fields.

Despite the agreement of our results with general expectations,  the details disagree with claims based on extrapolating recent analyses
\cite{Loges:2022nuw,andriolo:Axion_wormholes_massive_dilaton,Loges:2023ypl,Hertog:Axion-Saxion_Wormholes_Stable,aguilar:Axion_dS_Wormholes,loveridge:Axion_Wormhole_Alternate_Topologies}.  
There are several possible reasons for this disagreement.  First, with the exception of \cite{Loges:2022nuw}, the above references are intrinsically perturbative.  In particular, in our language they investigated a question analogous to setting the axion boundary values $\chi_L, \chi_R$ to be purely imaginary, so that the Euclidean axion wormhole saddle appears at real positive $L_0$, and asking if the saddle is a local minimum of the action along the real $L_0$ axis.  Whether or not this is the case,  we have seen that after analytically continuing $\chi_L, \chi_R$ to be purely imaginary, the integration contour for the problem we chose to study has in fact moved to the upper left quadrant of the complex $L_0$-plane.  As a result, even if it were to pass some naive test for ``perturbative stability,'' the saddle is at most marginally relevant (as a subdominant contribution) to our connected partition function.

On the other hand, in our analysis, we found even for $i(\chi_R-\chi_L) \in (\pi,2\pi)$ that the Euclidean wormhole is in fact {\it not} a local minimum of the Euclidean action along the real $L_0$ axis.  Indeed, it is a local maximum, while \cite{Loges:2022nuw,andriolo:Axion_wormholes_massive_dilaton,Loges:2023ypl,Hertog:Axion-Saxion_Wormholes_Stable,aguilar:Axion_dS_Wormholes,loveridge:Axion_Wormhole_Alternate_Topologies,Marolf:2025evo} 
found real fluctuations around their Euclidean axion wormholes to always increase the Euclidean action.  One possible reason for this difference stems from the above references' focus on the case of asymptotically flat wormholes instead of on the asymptotically locally AdS (AlAdS) studied here.  As a result, \cite{Loges:2022nuw}  chose to impose a {\it Neumann} boundary condition on their axion (fixing the axion charge $q$).  This choice of boundary conditions appears to be allowed in the asymptotically flat context while, as noted above, in the AlAdS case one must use Dirichlet boundary conditions to avoid ghosts.  This distinction may make it harder than might naively be expected to extrapolate their asymptotically flat results to the AlAdS context.

Some evidence for this point of view comes from formally repeating our work above for the case of fixed axion charge.  The computations are formal in the sense that, due to the expected ghosts, the Hamiltonian for this system should not be bounded below and the manipulations of section \ref{sec:Background} cannot be justified.  Nevertheless, the associated computations are summarized in appendix \ref{sec:EuclideanCWormholes}, where we find for imaginary charge $q$ (real Euclidean charge $q_E$) that the Euclidean action does indeed increase as one increases $L_0$ above the saddle-point value along the real positive $L_0$-axis.  This indicates some level of consistency with \cite{Loges:2022nuw,andriolo:Axion_wormholes_massive_dilaton,Loges:2023ypl,Hertog:Axion-Saxion_Wormholes_Stable,aguilar:Axion_dS_Wormholes,loveridge:Axion_Wormhole_Alternate_Topologies,Marolf:2025evo}, though the details are confusing.  Indeed, comparison of the results is difficult as it appears that, even when our conical singularity disappears (i.e., for $\alpha=1$), our constrained wormholes cannot be directly described in the framework used by the above works (again, with the potential exception of \cite{Loges:2022nuw}).  In particular, the above works find in their framework that there is simply no spherically-symmetric off-shell mode to analyze. It is thus also 
possible that their implicit choice of path integral contour does not even coincide with what we would call real positive $L_0$. It would be useful to better understand the relation between these approaches in the future.

Of perhaps more interest is comparison with \cite{Loges:2022nuw}, which also performed a Lorentzian-signature-based Picard-Lefschetz analysis, but which found the Euclidean axion wormhole to be a dominant saddle.  Here, again, the difference from our results may arise simply from the fact that \cite{Loges:2022nuw} studied the case of vanishing cosmological constant with asymptotically flat boundary conditions for the metric and Neumann (fixed charge) boundary conditions for the axion.

However, it is also possible that the real Lorentz-signature contour used to define the path integral in \cite{Loges:2022nuw} might differ in important ways from the one used in our work above.  The point here is that \cite{Loges:2022nuw} took the direction along the wormhole (called $r$ or $z$ above) to be timelike on their defining contour, while we took the timelike direction to run along the boundary; see again figure \ref{fig:WickRotations} from section \ref{sec:intro}. Said differently, the physical question analyzed in \cite{Loges:2022nuw} 
concerned potential contributions of axion wormholes to what are often called Lorentz signature amplitudes between spherical closed cosmologies of different radii\footnote{Though we prefer the discussion of parallel issues in \cite{Marolf:1996gb}, which considers these to be instead the inner product between the projections into the physical space of states of non-physical states defined by closed cosmologies of different radii.}.   In contrast, we investigated potential axion wormhole contributions to connected AdS/CFT-style partition functions associated with putting the CFT on two copies of the static patch of de Sitter space.  It is possible that these two questions are not mathematically related, and that they receive different contributions from wormholes.

While we followed the general rule of \cite{Marolf:Thermodynamics} that the $L_\gamma$ integral should be done last, this was unnecessary in retrospect. After performing the $T$-integral, for real $\chi_\infty$ we found the remaining integral over real positive $L_0,L_\gamma$ to converge absolutely.  These two integrals could thus have been performed in either order.  In particular, we may perform the $L_\gamma$ integral first using the saddle-point value $L_\gamma^*$ in \eqref{eq:Lgamma*}, which due to the comments below that equation always dominates the $L_\gamma$ integral at fixed $L_0$ when $\beta$ is positive and real.\footnote{\label{foot:gammasaddle} When the $L_\gamma$ integral is performed first, it is convenient to use the alternate parameterization of appendix \ref{sec:AlternateParamatrization} that fixes $\tilde L_0=\alpha L_0$ instead of $L_0$.  Doing so means that the term $S_\mu$ is a minimally-coupled matter term in the sense of \cite{Dong_2020}, and thus that the saddle for $L_\gamma$ is then the value that sets $\alpha=1$ and which thus removes the conical singularity.  One may then simply analyze the round Euclidean axion wormholes to evaluate the integral over $\tilde L_0$, bypassing many of the complications described in the main text.  This observation may be useful to simplify future explicit calculations.
}
The integral over $L_0$ must then proceed much as above.  Performing the integrals in the order $T,L_0,L_\gamma$ (with $T$ integrated first and $L_\gamma$ integrated last) thus gives the same result as the order $T,L_\gamma,L_0.$

Furthermore, while
for the reasons reviewed in section \ref{sec:overview} it may generally be important to perform the integral over $T$ before evaluating the integral over $L_\gamma$, at least in the current case we could have inverted this order in order to first perform the integral over $L_\gamma$.  In particular,  our $S_{CWH}$ was explicitly quadratic in $L_\gamma$ and, moreover, the coefficient of the quadratic term did not depend on $\chi_\infty$ (see comment below \eqref{eq:CWHAction_Explicit2}).    This integral is thus readily performed in a distributional sense even before integrating over $T$.  In particular, in this case we may proceed by rotating the $L_\gamma$ contour to one where $\mathrm{Im} (T L_\gamma^2) \rightarrow - \infty$ as the real part of $L_\gamma$ becomes large and positive (so that it is then dominated either by the Gaussian saddle or by the endpoint at $L_\gamma=0$).  After integrating over $L_\gamma$ we may then again evaluate the $T$-integral using the pole at $T=-i\beta$.  The resulting integral over $L_0$ is then identical to that obtained by integrating first over $T$ and then over $L_\gamma$, since this agreement is a requirement defining the distributional sense in which the $L_\gamma$ integral can be evaluated. 

We thus again obtain the same result when performing the integrals in the order $L_\gamma, T, L_0$ as was obtained in section \ref{sec:ConstrainedWormholes} by using the order $T,L_0,L_\gamma$ or, as argued above, by using the order $T, L_\gamma, L_0$.
Similarly, with enough care, it is plausible that by analogous arguments one may be able to show that we still obtain the same result when the $L_0$ integral is performed before integrating over $T, L_\gamma$, and perhaps even that the $T$ integral can be performed last.  If so, we would obtain the same result by performing the $L_0,L_\gamma,T$ integrals in any order.

One might also ask if the $T, L_0, L_\gamma$ integrals could have been performed even earlier.  In particular, one might ask if they could have been performed before some of the integrals over metric and axion degrees of freedom that we evaluated only by invoking the stationary phase approximation and using the constrained wormhole saddles of section \ref{sec:ConstrainedWormholes}.
A natural such procedure would be to write the Lorentzian path integral in canonical form and to first  integrate over Lagrange multipliers to enforce the constraints.  In cases of interest, this will make the Hamiltonian bounded below (though enforcing the Euclidean-signature constraints would not do so; see e.g. \cite{Horowitz:2025zpx}). One should then be able to integrate over $T$ as described in section \ref{sec:Background}, and we would expect the remaining integrals to converge absolutely.  Here it is important that we describe our constraint at the $\chi=0$ surface as fixing $L_0$ and not the full spacetime area of that surface so we fix a parameter ($L_0$) that enters only the spatial part of the metric and so that lapse and shift can be integrated earlier as described above\footnote{It is unclear whether the factor of $\alpha$ in \eqref{eq:Smu} will interfere with this argument, but it may in fact be better to use the alternate parameterizaton of appendix \eqref{sec:AlternateParamatrization} which instead fixes a different parameter $\tilde L_0$.}.  

Returning to our results for the connected partition function, it is important to comment further on our endpoint contribution.  Recall that this contribution was approximated by setting $L_0=0$ and then setting $L_\gamma$ to its value $L_\gamma^*$ at the stationary point of the remaining integral.  Fixing $L_0=0$ also fixes $\tilde L_0=\alpha L_0=0$, so by the argument in footnote \ref{foot:gammasaddle} setting $L_0=0,L_\gamma=L_\gamma^*$ must also set $\alpha=1$. Interestingly, for $\alpha =1$, the $L_0\rightarrow 0$ limit of our Euclidean constrained wormhole solution turns out to coincide exactly with the {\it disconnected} smooth saddle;  the wormhole connection has pinched off and disappeared.  
This observation raises the question of whether our endpoint contribution should be treated as truly separate from the disconnected saddle in an appropriate UV completion.  This also has the interesting effect that at leading semiclassical order the endpoint contribution from wormholes found here is of the same size as the full-disconnected quantity $Z[({\cal B}^\lambda,\beta)]^2$ (see again \eqref{eq:zeroaction2}), though this need not be true in UV completions that treat our endpoint as distinct from the disconnected saddle.

As a final remark, while our study of axion wormholes was non-perturbative, 
it is important to emphasize that it took into account only a certain limited set of deformations.  It would of course be useful to better consider other deformations as well, and in particular to develop better tools to definitely determine which deformations are the most important.  In particular, although they are believed to be important in other wormhole contexts (see e.g. \cite{Maldacena:2004rf,Marolf:2021kjc}) we have included no deformations that would probe the relevance of brane-nucleation processes. Indeed, to our knowledge the literature contains no studies at all of such processes in the context of axion wormholes.  It would therefore be interesting to return to this question in future work.

\acknowledgments
    We thank Vincent Chen, David Grabovsky, Maciej Kolanowski, Xiaoyi Liu, Sean McBride,  Douglas Stanford, Joaquin Turiaci, Chih-Hung Wu for helpful discussions.  JH was supported by U.S. Department of Defense through the National Defense Science and Engineering Graduate (NDSEG) Fellowship Program as well as the U.S. Air Force Office of Scientific Research under award number FA9550-19-1-0360, by NSF grant PHY-2408110, and by funds from the
    University of California.   DM was supported by NSF grants PHY-2408110
    and PHY-2107939, and by funds from the University of California. MK was supported by NSF grant PHY-2107939, by funds from the University of California, by the European Research Council (ERC) under the European Union’s Horizon 2020 research and innovation program (grant agreement No. 884762), and by the ERC under the QFT.zip project (grant agreement No. 101040260). ZW was supported by the DOE award number DE-SC0015655. 
\appendix
\addtocontents{toc}{\protect\setcounter{tocdepth}{1}}

\section{Picard-Lefschetz theory for semiclassical integrals}
\label{app:PL}

        We now provide an extremely brief review of Picard-Lefschetz theory that emphasizes results that are important for this paper. A broader pedagogical introduction can be found in \cite{witten2010analyticcontinuationchernsimonstheory}. Of particular interest here is what Picard-Lefschetz theory can tell us about evaluating complex, potentially oscillatory integrals. For simplicity, and because it is all we need in the main text, we will restrict discussion in this appendix to contour integrals associated with a single variable (i.e., to one-dimensional contours in $\mathbb{C}$).  However, similar ideas apply to integrals on higher-dimensional complex dimensional manifolds.

        Suppose then that we would like to perform a contour integral of the form
        \begin{equation}
            \mathcal{I}= \int_{\mathcal{C}_0}dz \,\exp[\lambda h(z)],
            \label{eq:testIntegral}
        \end{equation}
        where $\mathcal{C}_0$ denotes the contour of integration and $h(z)$ is a holomorphic complex function. Of course we are free to deform the contour of integration $\mathcal{C}_0$ to some other contour in the complex plane and the value of $\mathcal{I}$ will remain the same provided the deformation between the two contours can be performed without encountering any regions in which $h(z)$ is non-analytic. The main result of interest \cite{FAs,FP,AGV,BH,BH2,H} is that one such choice of integration contour is given by the sum over the `Lefschetz thimbles' associated with all of the stationary points of $h(z)$
        \begin{equation}
            \mathcal{C}_0\sim\sum_s n_s \,\mathcal{J}_s.
            \label{eq:thimbles}
        \end{equation}
        Here $s$ runs over the set of stationary points of the function $h(z)$ and $\mathcal{J}_s$ represents the associated Lefschetz thimble. In the present case, where $h(z)$ is a function on $\mathbb{C}$, these thimbles are simply the contours generated by gradient descent from the stationary point. Thus we will often refer to these thimbles as `steepest descent contours.'

        The reader will also note that each Lefschetz thimble in \eqref{eq:thimbles} appears with a weight $n_s$. This weight is determined by the `steepest \emph{ascent} contour' $\mathcal{K}_s$ associated with the stationary point $s$ (which is simply the contour generated by \emph{upward} gradient flow from $s$).  The coefficient $n_s$ is then just the signed intersection number between $\mathcal{K}_s$ and the original integration contour $\mathcal{C}_0$.

        In the limit where the parameter $\lambda$ appearing in equation \eqref{eq:testIntegral} is taken to be extremely large, we can evaluate $\mathcal{I}$ by saddle-point methods, approximating the integrand by the value taken at the stationary points $s$ of $h(z)$. If we imagine approximating $\mathcal{I}$ in this way, using the Lefschetz thimbles as the contour of integration we then find

        \begin{equation}
            \mathcal{I}\approx\sum_s n_s \exp[\lambda\, h(z_s)],
        \end{equation}
        where this expression neglects both perturbative corrections and any measure factors as we expect both to be small in the semiclassical limit $\lambda\gg1$. The key point is then that a stationary point of $h(z)$  only contributes to the integral if $n_s$ is non-zero. We will apply this to our path integral in contexts where $h(z)$ represents the action evaluated on some appropriate class of spacetime wormholes.

     A particularly useful corollary of this result arises  in contexts where the integrand is a pure phase everywhere on the defining contour.  In this case,  if a stationary point lies on the original contour of integration then it \emph{must} contribute to that integral semi-classically (assuming that the integral converges in an appropriate distributional sense). The point here is that, since the magnitude of the integrand is constant along the defining contour,  the ascent contour is necessarily transverse and gives a non-zero local intersection number.  Furthermore, the fact that the magnitude of the integrand  necessarily increases along the ascent contour as one moves away from the original saddle forbids the existence of further intersections with the defining contour (whose local intersection numbers might then have cancelled the above local contribution).  As a result, in this context $n_s$ must be non-zero.

It is important to note that the above analysis of stationary points via their associated steepest ascent/descent contours is typically discussed in settings in which the defining contour of integration extends to infinity in both directions. Cases (like the ones considered in the main text) in which the contour of integration has a finite endpoint are somewhat more subtle.
To discuss these subtleties,         consider an integral of the form \eqref{eq:testIntegral} where $\mathcal{C}_0$ is a semi-infinite path on $\mathbb{C}$. Suppose also that the integrand takes a finite value at the endpoint $p_0$ of $\mathcal{C}_0$. In such a case, it is no longer true that we could deform the contour of integration to a contour of the form \eqref{eq:thimbles} since it is not guaranteed (nor is it likely) that such a contour will end at $p_0$. 

However, it is still true that looking at the intersection number between the steepest ascent contour attached to  stationary point of $h(x)$ and $\mathcal{C}_0$ is the appropriate diagnosis of whether or not that stationary point contributes to the integral.  To see that this is the case, we can consider performing a conformal transformation $\phi$ of the complex plane which takes $p_0$  to infinity (and which also maps the original infinity to infinity\footnote{For example, if $\mathcal{C}_0$ runs from $0$ to $\infty$ along the positive real axis, we may use the map $z \rightarrow z + 1/z$.}).  The integral then takes the form
\begin{equation}
\label{eq:newI}
{\cal I} = \int_{\tilde {\cal C}_0} dz J(z) \exp{\lambda \tilde h(z)}
\end{equation}
where $J$ is the appropriate Jacobian, $\tilde C_0 = \phi^{-1}({\tilde C}_0)$, and $\tilde h = h \circ \phi$.

In this new representation \eqref{eq:newI} we may attempt to apply the above analysis from Picard-Lefschetz theory.   Let us first consider the case where $h(p_0) = -\infty$.  Then $\exp{[\lambda \tilde h]}$ will vanish at both ends of 
$\tilde {\cal C}_0$ as required by the standard analysis.  

The only subtlety is then that \eqref{eq:newI} contains an extra factor of $J$ relative to the standard form \eqref{eq:testIntegral}.  But if the conformal transformation $\phi$ is taken to be independent of $\lambda$, then $J$ is also independent of $\lambda$.  In that case, we may absorb it into the exponent and, at large $\lambda$, we may simply treat $\lambda^{-1} \ln J$ as a perturbation of $\tilde h$.  As a result, the Picard-Lefschetz results described above are unchanged at leading order in $\lambda$.

It now remains to address the more general case where $\exp{[\lambda h(p_0)]}$ is non-zero.  It will be convenient to study this case by considering a 1-parameter family of analytic functions $h_\epsilon$ that approach $h(z)$ as $\epsilon \rightarrow 0$, while for $\epsilon >0$ we have both $\exp{[\lambda h_\epsilon(p_0)]}=0$ and that $\lambda h_\epsilon(z)] \approx \lambda h(z)]$ at points far from $p_0$.  For example, if $p_0=0$ and if  
${\cal C}_0$ runs along the positive real axis, then we may simply take $h_\epsilon(z) = h(z) - \frac{\epsilon h(0)}{z}.$

We may then study the desired integral as the $\epsilon\rightarrow 0$ limit of the integrals ${\cal I}_\epsilon$ defined by replacing $\tilde h$ with $\tilde h_\epsilon : = h_\epsilon \circ \phi$ in \eqref{eq:newI}.  In particular, we may apply the standard Picard-Lefschetz theory to each ${\cal I}_\epsilon$ before taking the limit $\epsilon \rightarrow 0$.  Saddles of $h_\epsilon$ that are far from $p_0$ will then converge to saddles of $h$ in the limit $\epsilon \rightarrow 0$, and the corrections at finite $\epsilon$ may be treated perturbatively.  However, since $h(p_0)$ does not generally vanish, there can also be additional saddles that approach the point $p_0$ as $\epsilon \rightarrow 0$ and which are unrelated to saddles of $h$.  Such saddles must be included at each $\epsilon >0$ and, while they are not associated with saddles of the original integrand $h(z)$, their effects may be interpreted as yielding `endpoint contributions at $p_0$' for the original integral \eqref{eq:testIntegral}.  In this sense, both endpoint and saddle-point contributions can be described using the same language of Lefschetz thimbles and Picard-Lefschetz theory.

\section{Action Computation for Constrained Axion Wormholes}
\label{sec:details}

This appendix describes the details of our computation of the actions for our constrained wormholes.

    \subsection{Overview}

        Recall that the metric we consider in Lorentzian axion gravity takes the form 

        \begin{equation}
             ds^2=dr^2+\ell L(r)\big(d\theta^2-\alpha^2\sin^2(\theta)dt^2\big).
        \label{eq:Bgwa}
        \end{equation}
        We can perform a coordinate transformation to put this into a Fefferman-Graham gauge by taking $z=z_0 \,e^{-r/\ell}$.
        The constant $z_0$ is fixed by expanding the metric function $L(r(z))$ in the form

        \begin{equation}
            L(r(z)) = \frac{\sqrt{\ell^2-q^2} z_0^2}{4z^2} -\frac{\ell}{2}+\mathcal{O}(z^2).
        \end{equation}
        Comparing with our boundary conditions for the metric \eqref{eq:LMetricBC1}, we then immediately find $z_0 = 2\big(\frac{\ell^2}{\ell^2-q^2}\big)^\frac{1}{4}$. Solving for $r$ yields  $r= -\frac{\ell}{4} \log\big(\frac{z^4}{16}(1-\frac{q^2}{\ell^2})\big)$. 
        
        We emphasize that this $z$ is the Fefferman-Graham coordinate associated with the conformal frame in which the boundary metric is
        \begin{equation}
        \label{eq:alphamet}
            ds^2_{bndy}= d\theta^2-\alpha^2\sin^2(\theta)dt^2.
        \end{equation}
        This is the conformal frame in which the bulk solution is easiest to treat, since it is related to the dS-invariant solution by the simply compactifying the static patch time  coordinate $t$ and rescaling $t$ by $\alpha$.  However, we in fact wish to compute classical actions associated with the $\alpha$-independent boundary metric
        \begin{equation}
        \label{eq:noalpha}
            ds^2_{bndy}= d\tilde \theta^2-\sin^2(\tilde \theta)dt^2.
        \end{equation}  
        Since \eqref{eq:noalpha} can be obtained from \eqref{eq:alphamet} by a Weyl-rescaling and a change of coordinates, in practice we will simply compute actions associated with the boundary conformal frame of \eqref{eq:alphamet}, and we will then use  the well-known conformal anomaly to translate such results to the conformal frame of \eqref{eq:noalpha}. The Weyl factor associated with this conformal transformation and accompanying coordinate transformation were reported in \cite{marolf:EntanglementPhaseTransition}:

        \begin{equation}
            \begin{split}
                &e^{2\omega}\big(d\tilde{\theta}^2-\sin^2(\tilde{\theta})dt^2\big)=d\theta^2-\alpha^2\sin^2(\theta)dt^2\\
                &\qquad\quad e^{\omega} =\alpha\sin(\theta)\frac{1+\tan^{2/\alpha}(\theta/2)}{2\tan^{1/\alpha}(\theta/2)}\\
                &\qquad\qquad\theta=2\arctan(\tan^\alpha(\tilde{\theta}/2)).
            \end{split}
            \label{eq:S2toDefectS2App}
        \end{equation}

        An important regulator in our problem is the one which renders finite the area of the codimension-2 defect.  Recall that we denote this regulator by  $\lambda$. In the main text we mentioned that, on each boundary, $\lambda$ is the angular size of the region around each defect (as measured by the $\alpha$-independent boundary metric \eqref{eq:noalpha}) which we excise to render our boundary conditions smooth. 
        Taking $\tilde \theta = \lambda$ then gives $\theta = \lambda_\alpha$ for
        \begin{equation}
            \label{eq:lambdaalphaeqn}
\lambda_\alpha=2\arctan(\tan^\alpha(\lambda/2)).
        \end{equation}

        Recall also that (in 2+1 dimensions) this excision creates two holes on each boundary, and that we glued together the edges of these holes to define smooth boundary conditions for the $\lambda$-regulated problem; see again figure \ref{fig:LBCs} from section \ref{sssec:singreg}.
      In practice, we then constructed approximate solutions to this problem by taking bulk geometries that satisfy unregulated boundary conditions, choosing arbitrary extensions of the two regulating surfaces into the bulk, excising the regions that contain non-compact pieces of the axis (defined by either $\theta=0$ or $\theta = \pi$), and then identifying the pair of regulating surfaces surfaces that anchor to the same boundary component; see again figure \ref{fig:RegulatingIdentification3d}.
   Recall also that we refer to these identifications as the left and right $\lambda$-seams.

However, to be useful we must choose regulating surfaces such that we can argue that the resulting cut-and-paste wormholes approaches that of the desired solutions in the $\lambda \rightarrow 0$ limit with fixed wormhole parameters $\alpha, q,L_0$.  In particular, we will choose regulating surfaces that
retract to the asymptotic boundary in this limit and which are thus effectively immersed in empty AdS$_3$. We can then make the extrinsic curvature of our $\lambda$-seam small by taking it to lie on a surface for which the entire extrinsic curvature tensor $K_{ij}$ would vanish at small $\lambda$ in empty global AdS. We thus choose the surfaces 
        \begin{equation}
        \label{eq:RTS}
            \cos(\lambda_\alpha)=\pm\frac{4-z^2}{4+z^2}\cos(\theta),
        \end{equation}
 where $z$ is the Fefferman-Graham coordinate defined above (associated with the $\alpha$-dependent boundary metric \eqref{eq:alphamet}). On these surfaces, we have $\theta=\lambda_\alpha$ or $\theta=\pi-\lambda_\alpha$ at the asymptotic boundary $z=0$, and $z=2\sqrt{\frac{1-\cos\lambda_\alpha}{1+\cos\lambda_\alpha}} = \lambda_\alpha+\lambda_\alpha^3/12 + O(\lambda_\alpha^5)$ when $\theta=0$ or $\pi$. 
Explicit computations for $z\ll \lambda_\alpha$ then give the extrinsic curvature components 
\begin{eqnarray}
\label{eq:lseamK}
K_{zz}=\frac{q^2\cot(\lambda_\alpha)}{4\ell}z^3 +\mathcal{O}(z^5) ,\quad K_{tt} = -\frac{q^2\alpha^2\sin(2\lambda_\alpha)}{32\ell}z^3 +\mathcal{O}(z^5),\quad K_{zt}= 0
\end{eqnarray}
which yield $K =\frac{5q^2\cot(\lambda_\alpha)z^5}{\ell^3}=O(\lambda_\alpha^4)$ and 
$K^{ij}K_{ij} =  \frac{1025q^4\cot^2(\lambda_\alpha)z^{10}}{16384}=O(\lambda_\alpha^8)$ so that $K_{ij}$ does indeed vanish in the limit $\lambda \rightarrow 0$.  

For general $z$ on the $\lambda$-seam, we write $\tilde z= z/\lambda_{\alpha}$ and find 
\begin{eqnarray}
\label{eq:lseamK2}
K_{zz}=\frac{q^2 \tilde z^3 (4-3\tilde z^2)}{16\ell(1-\tilde z^2)}\lambda_{\alpha}^2 +\mathcal{O}(\tilde z^3 \lambda_{\alpha}^4) ,\quad K_{tt} = -\frac{q^2\alpha^2\tilde z^3(1-\tilde z^2)}{16\ell}\lambda_{\alpha}^4 +\mathcal{O}(\tilde z^3\lambda_{\alpha}^6),\quad K_{zt}= 0
\end{eqnarray}
which yield $K =\frac{q^2 \tilde z^5(5-3\tilde z^2)}{16 \ell^3}\lambda_{\alpha}^4+O(\tilde z^5 \lambda_\alpha^6)$ and 
$K^{ij}K_{ij} =  \frac{q^4\tilde z^{10} (1025-192 \tilde z^2(8-3\tilde z^2))}{16384}\lambda_\alpha^8+O(\tilde z^{10}\lambda_\alpha^{10})$, so that again $K_{ij}$ vanishes in the limit $\lambda_{\alpha} \rightarrow 0$.
We may think of the above $K_{ij}$ as parameterizing the perturbation required to take our approximate-saddle to the actual solution satisfying the desired boundary conditions. However, it is then useful to estimate the difference in action between our spacetimes and the actual saddle-points.  To do so, let us note that both spacetimes are in fact saddles for a more general class of problem in which we also specify the induced metric on an additional finite-distance boundary that we take to lie at our $\lambda$-seam (see again figure \ref{fig:RegulatingIdentification3d}).  In particular, for our approximate-saddles we can in principle simply read off the associated metric on this surface and then impose it as an additional boundary condition.  In the same way, if we had the explicit form of the actual saddle-points in hand we could then similarly read off the induced metric on the appropriate extremal surface and then impose that result as a boundary condition.  In this way the two spacetimes are both saddles of analogous variational principles but with {\it different} induced metrics specified at the $\lambda$-seam.  

A key point is now that the correct action for this Dirichlet problem includes an additional Gibbons-Hawking-York term on either side of the $\lambda$-seam.  However, in that problem one in fact treats the spacetime as having been cut open along the $\lambda$-seam, so that one ignores any potential codimension-1 delta-function in the Ricci scalar at the $\lambda$-seam.  But as discussed in section \ref{ssec:ConstrainedWormholesConstruction} the contribution of these Gibbons-Hawking-York is numerically equal to the would-be contribution of the Ricci scalar delta-function at the $\lambda$-seam.  As a result, the desired action for each spacetime is numerically equal to the action that gives a good variational principle for this Dirichlet problem.

The second key point is then the well-known fact that, at first order about any saddle, varying such a Dirichlet boundary condition changes the action by an amount proportional to the Brown-York stress tensor $T_{ij}$ \cite{Brown:1992br}; i.e., under a small change $\delta h_{ij}$ in the induced metric at the $\lambda$-seam we have
\begin{equation}
\label{eq:actionchangeBY}
\delta S \approx \int_{\lambda-\rm{seam}} d^2x \sqrt{|h|} T^{ij}\delta h_{ij}. 
\end{equation}

We will apply \eqref{eq:actionchangeBY} to the variation about our cut-and-paste wormhole.  If one wished to instead estimate $\delta S$ by varying around the actual saddle one would instead need to work to second order in the perturbation since the first-order variation around any actual saddle must vanish.

Using the AdS$_{2+1}$ Fefferman-Graham expansion, the results \eqref{eq:lseamK} for the extrinsic curvature of the $\lambda$-seam of our cut-and-paste wormhole, and the fact that the $\lambda$-seam is restricted to $z<\lambda_\alpha$
gives  $\delta h_{ij} \sim z^0$, $h\sim \lambda_\alpha^{-2} \tilde z^{-4}$, and 
\begin{equation}
    T^{zz} = \frac{q^2\tilde z^7 (1-\tilde z^2)}{16}\lambda_\alpha^6 + O(\tilde z^7\lambda_\alpha^8),\quad T^{tt} = \frac{q^2\tilde z^7(4-3\tilde z^2)}{512\alpha^2(1-\tilde z^2)}\lambda_\alpha^4 +O(\tilde z^7 \lambda_\alpha^6),\quad T^{zt}=0.
\end{equation}
Here the difference $\delta h_{ij}$ between the induced metric on the $\lambda$-seam of our cut-and-paste wormhole and the metric of the true saddle should also vanish as $\lambda_\alpha \rightarrow 0$ but, since we have not yet estimated the rate at which it vanishes, we conservatively allow it to be $O(\lambda_\alpha^0)=O(1)$ in this limit.
Putting these estimates together yields
\begin{equation}
\delta S \sim \int_{z < \lambda_\alpha} dtdz\  \sqrt{|h|} T^{tt} \delta h_{tt}  \sim \int_{\tilde z < 1} dt d\tilde z\    f(\tilde z) \lambda_\alpha^4 \sim \lambda_\alpha^4, 
\end{equation}
for some function $f(\tilde z)$ that only depends on $\tilde z$, and a simple analysis shows that $f(\tilde z)=O(\tilde z^5)$ for small $\tilde z$, so the above integral converges near the asymptotic boundary.  Thus $\delta S$ vanishes as desired in the limit $\lambda_\alpha \rightarrow 0$. 

        As a result, in the limit of small $\lambda$, the action of the desired constrained wormhole saddles will be well-approximated by the action of the cut-and-paste wormholes constructed in section \ref{sssec:singreg}.  We compute that action in the following subsections, addressing the bulk action in section \ref{app:bulk} and boundary terms in section \ref{ssec:bndyaction}.  In both cases, we work in the conformal frame of the $\alpha$-dependent boundary metric \eqref{eq:alphamet}. In section \ref{sssec:BCAS} we then separately compute the conformal anomaly associated with transforming to the conformal frame of the $\alpha$-independent boundary metric \eqref{eq:noalpha}.   In both sections \ref{app:bulk} and \ref{ssec:bndyaction}
        we will need to take due care to consider possible contributions from the $\lambda$-seam discussed above.

        However, the above identifications lead to two localized contributions to our  action.  The first appears in the boundary counter-terms due to  a delta function in the Ricci scalar of the of the two dimensional asymptotic boundary. The coefficient of this delta-function can be computed from the Gauss-Bonnet theorem since the boundary is topologically a torus but has locally positive constant curvature. The second appears in the Gibbons-Hawking-York boundary term due to a delta function in the extrinsic curvature of the asymptotic boundary.  This delta-function arises from a discontinuous change in the outward unit normal as one crosses the identification surface. We will discuss these effects more in detail when we calculate boundary terms in the constrained wormhole action in section \ref{ssec:bndyaction}.

    \subsection{The Bulk Action}
\label{app:bulk}

        We are now ready to calculate the bulk terms in the action \eqref{eq:LAction1}  of Lorentzian constrained axion wormholes.  We will divide this computation into four pieces:
\begin{enumerate}
\item{}  The contribution 
        \begin{equation}
        \label{eq:Sreg2}
            \begin{split}
                S_\text{\tiny reg. bulk} &= \frac{2}{16\pi\Gn}\int_{\M_{\tiny R}} d^3x\sqrt{|g|}\bigg(R_\text{\tiny reg.}+\frac{2}{\ell^2}-\frac{1}{2}\partial_\mu\chi\partial^\mu\chi\bigg)\\
                &=\frac{1}{8\pi\Gn}\int_{\M_\text{\tiny R}} d^3x\sqrt{|g|}\bigg(\frac{-4}{\ell^2}\bigg),
            \end{split}
        \end{equation}
        from the regular part of the bulk. Here $R_\text{\tiny reg.}$ is the   regular part of the Ricci scalar and second line follows by using the trace of Einstein's equations to relate $R_\text{\tiny reg.}$ and $\partial_\mu\chi\partial^\mu\chi$ to $\ell$.
        As in the main text, the symbol $\M_{\tiny R}$ indicates that \eqref{eq:Sreg2} integrates only over the right half-wormhole, though the factor of 2 in the first line is included to give the regular part of the action for the full wormhole.

\item{} The localized contribution from the codimnsion-2 defect $\gamma$.

\item{} A localized contribution at the $\chi=0$ surface.

\item{} A localized contribution from the $\lambda$-seam.
\end{enumerate}
In each case, we regulate any divergences by cutting off the integrals at a surface $z=\delta$ defined by the above Fefferman-Graham coordinate associated with the $\alpha$-dependent boundary metric \eqref{eq:alphamet}.  While this is not the conformal frame that we eventually wish to study, it is the regulator we should use to consistently first evaluate the action in the conformal frame of \eqref{eq:alphamet} and to then compute the action in the conformal frame of \eqref{eq:noalpha} by adding the appropriate anomaly term.

        It is straightforward to calculate \eqref{eq:Sreg2} in the conformal frame of \eqref{eq:alphamet}, though one must take care to integrate only over the region included in our cut-and-paste wormhole; see again figure \ref{fig:RegulatingIdentification3d}.  As noted above, for small $\lambda$ we expect the relevant region to be approximately given by \begin{equation}
        \label{eq:applseam}
            z\geq \text{max}\Bigg(\delta,2\sqrt{\frac{\cos(\theta)\mp\cos(\lambda_\alpha)}{\cos(\theta)\pm\cos(\lambda_\alpha)}}\Bigg).
        \end{equation}
    We calculate the bulk action using the above region and the cutoff $z\geq \delta$. We can do this in parts by first integrating over the portion of the bulk $\delta\geq2\sqrt{\frac{\cos(\theta)-\cos(\lambda_\alpha)}{\cos(\theta)+\cos(\lambda_\alpha)}}$ and then integrating over the region $\delta \leq2\sqrt{\frac{\cos(\theta)-\cos(\lambda_\alpha)}{\cos(\theta)+\cos(\lambda_\alpha)}}$. The first integral is straightforward and gives
        
        \begin{equation}
        \begin{split}
             S_\text{\tiny bulk}^{(1)} &= \frac{2}{16\pi\Gn}\int_0^T dt \int_{r_0}^{r(\delta)} dr \int_{\arccos\big(\cos(\lambda_\alpha)\frac{4+\delta^2}{4-\delta^2}\big)}^{\arccos\big(-\cos(\lambda_\alpha)\frac{4+\delta^2}{4-\delta^2}\big)} d\theta \sqrt{-g}\Big(\frac{-4}{\ell^2}\Big)\\ 
            &= \frac{2 T \alpha \cos(\lambda_\alpha)}{16\pi\Gn }\bigg[ \frac{-4\ell}{\delta^2}-4\ell \log(\delta/2) -4r_0 - \ell\log(1-q^2/\ell^2)+2\sqrt{\ell^2-q^2}\sinh(2r_0/\ell)\bigg] +\mathcal{O}(\delta).\\
        \end{split}
        \end{equation}
    In the above, we integrated over half of the wormhole, from $r_0$ out to the regulated surface $z=\delta$, and then doubled the result to get the answer for the full wormhole. We have also made use of the fact that $\lambda_\alpha \ll 1$. 

    We can then integrate over the bulk region $\delta \leq2\sqrt{\frac{\cos(\theta)-\cos(\lambda_\alpha)}{\cos(\theta)+\cos(\lambda_\alpha)}}$. We find
    
    \begin{equation}
         \begin{split}
             S_\text{\tiny bulk}^{(2)} &= \frac{4}{16\pi\Gn} \int_0^T dt \int_{0}^{\lambda_\alpha}d\theta\int_{r_0}^{r\Big(z=2\sqrt{\frac{\cos(\theta)-\cos(\lambda_\alpha)}{\cos(\theta)+\cos(\lambda_\alpha)}}\Big)} dr \sqrt{-g}\Big(\frac{-4}{\ell^2}\Big)\\
             &=\frac{8\ell \alpha T}{16\pi\Gn}\big(\log(\delta)-\log(\lambda_\alpha)\big) -\frac{4\ell\alpha T \lambda_\alpha^2}{16\pi\Gn}\log(\delta/2)+\mathcal{O}(\delta)+\mathcal{O}(\lambda_\alpha^2\delta^0),
        \end{split}
    \end{equation}
    where we have again integrated over only half the wormhole and then doubled the result to find the contribution to the full wormhole. We also only integrate over the region with $\theta<\pi/2$ and again double the result. The regular part of the bulk thus contributes  
        \begin{equation}
        \begin{split}
           S _\text{\tiny reg. bulk}= S_\text{\tiny bulk}^{(1)}+S_\text{\tiny bulk}^{(2)} &=\frac{2\alpha T \ell}{16\pi\Gn}\Bigg(-\frac{4}{\delta^2}-2(1+2\frac{r_0}{\ell})-\log\Big(1-\frac{q^2}{\ell^2}\Big)+4\alpha\log(\lambda/2)\\
           &+2\sqrt{1-\frac{q^2}{\ell^2}}\sinh\Big(\frac{2r_0}{\ell}\Big)\Bigg) +\mathcal{O}(\lambda_\alpha^2\delta^0)+\mathcal{O}(\delta).
        \end{split}
        \end{equation}
        The term of order $\delta^{-2}$ is expected and should be canceled by the usual boundary counter terms.

        With this in hand, let's now turn to the remaining three contributions to the bulk action in this conformal frame.  We begin with the delta-function contribution to the bulk Ricci scalar $R$ associated with the codimension-2 conical defect $\gamma$.  This contribution is given by summing the $\partial U_\epsilon$ term and the $A_\gamma$ term in \eqref{eq:defectSEH}.  Since the parameter ${\cal N}$ vanishes in our case, we find 
            \begin{equation}
              S_\gamma=  \frac{-L_\gamma}{8\pi\Gn}(\alpha T+2\pi i)=\frac{\Big(2\alpha\ell\log(\lambda/2)+\frac{1}{2}\log(1-\frac{q^2}{\ell^2})+2r_0\Big)}{8\pi\Gn}(\alpha T+2\pi i),
            \end{equation}
            where $L_\gamma$ is the proper length of $\gamma$, the final step uses \eqref{eq:alphasolve}, and where our convention is that the defect contribution to the action is in fact $2S_\gamma$.  

We next consider the localized contribution from the $\chi=0$ surface.  As we  have seen, this surface is associated with a codimension-1 delta-function in the Ricci scalar.  Furthermore, as we have also discussed, this contribution may be equivalently computed using a Gibbons-Hawking-York term at the $\chi=0$ surface. 
On the smooth part of the right half-wormhole ${\cal M}_R$, the extrinsic curvature of constant-$r$ surfaces takes the simple form
        \begin{equation}
            K_\pm(r) =\pm \partial_r\log\big(L(r)\big),
        \label{eq:Kpm}
        \end{equation}
        where the sign is determined by whether we take the normal to the surface to be in the direction of increasing or decreasing $r$. In particular, the $(-)$ sign is appropriate when thinking of the $\chi=0$ surface as a boundary of ${\cal M}_R$ (using ${\cal M}_L$ would lead to the same result after two changes of sign).
        We thus find
        \begin{equation}
            \frac{1}{8\pi G_N}\int_{\chi=0}d^2x\sqrt{|h|}K_-(r_0)=-\frac{2}{8\pi\Gn}\alpha T \sqrt{\ell^2-q^2}\sinh\Big(\frac{2r_0}{\ell}\Big).
        \end{equation}
        However, desired Ricci scalar contribution is given by the sum of the above term and a corresponding term evaluated on the left side of the $\chi=0$ surface.  Due to the ${\mathbb Z}_2$ symmetry, we should thus include 2 times the above result in our action.

Finally, we should consider the corresponding contribution from the $\lambda$-seams.  Here we may again use the result \eqref{eq:lseamK2} for the extrinsic curvature at the $\lambda$-seam which yields
      
\begin{equation}
 \int_{{\text{right}}\ \lambda\text{-seam}} d^2x\sqrt{|h|}K= \frac{3 T q^2\alpha \lambda_\alpha^3}{64\ell} +O(\lambda_\alpha^5),
\end{equation}

where one factor of $\lambda_\alpha \approx \sin(\lambda_\alpha)$ and the factor of $T\alpha$ both come from the size of the $t$-circle.  As usual, the above should be multiplied by 2 to include the contribution from the left $\lambda$-seam as well.
 We thus see that this contribution is negligible when $\lambda$ is small.
This concludes our discussion of bulk contributions to the constrained wormhole action.

    \subsection{The Boundary Action}
    \label{ssec:bndyaction}

        We now turn to contributions to the action from asymptotic boundaries.  As noted above, we will first compute the boundary contributions as defined by the boundary conformal frame in which the metric takes the form \eqref{eq:alphamet}.  The conformal anomaly associated with transforming to the conformal frame of \eqref{eq:noalpha} will then be computed separately in section \ref{sssec:BCAS} below.

        The contribution to the action from the boundaries is perhaps the most involved part of the calculation.  There are several pieces to consider:
        
        \begin{enumerate}
\item{}  Counter-term contributions:  The standard volume counter-term of \cite{Henningson:1998gx}   (or see e.g. \cite{Chapter19} for a review matching our current conventions) yields
        \begin{equation}
            S_\text{\tiny CT} = \frac{-2\alpha T\ell\cos(\lambda_\alpha)}{16\pi\Gn}\frac{4}{\delta^2},
        \end{equation}
        where the factor of $2$ in the numerator comes from summing the contributions from both boundaries.   
        This is the only counter-term that we require.  In particular, there is no $\ln \delta$ counter-term since, after the surgery described in section \ref{sssec:singreg}, each boundary is a torus.  As a result, the (here, Lorentzian, or complex) Gauss-Bonnet theorem requires the integral of the boundary Ricci scalar-density to vanish, so that the coefficient of the log-divergence vanishes as well.

\item{} The Gibbons-Hawking-York term on the smooth part of the asymptotic boundary: Noting that the outward-directed normal at the right asymptoti boundary $\partial {\cal M}_R$ is in the direction of increasing $r$, we may use 
\eqref{eq:Kpm} to find the contribution 

        \begin{equation}
            \frac{2}{8\pi G_N}\int_{\partial\M_{\tiny R}}d^2x\sqrt{|h|}K_+(r_\infty)=\frac{2\ell\alpha T\cos(\lambda_\alpha)} {8\pi\Gn}\frac{4(1+\delta^2/2)}{\delta^2},
        \end{equation}
        where the above result makes use of the $\mathbb{Z}_2$ symmetry of the boundary conditions to obtain the full two-boundary GHY term from the one boundary computation.
        As expected, this precisely cancels the divergent $\mathcal{O}(\delta^{-2})$ terms in the bulk action and the counter term.
        
\item{} A further contribution to the Gibbons-Hawking-York term localized at the $\lambda$-seam on the  asymptotic boundary.

\end{enumerate}
        
It thus remains only to compute the third contribution above.  By definition, the asymptotic Gibbons-Hawking-York boundary term is computed on a surface $z=\delta$ in the limit $\delta\rightarrow 0$.  In particular, we take this limit for each fixed $\lambda, \alpha$ so that $\delta \ll \lambda\alpha$.  In this regime, we are far from the defect $\gamma.$  It is thus sufficient to compute delta function contributions to the extrinsic curvature on a slice of constant Killing time and to then later integrate over $t$. 

\begin{figure}
                \centering
                \includegraphics[width=0.4\linewidth]{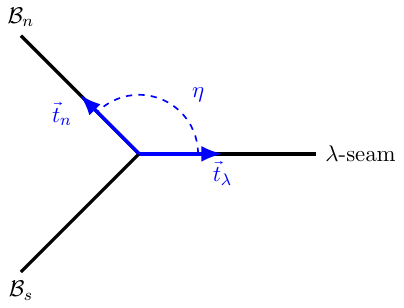}
                \caption{A region $\mathcal{B}_n$ of an asymptotic boundary with $\theta\sim \lambda$ (near the would-be north pole at $\theta=0$) is glued across the $\lambda$-seam to a region $\mathcal{B}_n$ of the same boundary with $\theta\sim \pi-\lambda$.  This gluing creates a kink in the asymptotic boundary at the $\lambda$-seam.   The contribution of this kink to the Gibbons-Hawking-York term can be computed from the difference in the angle between the unit normal of $\mathcal{B}_n$ and that of $\mathcal{B}_s$. Equivalently, using the ${\mathbb Z}_2$ symmetry $\theta\rightarrow \pi-\theta$ that exchanges ${\cal B}_n$ and ${\cal B}_s$, it can be computed from the angle $\eta$ between the unit tangent vectors which we've denoted as $2\eta$ here.}
                \label{fig:placeholder2}
            \end{figure}

             This procedure reduces the problem to that of  computing the extrinsic curvature of a non-smooth one-dimensional curve in a two-dimensional space of Euclidean signature.  We can thus use the standard result that, in {\it flat} two-dimensional Euclidean plane, integrating the extrinsic curvature along such a curve would give the change in direction of the unit normal vector (measured as an angle as in the Gauss-Bonnet theorem for spaces with boundaries that have corners).  The desired contribution is thus given by the change across the $\lambda$-seam in the angle of the normal to the asymptotic boundary; i.e., the change when we pass from the relevant surface anchored at  $\theta=\lambda_\alpha$ to the one at $\theta=\pi-\lambda_\alpha$. Furthermore, since our boundary has a symmetry under $\theta\rightarrow\pi-\theta$, this jump in angle can be computed from the angle $\eta$ between the asymptotic boundary on one side of the seam and the seam itself. From this we see that $\Delta\theta =2\eta-\pi$; see figure \ref{fig:placeholder2}.

            To compute this angle, we must first express the tangent vector to the regulated boundary ($\vec{t}_n$) and the tangent vector to the $\lambda$-seam ($\vec{t}_\lambda$) in terms of the parameters $\lambda_\alpha$ and $\delta$ controlling our regulating surfaces. Using the fact that, near the $\lambda$-seam and for small $\lambda$, the metric on the $z\theta$ plane can be approximated by empty AdS, we find that at the locus where the $\lambda$-seam intersects the surface $z=\delta$, we have

            \begin{eqnarray}
                t_{n}^{\ i}&&= \frac{\delta}{\ell}\delta^i_{\ \theta} +\mathcal{O}(\delta^3),\\
                t_{\lambda,\ i}&&=\Big(-\frac{\ell}{\delta}+\frac{\ell}{2}\cot^2(\lambda_\alpha)\Big)\delta_i^{\ z}+\ell\cot(\lambda_\alpha)\delta_i^{\ \theta} +\mathcal{O}(\delta^3).
            \end{eqnarray}
            We can now extract the angle $\eta$ and, consequently, the contribution to the extrinsic curvature from the $\lambda$-seam directly from the above results:

            \begin{eqnarray}
                \eta &&= \arccos(\vec{t}_n \cdot \vec{t}_\lambda) = \frac{\pi}{2}-\delta\cot(\lambda_\alpha)+ \dots,\\
                \int_{{\cal B} \ {\rm near} \ \lambda\text{-seam}}d^2x\sqrt{|h|}K &&= \int_0^T dt \frac{\alpha\ell(2\eta-\pi)}{\delta}\sin(\lambda_\alpha) (1+O(\delta))=-2\ell\alpha T +O(\lambda)+O(\delta).
            \end{eqnarray}
            In the first line the $\dots$ represent higher order corrections in $\lambda$ and $\delta$.  In the last line, the integral on the left is performed over a tiny region of the asymptotic boundary ${\cal B}$ around the $\lambda$-seam and we have used the fact that $1\gg\lambda_\alpha$ to write $\sin(\lambda_\alpha)\approx\lambda_\alpha$. This is the contribution to the boundary action from one $\lambda$-seam, but as can be seen in figure \ref{fig:RegulatingIdentification3d} there are two such seams, so the total contribution from the action from these seams is,

            \begin{equation}
                \frac{2}{8\pi\Gn}\int_{{\cal B} \ {\rm near} \ \lambda\text{-seam}}d^2x\sqrt{|h|}K=\frac{-4\ell\alpha T}{8\pi\Gn}.
            \end{equation}

Summing the above contributions gives the action of our constrained wormholes as defined by the boundary conformal frame associated with \eqref{eq:alphamet}.  The translation to the desired boundary conformal frame (associated with \eqref{eq:noalpha}) will be discussed below.

        \subsection{The Contribution from the Conformal Anomaly}
        \label{sssec:BCAS}

            We now compute the conformal anomaly associated with changing the boundary conformal metric from \eqref{eq:alphamet} to \eqref{eq:noalpha}.  This will be the final piece in our calculation of the desired constrained wormhole action.   
            
In two boundary dimensions, it is well known \cite{POLYAKOV1981207} that the conformal anomaly associated with changing from the conformal frame defined by a line element ${ds}^2$ to a line element $\widetilde{ds}^2 = e^{-2\omega}{ds}^2$ is computed by the Liouville action (with zero cosmological constant) of the (log of the) associated Weyl factor $\Omega=e^\omega$,   where the conformal charge is taken to be $c=3\ell/2G$ \cite{Brown:1986nw}; i.e.,

                \begin{equation}
                    S_\text{\tiny Anomaly} = \frac{\ell}{16\pi\Gn}\int_{\tiny{\partial\M}} d^2{x} \sqrt{|{h}|} \Big(\partial_{{\mu}}\omega\partial^{{\mu}}\omega -{ {\mathcal{R}}}^{(2)}\omega\Big),
                    \label{eq:Sanom}
                \end{equation}
                where  the volume element $|h|$ and Ricci curvature ${\mathcal {R}}$ are computed using the original line element ${ds}^2$ (for us, \eqref{eq:alphamet}).  Since we need to perform this Weyl transformtion on both the left and right asymptotic bouundaries, we will in fact need to include a contribution $2S_\text{\tiny Anomaly}$ in the total action of our wormhole.
                
\begin{figure}
                    \centering
                    \includegraphics[width=0.6\linewidth]{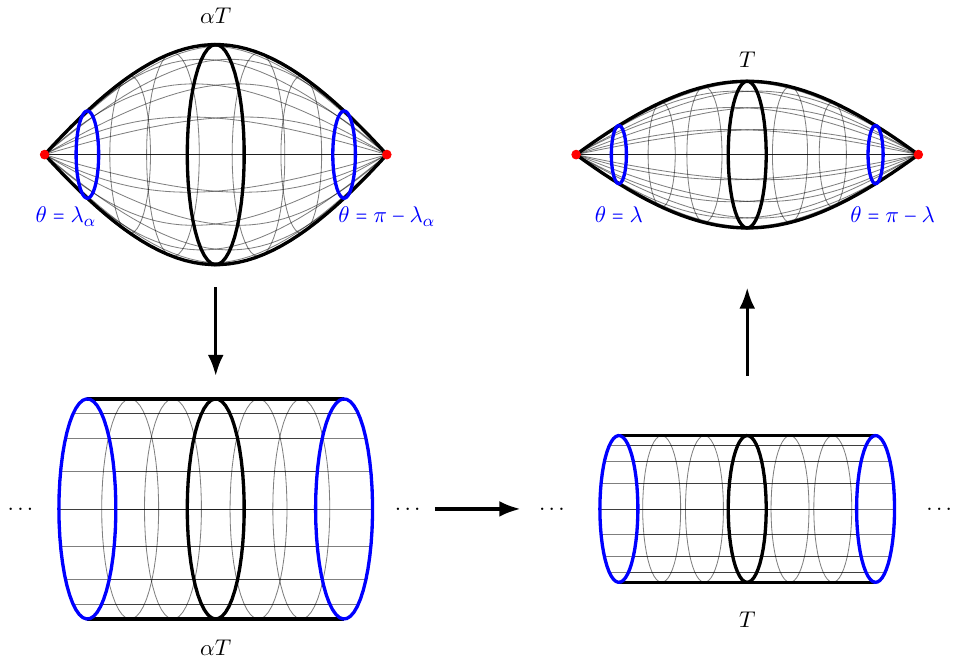}
                    \caption{The three steps in our Weyl transformation from \eqref{eq:alphamet} to \eqref{eq:noalpha}.  In the initial boundary metric \eqref{eq:alphamet} (upper left), the proper time along the maximal closed timelike curve is $\alpha T$.  We Weyl-transform this to a cylinder of circumference $\alpha T$ (lower left), which we then rescale to a cylinder of circumference $T$ (lower right).  A final Weyl transfrormation takes us to the desired metric \eqref{eq:noalpha} (upper right), where the  the proper time along the maximal closed timelike curve is again $T$.  In each case, the blue circles denote the cutoff associated with the UV regulator $\lambda$.}
                    \label{fig:BoundaryWeyl}
                \end{figure}

                Evaluating the Liouville action using the desired Weyl factor \eqref{eq:S2toDefectS2App} is rather messy.  However, the calculation can be simplified by breaking up the Weyl transformation into simpler steps and summing the associated anomaly contributions.  In particular, we proceed in three steps as shown in figure \ref{fig:BoundaryWeyl}:

                \begin{enumerate}
                    \item We first perform a Weyl transformation which takes the boundary metric 
                    \begin{equation}
                    \label{eq:alphamet2}
                                ds^2_{bndy}= d \theta^2-\alpha^2 \sin^2( \theta)dt^2,
                                \end{equation}
                     (identical to \eqref{eq:alphamet} above) to a flat Lorentzian cylinder with periodic time. In particular, we take $t$ to be periodic with period $T$.  The scale of the cylinder is set by taking the proper time around its circumference to be $\alpha T$.  Note that the regulator $\lambda$ will render the length of the cylinder finite by restricting us to $\theta \in [\lambda_\alpha, \pi-\lambda_\alpha]$ with $\lambda_\alpha$ given by \eqref{eq:lambdaalphaeqn}. In fact, as explained above, the ends of the cylinder at $\theta = \lambda_\alpha, \theta = \pi-\lambda_\alpha$ should be identified to form a flat Lorentzian torus.  As in the main text, it will be useful to refer to this identification as `the $\lambda$-seam'.  
                     \item We next perform a second Weyl transformation which simply scales all lengths on the boundary torus by a constant factor of $1/\alpha$.  The proper time around the $t$-cycle thus becomes $T$.  Furthermore, as will be clear in a moment, the proper distance around the $\theta$-cycles becomes just $\pi-2\lambda$. 
                    \item Finally, we  Weyl transform this rescaled cylinder to 
                    \begin{equation}
                    \label{eq:noalpha2}
                                \widetilde{ds}^2_{bndy}= d\tilde \theta^2-\sin^2(\tilde \theta)d t^2,
                    \end{equation}
(identical to \eqref{eq:noalpha} above) as depicted on the right side of figure \ref{fig:BoundaryWeyl}.  After doing so, the $\lambda$-seam identifications are described as identifying the $\alpha$-independent surfaces $\theta = \lambda_\alpha$ and $\theta = \pi-\lambda_\alpha$.  Indeed, the original cutoff $\lambda_\alpha$ \eqref{eq:lambdaalphaeqn} on \eqref{eq:alphamet2} was designed to ensure that this is the case, and it is this fact that sets the proper distances around the $\theta$-cycles of the tori above.  Note also that the coordinate $t$ is the same as in \eqref{eq:alphamet2} and, as a result, its period is again  $T$.
                \end{enumerate}

The Weyl factor of the first Weyl transformation is simply $e^{-\omega} = \frac{1}{\sin \theta}$, or $\omega = \ln \big( \sin \theta \big)$.  The associated computation of \eqref{eq:Sanom} is then largely straightforward.  In particular, since the factor of $\alpha$ in \eqref{eq:alphamet2} can in principle be absorbed into a rescaling of the coordinate $t$, the smooth part of the boundary has locally constant positive curvature ${\mathcal{R}}^{(2)}=2$.  However, due to the  identification of $\theta=\lambda_\alpha$ with $\theta=\pi-\lambda_\alpha$, the boundary Ricci scalar also has a delta function contribution at the $\lambda$-seam. That this must be the case can be seen from the fact that the Ricci scalar must integrate to zero by the Gauss-Bonnet Theorem. Using this result to compute the delta-function contribution gives 
                \begin{equation}
                    \sqrt{|h|}{\mathcal{R}}^{(2)}_\delta =-\delta(\theta-\lambda_\alpha)\bigg[2 \int_{\overline{\partial\mathcal{M}}_{\tiny R}}d^2x\,\sqrt{|{h}|}\bigg]=-4\alpha T \cos(\lambda_\alpha)\delta(\theta-\lambda_\alpha).
                \end{equation}
                
Including these two curvature contributions as well as the contribution from $\partial_\mu \omega \partial^\mu \omega$,   we find that, with conventions defined in equations \eqref{eq:Sanom} and \eqref{eq:LSbndy} and to leading order in $\lambda$) the anomaly action $S_\text{\tiny Anomaly}^{(1)}$ for the first step in our Weyl transformation  takes the form 
                \begin{equation}
                    S^{(1)}_\text{\tiny Anomaly}= \frac{\ell\alpha T}{8\pi\Gn} \bigg(1+\alpha\log(\lambda/2)\bigg).
                    \label{eq:anom1}
                \end{equation}
The second step involves rescaling a flat torus by the constant Weyl factor $\omega = \ln \alpha$.  As a result, we have both $\partial_\mu \omega =0$ and $ \tilde{\cal R} =0$.  Thus both terms in \eqref{eq:Sanom} vanish explicitly and our second step gives $S^{(2)}_\text{\tiny Anomaly}=0$.

Let us now observe that the Weyl transformation in the third step is just the inverse of that for the first step evaluated for the special case $\alpha=1$.  We thus have $S^{(3)}_\text{\tiny Anomaly} =-S^{(1)}_\text{\tiny Anomaly}|_{\alpha=1}$.  Putting all of these together then writes the full anomaly contribution (from both the left and right asymptotic boundaries) to the action of our wormhole in the form
                \begin{equation}
                    2S_\text{\tiny Anomaly}= \frac{\ell(\alpha-1) T}{4\pi\Gn} \bigg(1+(\alpha+1)\log(\lambda/2)\bigg).
                \end{equation}
Adding this to the contributions computed in previous sections then yields \eqref{eq:CWHAction_Explicit}.

\section{Spurious saddles in the $L_0$-parameterization}
\label{app:spurious}

As mentioned in section \ref{ssec:SPCD}, at fixed $L_\gamma, \beta, \chi_\infty$ the Euclidean action $S_E$ will be stationary with respect to $L_0$ when the traced extrinsic curvature is continuous across the $\chi=0$ surface.  However, it is also potentially stationary on other constrained wormholes.  We explain this result below, describe why the latter class of saddles is likely to be a spurious result of a degenerate parameterization, and verify that such spurious saddles do not affect the discussion in the main text.  

Let us begin by
recalling from \cite{Dong_2020} that the Einstein-Hilbert action is a good variational principle when the length $L_\gamma$ of our defect is fixed and our defect parameter $\alpha$ is free to vary.  Furthermore, as also described in \cite{Dong_2020}, this property remains intact under the addition of matter terms that are algebraic in the spacetime metric.  Now, the term that we in fact need to add to the Einstein-Hilbert action is an appropriate version of the Lagrange multiplier term \eqref{eq:Smu}.  Recalling that the purpose of this term is to fix the area of the surface where $\chi=0$, let us write this term in the form
        \begin{eqnarray}
            S_\mu &=& \frac{\mu}{16\pi\Gn}\Bigg[\frac{1}{2}\int_{\cal M} \sqrt{|h|}\, d^2x\wedge d \left[\text{sgn}(\chi)\right]  - {\cal A}_0\Bigg] \\ &=& \frac{\mu}{16\pi\Gn}\Bigg[\int_{-\infty}^\infty dr\int_{\tiny\Sigma_r}d^2x\sqrt{|h|}\,\frac{1}{2}\partial_r\text{sgn}(\chi)  -{\cal A}_0\Bigg],
        \label{eq:Smu3}
        \end{eqnarray}
which is indeed algebraic in the metric.  Taking $S$ to be the usual gravitation-plus-axion action (as in \eqref{eq:LAction1}), it follows that the desired constrained wormholes are stationary points of $S+S_\mu$ when we hold fixed the boundary conditions, $L_\gamma$, and the parameter ${\cal A}_0$ in \eqref{eq:Smu3}.\footnote{Recall, however, that the actual cut-and-paste wormholes constructed in section \ref{sssec:singreg} do not quite satisfy this condition due to a discontinuity at the $\lambda$-seam.  We will comment on this further below.}     We may thus write
        \begin{eqnarray}
        \label{eq:varywrtA0}
        \frac{\partial S_{E}}{\partial {\cal A}_0}\Big|_{L_\gamma,\chi_\infty,\beta} &=& -i        \frac{\partial S_{CWH}}{\partial {\cal A}_0}\Big|_{L_\gamma,\chi_\infty,\beta} =  -i        \frac{\partial (S+S_{\mu})}{\partial {\cal A}_0}\Big|_{g,\chi,L_\gamma} \\ &=&
        -i  \frac{\mu }{16\pi G_N} = i  \frac{K_R }{8\pi G_N} ,
        \end{eqnarray}
 where the equality at the end of the first line follows from the fact that (as noted above) we evaluate the derivative at a spacetime metric $g$ and axion field $\chi$
that would define a stationary point of $(S+S_{\mu})$ if the parameters
$L_\gamma, {\cal A}_0$ were both held fixed.  The 2nd line then follows immediately since the only term in our action that contributes to $\frac{\partial (S+S_{\mu})}{\partial {\cal A}_0}$ is the final term in \eqref{eq:Smu3} (which is explicitly linear in ${\cal A}_0$); all other terms are constants when $g$ and $\chi$ are held fixed.  In the final step we have also used \eqref{eq:musolve} and the ${\mathbb Z}_2$ symmetry that sets $K_R=K_L.$

As a result, and as one might expect, the Euclidean action $S_E$ will be stationary with respect to ${\cal A}_0$ precisely when $K_R=0.$  Recall, however, that we wish to study stationarity with respect to $L_0$ at fixed $L_\gamma$.  Having set the period of the boundary time coordinate $t$ to $T=-i\beta$, the metric ansatz \eqref{eq:L2Ansatz} gives an area (see \eqref{eq:AL001})
\begin{equation}
{\cal A}_0 = -i2\alpha \beta \ell L_0.
\end{equation}
We thus compute 
        \begin{equation}
        \frac{\partial S_{E}}{\partial L_0}\Big|_{L_\gamma,\chi_\infty,\beta} = \frac{\partial S_{E}}{\partial {\cal A}_0}\Big|_{L_\gamma,\chi_\infty,\beta}\frac{\partial {\cal A}_0}{\partial L_0}|_{L_\gamma,\chi_\infty,\beta} 
        = \frac{K_R }{4\pi G_N} \left(\alpha \ell + \ell L_0 \frac{\partial \alpha}{\partial L_0}|_{L_\gamma,\chi_\infty}  \right).
        \end{equation}

As a result, we should be aware that our parameterization in terms of $T, L_0, L_\gamma$ may introduce additional saddles at zeros of the quantity  $\alpha + L_0 \frac{\partial \alpha}{\partial L_0}|_{L_\gamma,\chi_\infty}$.  We expect that such saddles are spurious in that they will contribute to the path integral with strictly-vanishing one-loop determinant.  This expectation is based on the assumption that the path integral measure is smooth with respect to the natural variables defined by the $\chi=0$ area ${\cal A}_0$ and the defect length $L_\gamma$.  In terms of $L_\gamma, L_0$ we thus find
\begin{equation}
d{\cal A}_0 \ dL_\gamma = \frac{\partial {\cal A}_0}{\partial L_0}|_{L_\gamma,\chi_\infty,\beta}  dL_0 \ dL_\gamma   = 2\beta \ell \left(\alpha + L_0 \frac{\partial \alpha}{\partial L_0}|_{L_\gamma,\chi_\infty}  \right)dL_0 \ dL_\gamma,
\end{equation}
so that the coefficient of $dL_0 \ dL_\gamma$ vanishes at zeros of 
$\alpha + L_0 \frac{\partial \alpha}{\partial L_0}|_{L_\gamma,\chi_\infty}$ and, as a result, any one-loop determinant should vanish as well.

This subtlety can clearly be avoided by using the non-degenerate parameterization defined by $L_\gamma$ and $\tilde L_0 = \alpha L_0$.  For the interested reader, we present an analysis using this parameterization in appendix \ref{sec:AlternateParamatrization}, which finds results that agree with those of sections \ref{ssec:SPCD} and \ref{ssec:AnalyticContinue} but which requires slightly more numerics.

Luckily, however, one can also show directly that such spurious saddles do not contaminate the analyses of sections \ref{ssec:SPCD} and \ref{ssec:AnalyticContinue}.  We have verified this in several ways.  First, we have checked numerically that the saddles discussed there have vanishing $K_R$ in the limit where our spacetimes are indeed saddle points of the action.  Recall that, due to our cut-and-paste construction this occurs only in the limit where the $\lambda$-seam is far from the throat of the wormhole so that the discontinuity in extrinsic curvature at the $\lambda$-seam is small.  As a result, the correct limit to study is the one that takes $\lambda \rightarrow 0$ while holding fixed $\alpha, L_0, \beta, \chi_\infty$ and thus where $L_\gamma$ becomes large.

\begin{figure}
            \centering
            \begin{minipage}{0.4\textwidth}
                \includegraphics[width=\linewidth]{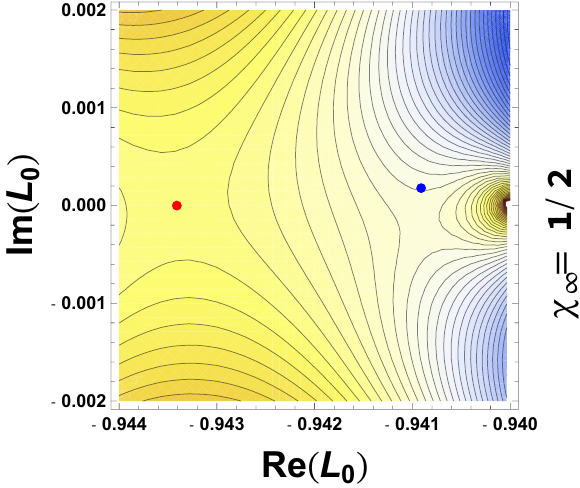}
            \end{minipage}
            \begin{minipage}{0.4\textwidth}
                \includegraphics[width=\linewidth]{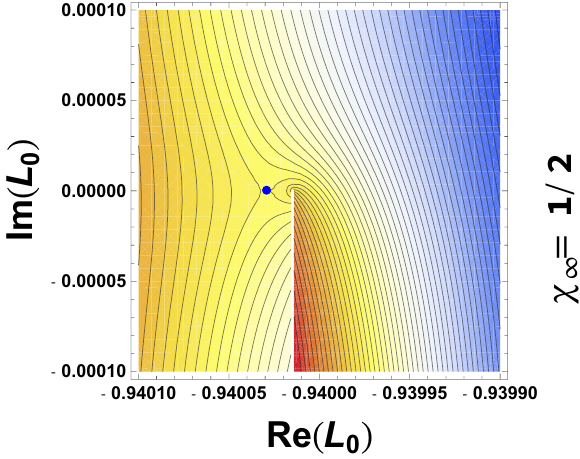}
            \end{minipage}
            \begin{minipage}{0.35\textwidth}
                \includegraphics[width=\linewidth]{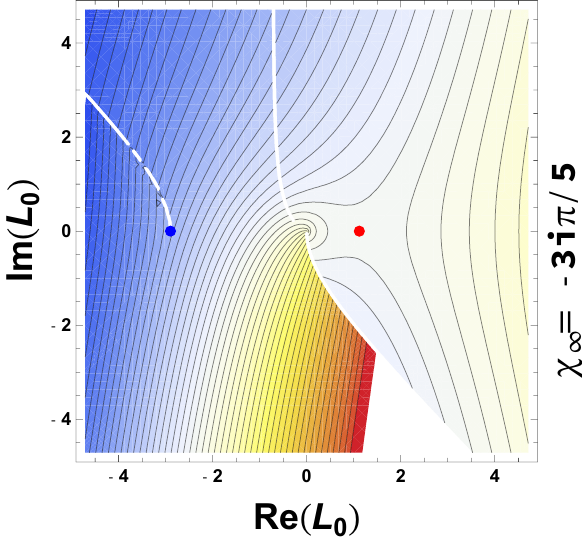}
            \end{minipage}
            \begin{minipage}{0.4\textwidth}
                \includegraphics[width=\linewidth]{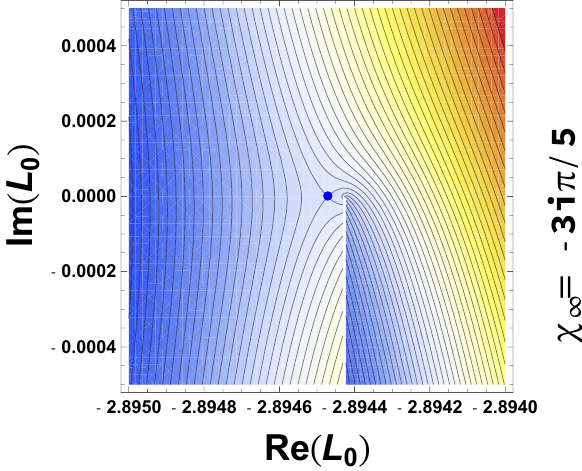}
            \end{minipage}

\caption{Contour plots of $\text{Re}(iS)$ in the complex $L_0$ plane. In all panels we have taken $\ell =1$, $8\pi\Gn=1$, $\lambda = 10^{-2^9}$, and $\beta=2\pi$. The plots on the left have $L_\gamma = 2^5$, while $L_\gamma =2^8$ in the plots on the right. The top panels have real $\chi_\infty =1/2$ while the bottom two have imaginary $\chi_\infty = -3i\pi/5$. The red point denotes the stationary point discussed in the main text, while the blue point  denotes the numerically determined root of $\alpha +L_0\frac{d\alpha}{dL_0}$. In the top left panel one can see from the contours that the blue point does indeed lie close to, though slightly above, a second (spurious) saddle of the action near to the root.  As can be seen in the top right panel, the blue point and the spurious saddle coincide at large $L_\gamma$ as they should. Similarly, the bottom right panel shows that this second stationary point persists for complex values of $\chi_\infty$, although in that case the spurious saddle (blue) is well-separated from the $K\Big|_{\chi=0}=0$ saddle (red) as shown in the bottom left panel. Attentive readers will note that the branch cut around the spurious saddle in the bottom right panel is in a different orientation than in the panel to its left. This is because, while the location of the spurious saddle is marked in the left panel, it actually lies on a different sheet from the one shown there. We have thus moved the branch cut to uncover the relevant sheet in the right figure.  In this case, the two saddle points lie on entirely different sheets of the Riemann surface for  the action.}
\label{fig:SpurSaddles}
\end{figure}

Second, we can numerically look for zeros of 
$\alpha + L_0 \frac{\partial \alpha}{L_0}|_{L_\gamma,\chi_\infty}$.  For the parameters studied in sections \ref{ssec:SPCD} and \ref{ssec:AnalyticContinue}, we find no such zeros whose ascent curves can intersect our contour of integration; see figure \ref{fig:SpurSaddles}.

\section{An Alternate Parametrization for Constrained Axion Wormholes}
\label{sec:AlternateParamatrization}

In section \ref{sec:ConstrainedWormholes}, we took the area of the $\chi=0$ surface in our constrained wormholes to be $2 \alpha T \ell L_0$.  The factor of $\alpha$ in this expression led to certain simplifications in the analytic evaluation of the constrained wormhole action $S_{CWH}$, but it also led to potential extra saddles in the $L_0$ integral as discussed in section \ref{app:spurious}.  While this was not a problem in principle, or even in practice, it may seem somewhat more natural to parameterize the constrained wormholes by replacing $L_0$ with ${\widetilde L}_0=\alpha L_0$.  In particular, if we fix ${\widetilde L}_0=\alpha L_0$ then the $S_\mu$ term \eqref{eq:Smu} qualifies as a minimally-coupled matter term in the sense of \cite{Dong_2020}, whence we can use the result derived there that the saddle-point value of $L_\gamma$ (at fixed $T,\tilde L_0$) is the value that sets $\alpha=1$.    

In any case, as a check on our work above we have also analyzed $Z^\text{\tiny c}[{\cal B}_T^\lambda \sqcup {\cal B}_T^\lambda]$ using this alternate approach.  The current appendix provides a brief summary of the results so obtained and, in particular, shows agreement with the results described in the main text.

\begin{comment}
As further motivation for this alternate approach we note that, after replacing $L_0$ by ${\widetilde L}_0/\alpha$,
the metric ansatz and the Lagrange multiplier term in the action become
    \begin{equation}
        ds^2=dr^2+\frac{\ell\widetilde{L}(r)}{\alpha}\big(d\theta^2-\alpha^2\sin^2(\theta)dt^2\big), \ \ \ {\rm and} 
    \end{equation}
        \begin{equation}
        \label{eq:Smualt}
            S_\mu =\frac{\mu}{16\pi\Gn}\Bigg[\int dr \int_{\Sigma_r}d^2x\,\sqrt{|h|}\,\frac{1}{2}\partial_r\sgn(\chi)-2\ell T\widetilde{L}_0\Bigg].
        \end{equation}
In particular, $\alpha$ no longer appears explicitly in the Lagrange multiplier term \eqref{eq:Smualt}. As a result, at fixed $T\neq 0, L_\gamma$ the action is stationary with respect to $\widetilde{L}_0$ {\it precisely} when $K_{r_0}$ vanishes; the potential extra stationary points that were discussed in appendix \ref{sec:alpha0} have now disappeared.   In addition, one correspondingly avoids the subtle issues discussed in section \ref{app:spurious}.  Using this alternate approach thus provides a useful check on the reasoning of sections \ref{ssec:SPCD} and \ref{ssec:AnalyticContinue}.  
\end{comment}

However, the above simplifications come with an associated cost.  In particular, we will no longer be able to obtain a closed form expression for the action in terms of $\widetilde{L}_0, T, L_\gamma$. While $r_0$ is still given by \eqref{eq:r0solve}, trying to solve for the axion flux $q$ in terms of $\widetilde{L}_0$ using boundary conditions and constraints leads to a transcendental equation which we can only solve numerically. This has the unfortunate side effect of removing explicit analytic control over various branch cuts in the on-shell action as a function of $\widetilde{L}_0$, and thus of complicating the analysis of various contour deformations in the $\widetilde{L}_0$ integral.   Nevertheless, with sufficient care one may simply treat such issues numerically as well.

\begin{figure}
            \centering
            \includegraphics[width=0.5\linewidth]{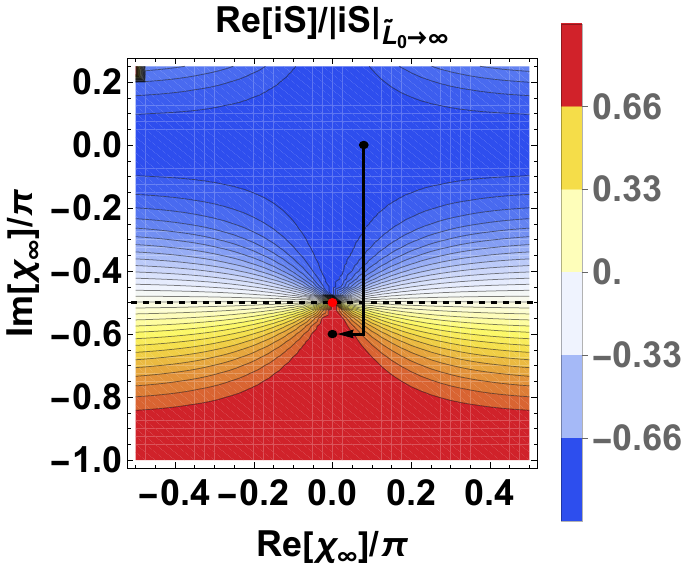}
            \caption{A numerically generated contour plot showing $\text{Re}(iS)$ normalized by $|iS|$ sampled at $\widetilde{L}_0=3\times10^4\approx\infty$ in the complex $\chi_\infty$ plane. The black line represents the path through the complex $\chi_\infty$ plane taken across the sequence of images in figure \ref{fig:NumActionProgression} below. We've taken $\ell=1,\,8\pi\Gn=1,\, \lambda=10^{-2^8},\,L_\gamma=2^8,\,\rm{and}\,\beta=2\pi $ for this plot.}
            \label{fig:NumActionAsymptotics}
        \end{figure}

We will shortly display results for various complex choices of the boundary condition $\chi_\infty$.  
 As in our original parameterization, one must deform the contour of integration as one changes $\chi_\infty$.  The need for this deformation is clear from the asymptotics of the on-shell action shown in figure \ref{fig:NumActionAsymptotics}.  This figure also shows the contour along which we will analytically continue $\chi_\infty$ (starting on the positive real axis).
        
 \begin{figure}[!t]
            \centering
            \begin{minipage}{0.34\textwidth}
\includegraphics[width=\linewidth]{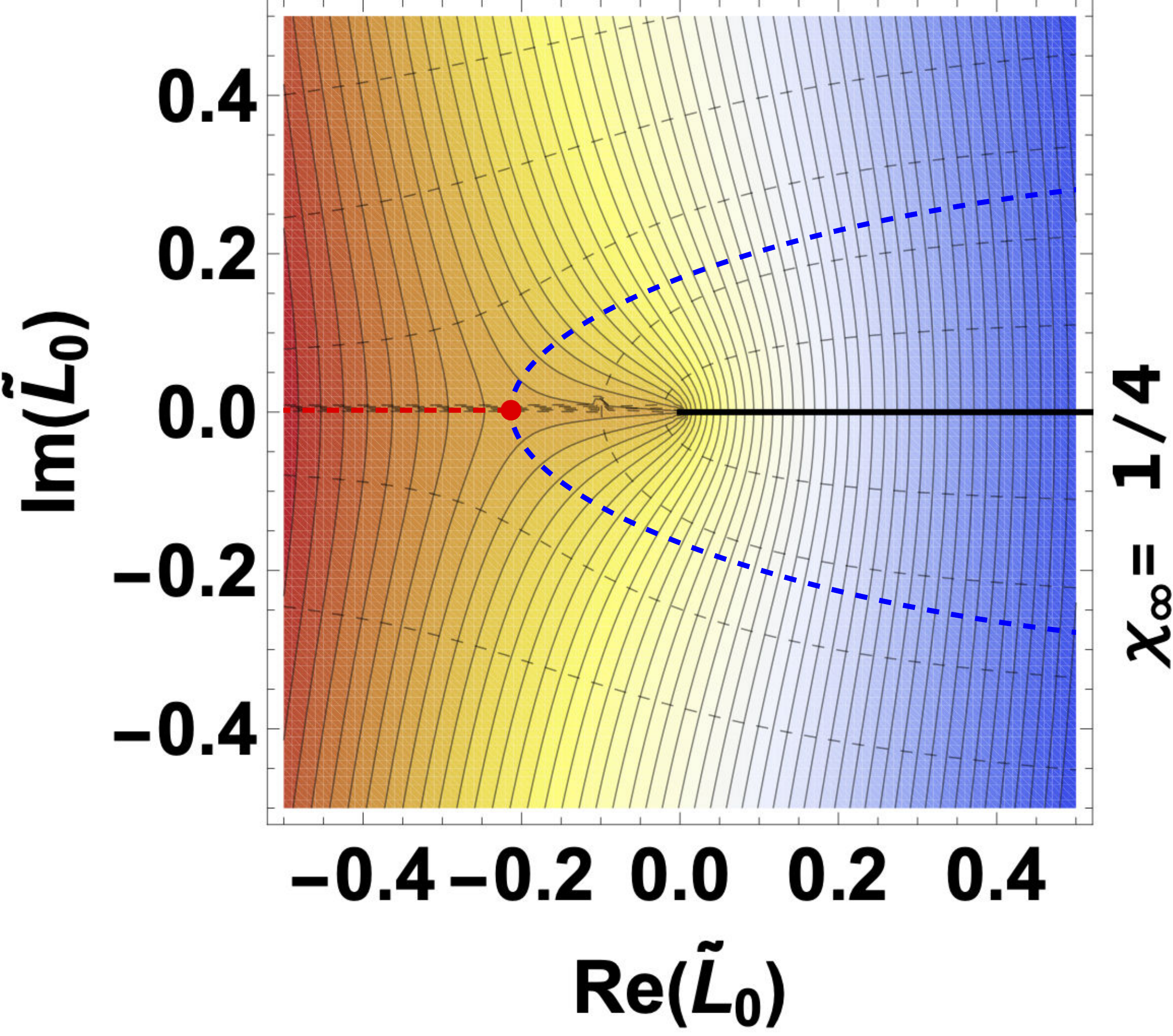}
            \end{minipage}
            \begin{minipage}{0.34\textwidth}
\includegraphics[width=\linewidth]{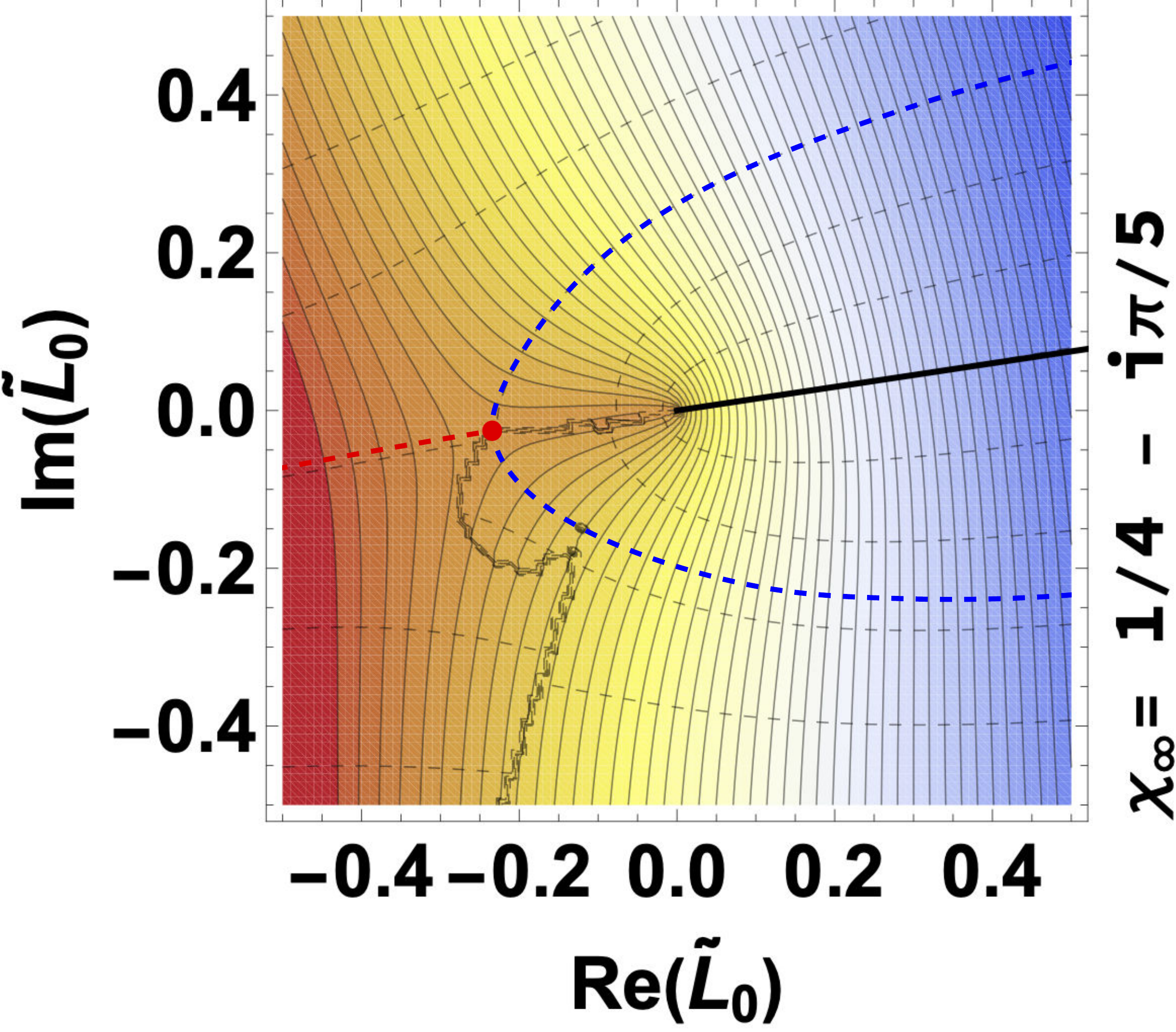}
            \end{minipage}
            \begin{minipage}{0.34\textwidth}    \includegraphics[width=\linewidth]{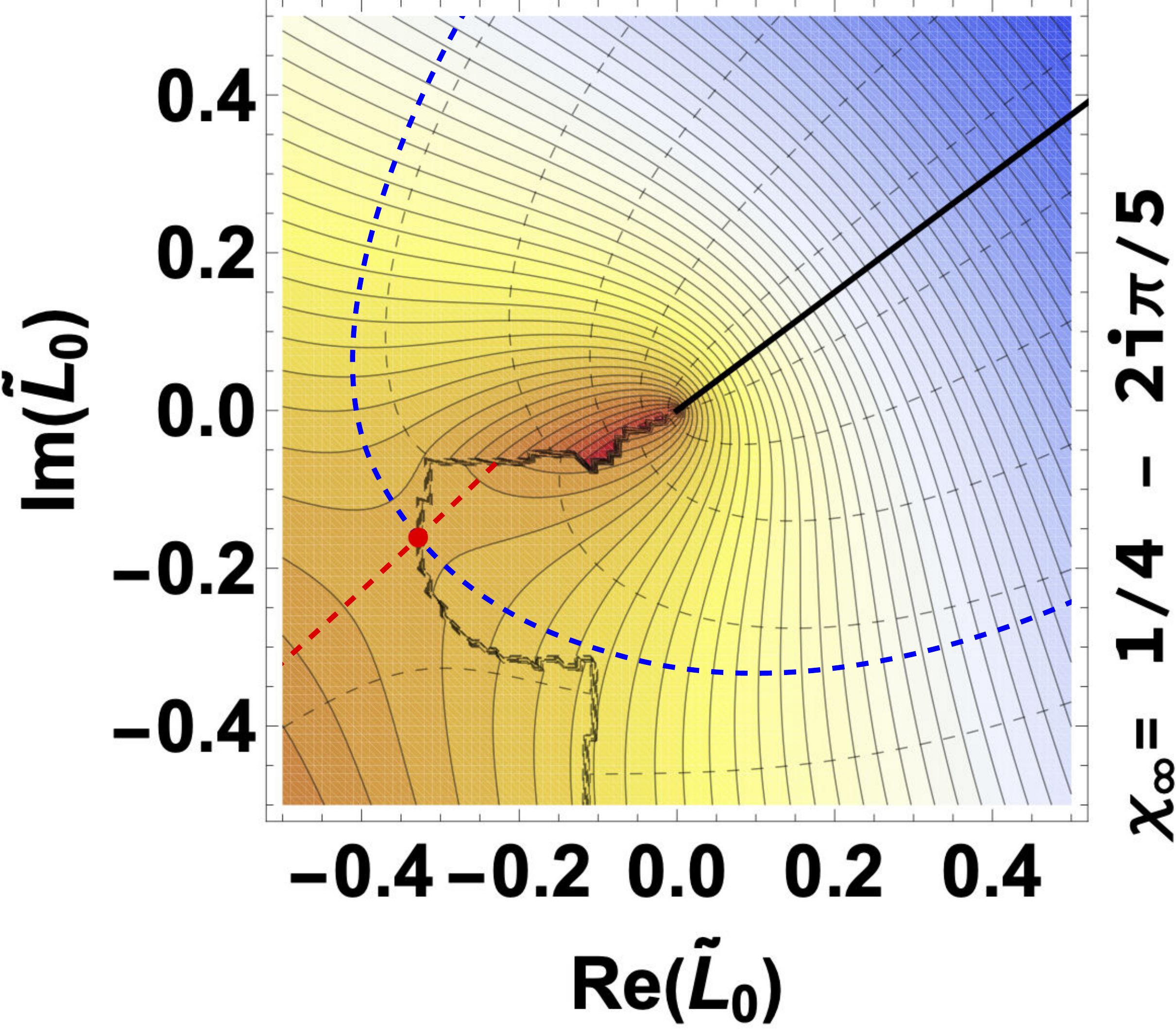}
            \end{minipage}
            \begin{minipage}{0.34\textwidth}
\includegraphics[width=\linewidth]{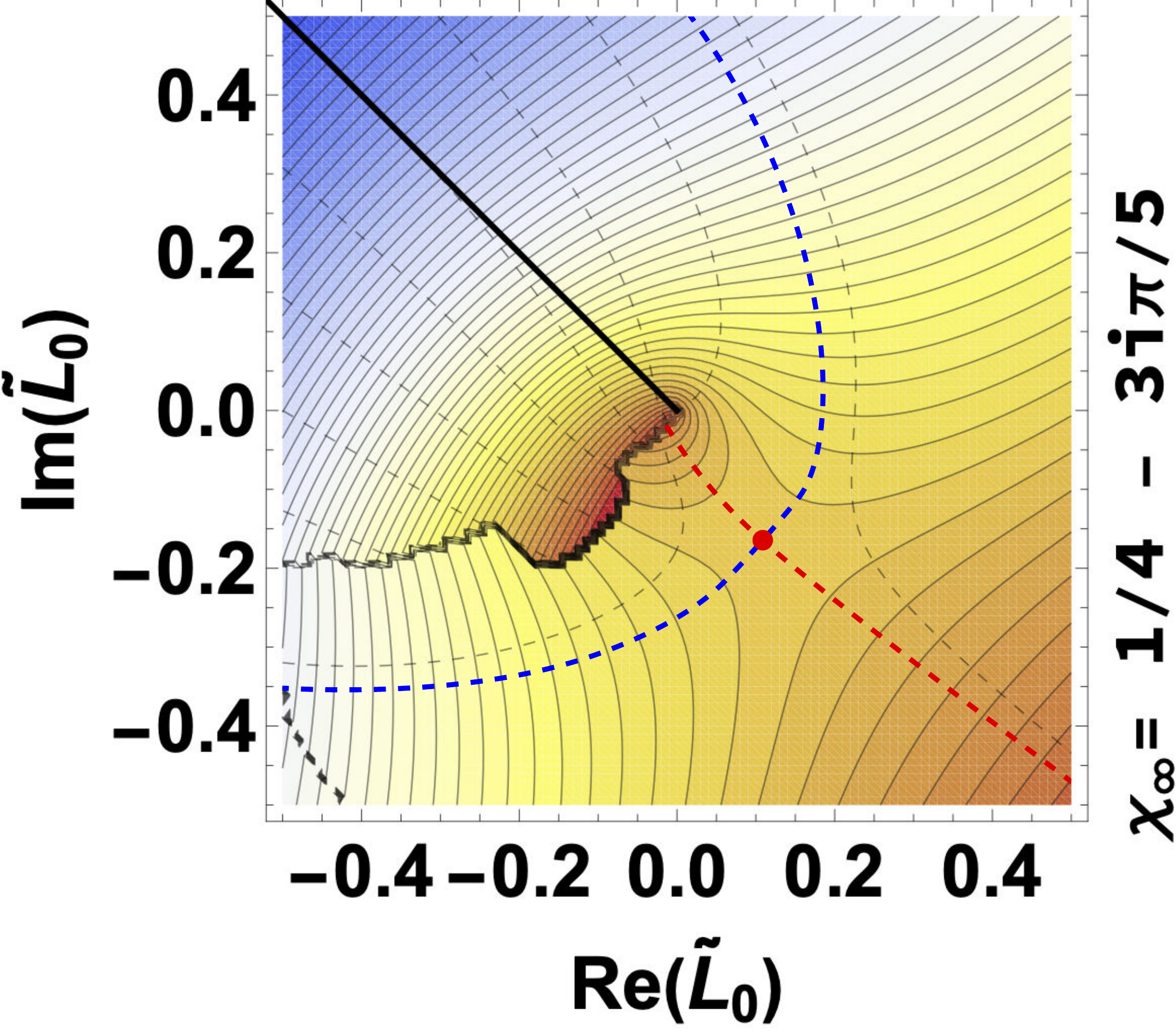}
            \end{minipage}
            \begin{minipage}{0.34\textwidth}\includegraphics[width=\linewidth]{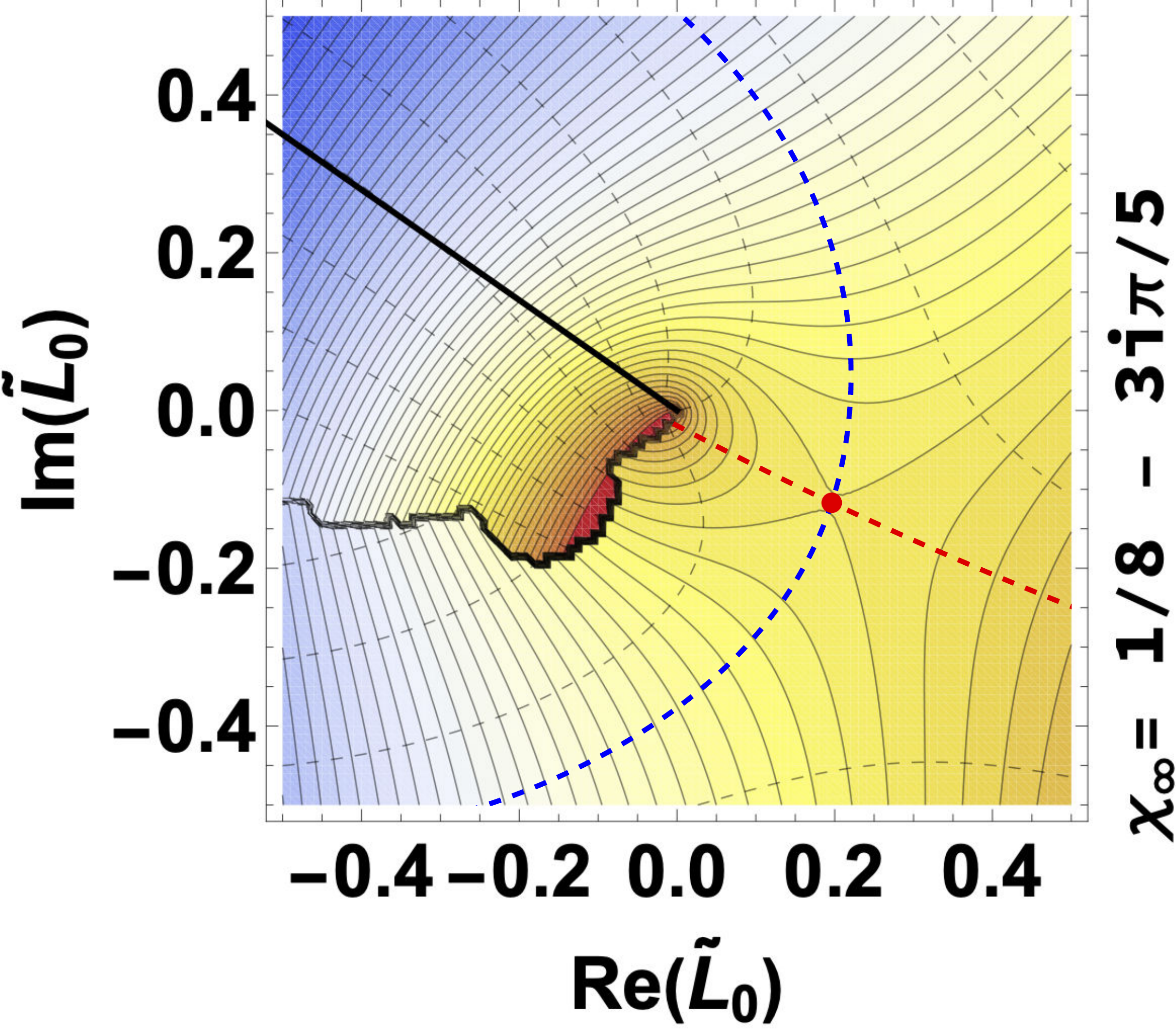}
            \end{minipage}  
            \begin{minipage}{0.34\textwidth}\includegraphics[width=\linewidth]{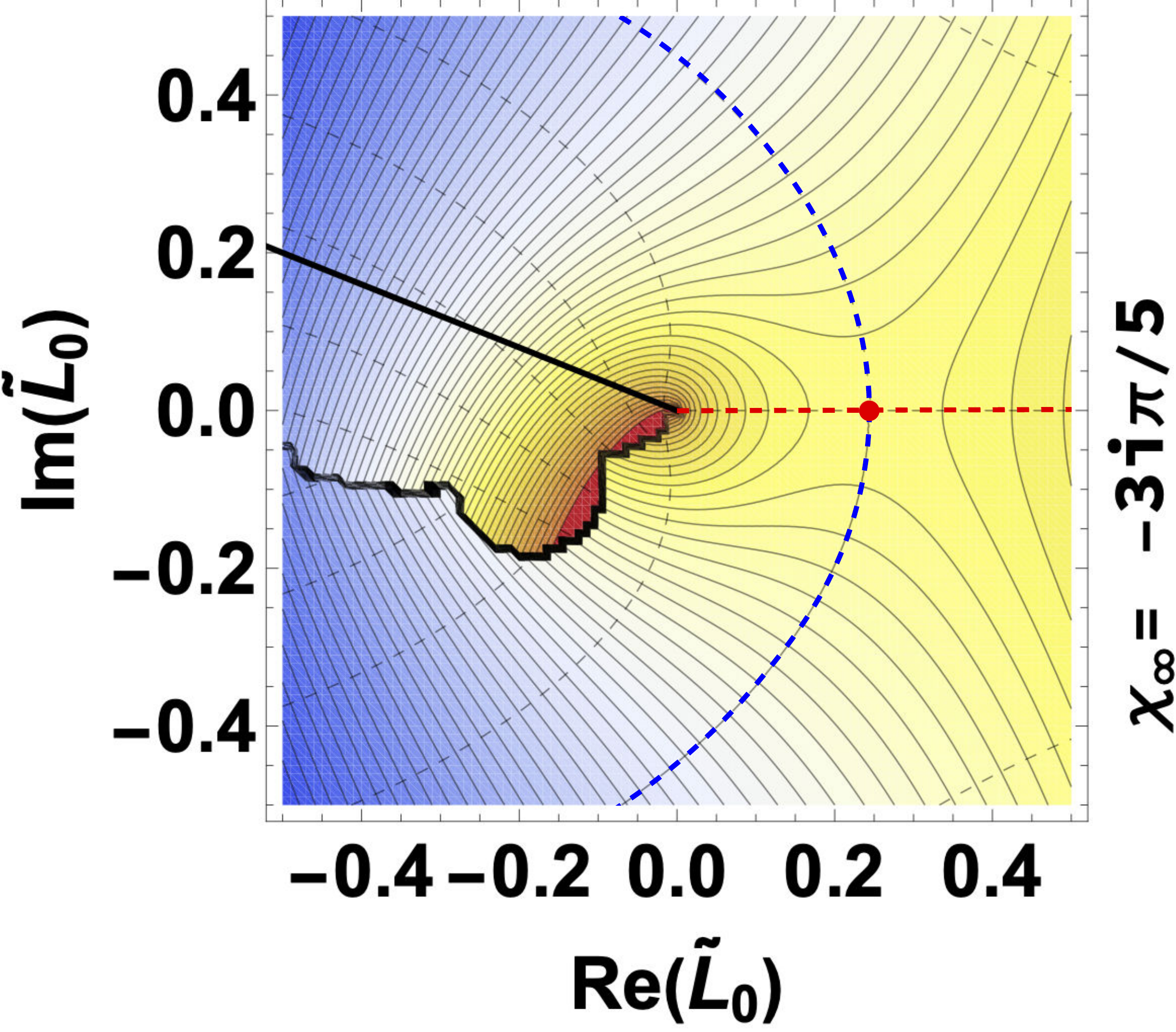}
            \end{minipage}  
            \caption{A sequence of numerically generated contour plots showing $\text{Re}(iS)$ in the complex $\widetilde{L}_0$ plane for various complex values of $\chi_\infty$. In all of the above we have taken $\ell=1,\,8\pi\Gn=1,\, \lambda=10^{-2^8},\,L_\gamma=2^8,\,\beta=2\pi $. The stationary point (which we have verified numerically to be associated with $K|_{\chi=0}=0$) is denoted by the red dot. The dotted lines represent contours of constant $\text{Im}(iS)$, and the grey (and sometimes erratic-looking) curve at which many such contours coalesce represents a branch cut in the on-shell action. The solid black line in each panel represents an appropriate choice of integration contour. Warmer/cooler colors indicate larger/smaller values of $\text{Re}(iS)$. Since steepest ascent/descent curves (dashed red/blue lines) have constant phase, they have been sketched so as to run parallel to the other dashed curves and orthogonal to the solid contours. Note that, in the first two panels, ascent contours running from the saddle toward the right disappear quickly behind the branch cut.}
            \label{fig:NumActionProgression}
        \end{figure}

 Our results are summarized in the six panels of figure \ref{fig:NumActionProgression}, where each panel corresponds to a different value of $\chi_\infty$ along the contour shown in figure \ref{fig:NumActionAsymptotics}. We see that for each value of $\chi_\infty$ there is a stationary point in the complex $\widetilde{L}_0$ plane (and at which we have checked numerically that $K|_{\chi=0}=0$). This stationary point lies on the negative-real $\widetilde{L}_0$-axis for real $\chi_\infty$, though it moves to the positive-real $\widetilde{L}_0$-axis as we move to negative-imaginary $\chi_\infty$.  
%However, unlike in our original parametrization, there is no second stationary point associated with $\alpha=0$ restores the convergence of the integral. 
However, the integral over the positive-real $\widetilde{L}_0$ contour diverges whenever  ${\rm Im} \ \chi_\infty < -\pi/2$. This can again be seen in figure \ref{fig:NumActionAsymptotics}. 
  
As for the parameterization studied in the main text, at each stage of the analytic continuation in $\chi_\infty$ shown in figure \ref{fig:NumActionProgression} we conclude that the stationary point either does not contribute to the integral (again employing appropriately distorted contours of integration to maintain convergence of the integral), or that it lies on a Stokes' ray (so that it is a choice whether to take it to contribute). In the first three panels, this can again be seen by noting in each case that the magnitude of the integrand at the stationary point exceeds the magnitude at the integrand at any point along the integration contour (so that it cannot possibly contribute).  In contrast, in the fourth and fifth panels, the magnitude at the saddle has become smaller than the magnitude at the endpoint $\widetilde{L}_0=0$, but the ascent contour from the saddle still does not intersect our contour of integration.

Finally, in the 6th panel the ascent contour runs directly along the positive-real $\widetilde{L}_0$-axis to arrive at the endpoint of our integration contour.  In this case, as with the original parameterization of section \ref{ssec:AnalyticContinue}, we find that the system lies on a Stokes' ray.  It is thus again a matter of choice whether one takes the saddle to contribute in this case and, if one does so,  the contribution is again subleading relative to the contribution from the UV-sensitive endpoint at $\widetilde{L}_0=0$.  

\begin{figure}
            \centering
            \includegraphics[width=0.5\linewidth]{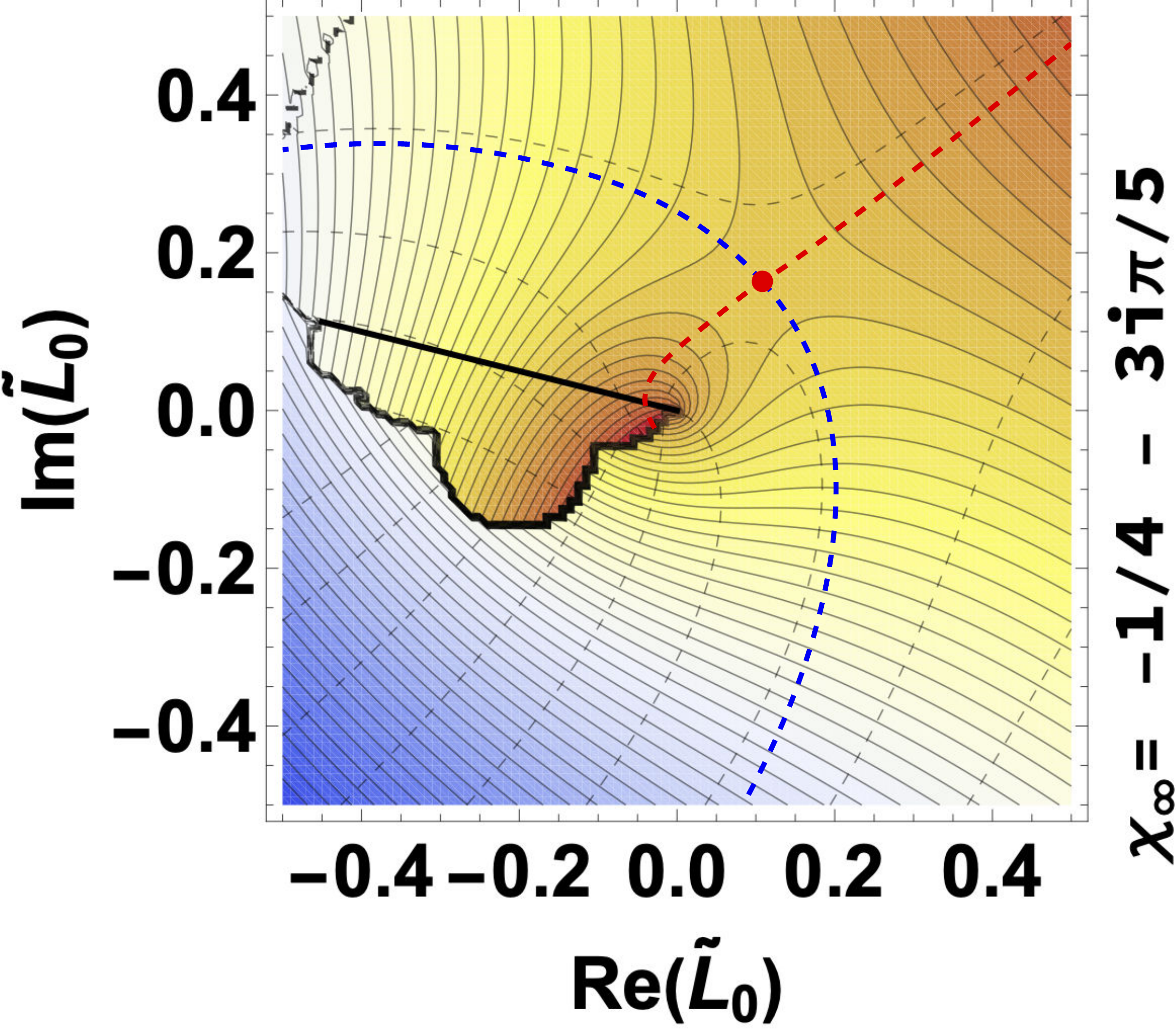}
            \caption{The analogue of the plots in figure \ref{fig:NumActionProgression} for the case $\chi_\infty =-1/4-3\, i \pi/5$. Again, steepest ascent/descent curves (dashed red/blue lines) associated with the saddle point have been sketched so as to run parallel to the other dashed curves and orthogonal to the solid contours. In fact, the steepest ascent curve (dashed red line) runs between two dashed curves that intersect the contour of integration (straight black line running from origin toward upper left). It is thus unambiguous that the steepest ascent contour intersects the contour of integration. In particular, this intersection is close to the origin.  This indicates that the stationary point contributes to the path integral although it remains subdominant to the endpoint contribution in this regime.}
            \label{fig:LargeLambda}
        \end{figure}

As for the parameterization used in the main text, further analytic continuation to $\chi_\infty$ with negative real part yields a case where the saddle unambiguously contributes; see figure \ref{fig:LargeLambda}.  Based on the results of section \ref{ssec:AnalyticContinue}, we expect that further analytic continuation would in fact lead to situations where the wormhole saddle dominates the integral.  However, the numerics is rather complicated and our implementation becomes unstable before we can convincingly show this directly.

\section{Constrained Axion Wormholes in Euclidean Gravity}
\label{sec:EuclideanCWormholes}

The main body of this paper, we outlined a computation of the connected partition function $Z^\text{\tiny c}[{\cal B}_T^\lambda \sqcup {\cal B}_T^\lambda]$ in Lorentzian 2+1 dimensional axion gravity with Dirichlet boundary conditions.  The computation makes critical use of constrained wormholes and carefully considered the impacts of analytically continuing the axion boundary value $\chi_\infty$ from real to imaginary values (and to more general complex values).  

Historically, studies of axion wormholes have taken Euclidean signature formulations as a 
starting point.  As we have seen, they also work in terms of the parameter $\chi_{E,\infty}=i\chi_\infty.$  If one does not attempt to analytically continue from a formulation at real $\chi_\infty$ based on the real Lorentzian contour, one will not be able to track the deformations of the contour of integration shown in figure \ref{fig:ActionProgression} that move it away from the real $L_0$ axis.  However, an attempt to analyze the problem directly in Euclidean signature will still uncover related features.  In particular, for real $\chi_{E,\infty}$ the final panel of \ref{fig:ActionProgression} indicates that the saddle is a local {\it minimum} of $e^{-S_E}$ along the real $L_0$ axis, and also that the integral over real $L_0$ will diverge.  

\begin{figure}
    \centering
\includegraphics[width=0.6\linewidth]{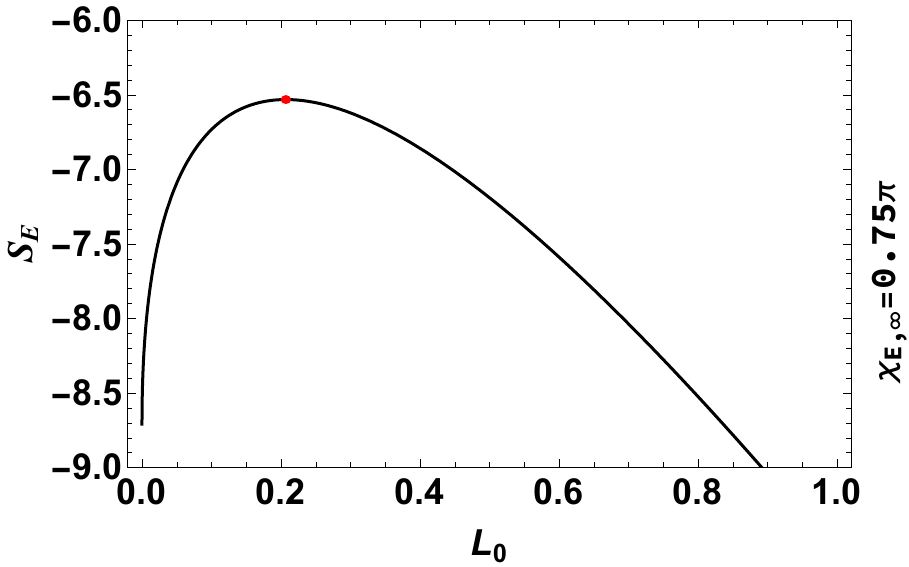}
    \caption{ The Euclidean action $S_E$ along the real positive-$L_0$ contour with $\ell=8\pi\Gn=1$ and  $\alpha=1$ for real $\chi_{E, \infty} = \frac{3\pi}{4}$. The red dot is the saddle point, corresponding to the on-shell Euclidean wormhole.  It is clearly an ``unstable'' point along this contour in the sense that it maximizes $S_E.$}   \label{fig:EuclideanWormholeAction}
\end{figure}

These results, as well as other results with $\beta=iT$, are clearly reproduced by a Euclidean analysis.  In particular, as in the main text, for $\beta=iT$ one finds that the integral  over positive $L_0, L_\gamma$ converges absolutely.  One is therefore free to perform the $L_\gamma$ integral first and, as discussed in section \ref{ssec:SPCD} and appendix \ref{app:spurious}, that integral is dominated by a saddle at which $\alpha=1$.  Using the associated saddle-point approximation, It is then particularly easy to study the remaining $L_0$ integral.  A plot of $S_E$ vs. $L_0$ for such saddles is shown in figure 
\ref{fig:EuclideanWormholeAction}.  The smooth Euclidean wormhole is the saddle-point with respect to $L_0$-variations.  It is clearly a local maximum and thus an ``unstable'' saddle in the classic sense of Euclidean gravity.  In particular, a direct Euclidean analysis of our constrained wormholes would certainly suggest that this saddle is not relevant.

It is notable that even this Euclidean analysis gives results that differ substantially from those of the existing literature on Euclidean axion wormholes \cite{Hertog:AxionWormholeUnstable, Loges:2022nuw, andriolo:Axion_wormholes_massive_dilaton,Loges:2023ypl,aguilar:Axion_dS_Wormholes,Hertog:Axion-Saxion_Wormholes_Stable,loveridge:Axion_Wormhole_Alternate_Topologies,Marolf:2025evo}.  However, as discussed in the introduction, those works considered asymptotically flat wormholes with Neumann boundary conditions for the axion, while we have studied Dirichlet axion boundary conditions and AdS asymptotics above.  It is therefore interesting to ask what one would find if one studies Neumann boundary conditions in AdS.

Unfortunately, the Neumann analogues of the computations given in the main text are far from straightforward and, in fact, some steps are likely ill-defined.  First, while Lorentzian constrained wormholes with fixed real charge $q$ will certainly exist, the boundary-conformal-transformations of section \ref{sssec:singreg} will not suffice to find the desired family of solutions with general $L_0$.  The issue is simply that while the Dirichlet boundary data $\chi_\infty$ is invariant under Weyl rescalings of the boundary metric, the Neumann data is not. In particular, the Dirichlet data $\chi_\infty$ has conformal dimension zero, but the associated Neumann boundary data has conformal dimension equal to the dimension of the boundary spacetime; i.e., $\Delta = d=2$ for wormholes locally asymptotic to AdS$_3$.  Thus the Weyl rescaling of the boundary in section \ref{sssec:singreg} maps an $\alpha=1$ wormhole satisfying our boundary conditions to an $\alpha\neq 1$ wormhole satisfying {\it different} boundary conditions.  Thus, for Neumann boundary conditions on the axion, the AdS$_3$ case is not in fact simpler than the case of general dimensions.

Furthermore, as noted in the introduction, Neumann boundary conditions for the axion at an AdS boundary are associated with ghost-modes \cite{andrade:BeyondUnitarityBound} which have a negative kinetic term (in standard Lorentz-signature conventions).  For real $q$ we thus expect such modes to make the energy unbounded below, forbidding the sort of argument used in section \ref{sec:Background} and presumably making our partition function fail to converge.

Nevertheless, since such ghost modes require non-zero derivatives along the boundary \cite{andrade:BeyondUnitarityBound}, they will be invisible if we restrict to asymptotically AdS$_{d+1}$ spacetimes with $SO(d+1)$ symmetry.  If we were to ignore the likely failures of the $T$ and $L_\gamma$ integrals to converge, we might then proceed as described for the Dirichlet case above and for real $q_E$ (imaginary $q$) we could study the Euclidean action of $SO(d+1)$-invariant constrained wormholes in which the value of $L_0$ has been fixed.

For the Neumann case,  we will fix $\partial_r \chi_E$ at each of the two asymptotic boundaries. Since we now study wormholes that preserve SO($d+1$), This amounts precisely to fixing the Euclidean axion flux $q_E$.
In order to ensure we have a good variational principle for this calculation, we must modify our boundary terms in the action \eqref{eq:ESbndy} by adding an additional term such that $\frac{\delta S_E}{\delta  \chi}$ vanishes at the boundary. The full set of boundary terms appropriate to this case is thus

    \begin{equation}
        S_{\tiny E, \partial\M} = \frac{1}{8\pi\Gn}\int_{\tiny \partial\M}d^2x\,\sqrt{|h|}\Big(\,K -\frac{1}{2}\chi\partial_r\chi-\frac{1}{\ell} + \frac{\delta^2}{2\ell}\log(\delta)\mathcal{R}^{(2)}\Big).
        \label{eq:ESbndyNeumann}
    \end{equation}
    
To construct constrained wormholes, we can add the Euclidean version of the constraint \eqref{eq:Smu} to the action \eqref{eq:EAction1}. Just as in the Lorentzian, calculation we can find  stationary points of this constrained action by cutting and pasting together different solutions to the unconstrained equations of motion, ordinary Euclidean axion wormholes such as those described in section \ref{ssec:EOSWormholes}. The key difference is that we will not fix the value of the axion at the boundary, rather the axion flux $q$.

In order to solve the constraint, the Euclidean wormholes we cut and paste together must be such that $\chi_E$ vanishes on a surface such that $L(r)$ given by equation \eqref{LEuclidean} takes the value $L_0$. Of course, with Neumann boundary conditions, we can freely shift the value of the axion by any desired constant.  Furthermore,  on any unconstrained solution the function $L(r)$ takes all values greater than the throat size of the unconstrained saddle:
\begin{equation}L_{0,min}=\frac{\ell}{2}\Big(\sqrt{1+q_E^2/\ell^2}-1\Big)
\label{eq:L0min}
\end{equation}
and, in fact, it does so \emph{twice} on any given solution. However it is never smaller than \eqref{eq:L0min}.

\begin{figure}[h!]
    \centering
    \includegraphics[width=0.6\linewidth]{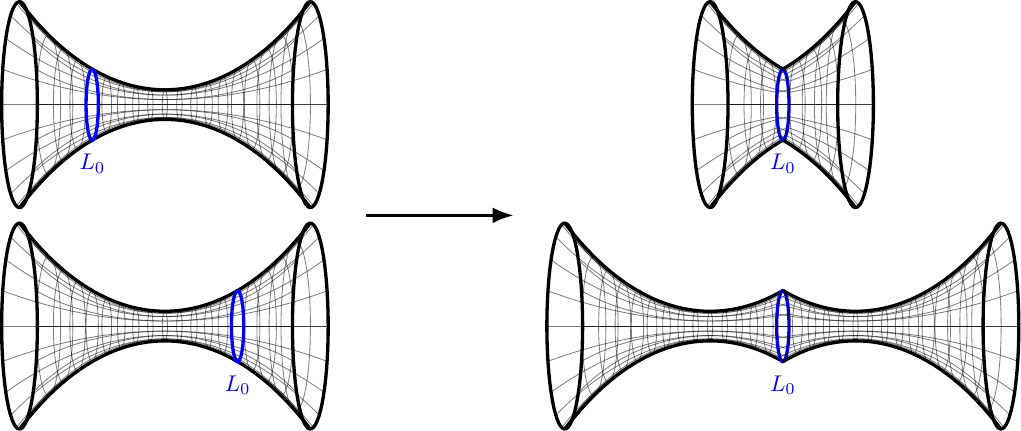}
    \caption{At fixed $q_E$, our cut-and-paste generates both `short' (top at right) and `long' (bottom at right) constrained wormholes.  In the former, $L_0$ is the minimum value of $L(r)$, while it is a local maximum in the latter.}
    \label{fig:NeumannConstrainedWormholes}
\end{figure}

Due to this fact, for $L_0\ge L_{0,min}$,  there are actually two obvious solutions\footnote{\label{foot:add}There are additional generalizations if we fix $L$ at more than one surface, but these are the only saddles we have identified with a single constrained-$L$ surface.} to the constrained equations of motion which we can obtain via cut and paste procedures on smooth Euclidean wormholes, one short and one long. The construction of these two solutions follows very closely the construction of the Lorentzian axion wormholes described in the main text and is depicted in figure \ref{fig:NeumannConstrainedWormholes}.  However, we note that since Lorentz-signature  wormholes have $L(r)$ monotonic in $r$, there is no analogue of the long wormholes in Lorentz signature (though see again footnote \ref{foot:add}).  One might be inspired by this observation to prefer the short wormholes despite the above-mentioned likely lack of a convergent Lorentzian treatment.  Indeed, the long wormholes can be obtained from the short wormholes by 
analytically continuing the short wormhole action around the branch point at $L_{0,min}$.
it is tempting to regard the long wormhole branch as an artifact of analytically continuing around the branch point at $L_{0,min}$, and is thus tempting to regard them as simply an artifact of this analytic continuation.  One might therefore expect them to lie on a different sheet of the Riemann surface from the sheet that  contains the desired contour of integration.

\begin{figure}[h!]
    \centering
    \includegraphics[width=0.6\linewidth]{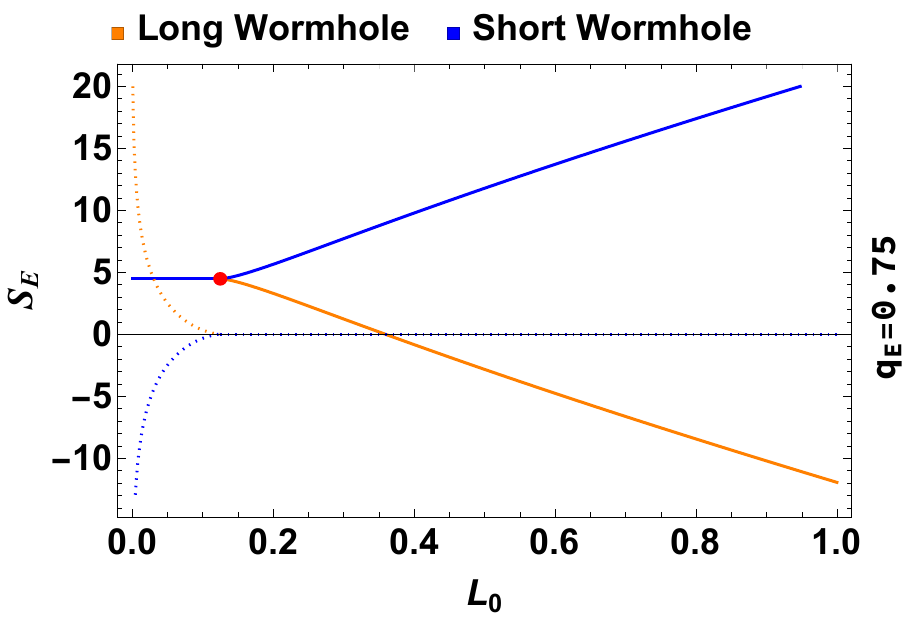}
    \caption{The real (solid lines) and imaginary (dotted lines) parts of $S_E$ for long (orange) and short (blue) constrained wormholes.  The long and short branches merge at $L_0=L_{0,min}$ (red dot), which is the smallest $L_0$ for which our Euclidean constrained wormholes are real.  The action thus develops an imaginary part at smaller values of $L_0$.  The long and short wormhole actions are different branches of the same analytic function, and are related by analytic continuation around the branch point at $L_{0,min}.$  For $L_0 \le L_{0,min}$ the real parts of the long and short wormhole actions coincide, though only the blue line is visible in that regime as it exactly covers the corresponding orange line.}
    \label{fig:NeumannEWHAction}
\end{figure}

The Euclidean action of our constrained wormholes is plotted for different values of $L_0$ in figure \ref{fig:NeumannEWHAction}.  Here we have also analytically-extended the function to values of $L_0$ smaller than the value $L_{0,min}$. For such $L_0< L_{0,min}$ the action becomes complex.

While the on shell wormhole at $L_0=L_{0,min}$ is a stationary point for both the long and short wormhole branches, the behavior of the action at larger $L_0$ is quite different between the two families of constrained solutions. 
The action of the long wormholes decreases monotonically for $L_0$ larger than the saddle point value, while the action of the short wormholes increases.  This suggests that the path integral restricted over short constrained wormholes is dominated by the value at the saddle point given by the usual Euclidean wormhole while the path integral over the Long wormholes will, in-fact, diverge. 

The behavior of the short wormhole branch is thus very different from that of the Dirichlet case and is more in line with what was found in \cite{Hertog:AxionWormholeUnstable, Loges:2022nuw, andriolo:Axion_wormholes_massive_dilaton,Loges:2023ypl,aguilar:Axion_dS_Wormholes,Hertog:Axion-Saxion_Wormholes_Stable,loveridge:Axion_Wormhole_Alternate_Topologies,Marolf:2025evo}.  This supports the idea that the difference in boundary conditions played a large role in our finding very different results in the main text.  Nevertheless, we caution the reader that direct comparison with \cite{Hertog:AxionWormholeUnstable, Loges:2022nuw, andriolo:Axion_wormholes_massive_dilaton,Loges:2023ypl,aguilar:Axion_dS_Wormholes,Hertog:Axion-Saxion_Wormholes_Stable,loveridge:Axion_Wormhole_Alternate_Topologies,Marolf:2025evo} remains difficult as the approach used in those works claims that there are simply no deformations of the unconstrained wormhole that maintain SO($d+1$) symmetry and which one should analyze.  In particular, all of our constrained wormholes are excluded (whether short or long).  It thus appears that the approach used in those references must be associated with a very different contour, though it would be useful to better understand this difference in the future.

\bibliographystyle{jhep}
\bibliography{ref}

\end{document}